\documentclass[]{aa}
\usepackage{amsmath}
\usepackage{longtable}
\usepackage{natbib}
\usepackage{graphicx}
\usepackage{txfonts}
\bibpunct{(}{)}{;}{a}{}{,}

\newcommand{\hipp}{{\em Hipparcos}}
\newcommand{\sco}{\object{Sco~OB2}}
\newcommand{\ob}{OB~association}
\newcommand{\obs}{OB~associations}
\newcommand{\msun}{~{\rm M}_\odot}
\newcommand{\rsun}{~{\rm R}_\odot}
\newcommand{\rverr}{\epsilon_{\rm RV}}
\newcommand{\sigmaoned}{\sigma_{v,{\rm 1D}}}
\newcommand{\sigmathreed}{\sigma_{v,{\rm 3D}}}
\newcommand{\binfrac}{F_{\rm M}}
\newcommand{\binfracm}{F_{\rm M}(M_1)}
\newcommand{\binfracobs}{\tilde{F}_{\rm M}}
\newcommand{\binfracobsm}{\tilde{F}_{\rm M}(M_1)}
\newcommand{\frhoobs}{\tilde{f}_\rho(\rho)}
\newcommand{\fqobs}{\tilde{f}_q(q)}
\newcommand{\amin}{a_{\rm min}}
\newcommand{\amax}{a_{\rm max}}
\newcommand{\pmin}{P_{\rm min}}
\newcommand{\pmax}{P_{\rm max}}
\newcommand{\kms}{km\,s$^{-1}$}

\begin{document}

\renewcommand{\topfraction}{1.00}
\renewcommand{\floatpagefraction}{1.00}
\renewcommand{\textfraction}{0.00}
\renewcommand{\dbltopfraction}{1.00}
\renewcommand{\dblfloatpagefraction}{1.00}

\title{The primordial binary population II: Recovering the binary population for intermediate mass stars in \sco{}}

 \author{M.B.N. Kouwenhoven\inst{1,2}
          \and
          A.G.A. Brown\inst{3}
          \and
          S.F. Portegies Zwart\inst{2,4}
          \and
          L. Kaper\inst{2}
          }

 \offprints{M.B.N. Kouwenhoven \email{t.kouwenhoven@sheffield.ac.uk}}

 \institute{
   Department of Physics and Astronomy, University of Sheffield,
   Hicks Building, Hounsfield Road, Sheffield S3~7RH, United Kingdom
   \\\email{t.kouwenhoven@sheffield.ac.uk}
   \and
   Astronomical Institute Anton Pannekoek,
   University of Amsterdam,
   Kruislaan 403, 1098 SJ Amsterdam, The Netherlands
   \\\email{lexk@science.uva.nl}
   \and
   Leiden Observatory, University of Leiden,
   P.O. Box 9513, 2300 RA
   Leiden, The Netherlands \\\email{brown@strw.leidenuniv.nl}
   \and
   Section Computer Science, University of Amsterdam,
   Kruislaan 403, 1098 SJ Amsterdam, The Netherlands
   \\ \email{spz@science.uva.nl} }

 \date{Received ---; accepted ---}

 \authorrunning{Kouwenhoven et al.}

 \titlerunning{Recovering the binary population for intermediate mass stars in \sco{}}

% ====================================================================
% ====================================================================
% ====================================================================
% ==ABSTRACT==========================================================
% ====================================================================
% ====================================================================
% ====================================================================

\abstract{

We characterize the binary population in the young and nearby \ob{} \object{Scorpius~OB2} (\sco{}) using available observations of visual, spectroscopic, and astrometric binaries with intermediate-mass primaries. We take into account observational biases by comparing the observations with simulated observations of model associations. The available data indicate a large binary fraction ($> 70\%$ with $3\sigma$ confidence), with a large probability that all intermediate mass stars in \sco{} are part of a binary system. The binary systems have a mass ratio distribution of the form $f_q(q) \propto q^{\gamma_q}$, with $\gamma_q \approx -0.4$. \sco{} has a semi-major axis distribution of the form $f_a(a) \propto a^{\gamma_a}$ with $\gamma_a \approx -1.0$ (\"{O}pik's law), in the range $5\rsun \la a \la 5\times 10^6 \rsun$. The log-normal period distribution of \cite{duquennoy1991} results in too few spectroscopic binaries, even if the model binary fraction is 100\%. \sco{} is a young association with a low stellar density; its current population is expected to be very similar to the primordial population. The fact that practically all stars in \sco{} are part of a binary (or multiple) system demonstrates that multiplicity is a fundamental factor in the star formation process, at least for intermediate mass stars.


}

\maketitle

% ====================================================================
% ====================================================================
% ====================================================================
% ==INTRODUCTION======================================================
% ====================================================================
% ====================================================================
% ====================================================================

\section{Introduction}

Over the past decades observations have indicated that a large fraction of stars are part of a binary or multiple system. Apparently, multiplicity is an important aspect of the star formation process. Binaries also play a vital role in explaining many spectacular phenomena in astrophysics, e.g. supernovae type Ia \citep{yungelson1998,hillebrandt2000},
%supernovae type Ib/c and blue-type supernovae \citep{podsiadlowski1992}, 
short and long gamma-ray bursts \citep{fryer1999}, OB~runaway stars \citep{blaauw1961,hoogerwerf2001,gualandris2004}, and binary systems with compact remnants such as X-ray binaries, millisecond pulsars, and double neutron stars \citep{vandenheuvel1994,fryer1997}. Binary systems are also known to strongly affect the dynamical evolution of dense stellar clusters \citep{hut1992,ecology4,ecology7}. 
This is an excellent motivation to characterize the outcome of the star forming process in terms of multiplicity and binary parameters.

In this paper we aim to recover the properties of the population of binaries that result from the formation process: the primordial binary population, which is defined as {\em the population of binaries as established just after the gas has been removed from the forming system, i.e., when the stars can no longer accrete gas from their surroundings} \citep{kouwenhoven2005}. 
The dynamical evolution of stars of a newly born stellar population is influenced by the presence of gas. After the gas has been removed $1-2$~Myr after the formation of the first massive stars, the binary population is only affected by stellar evolution and pure N-body dynamics. From a numerical point of view, the primordial binary population can be considered as a boundary between hydrodynamical simulations and N-body simulations. Hydrodynamical simulations of a contracting gas cloud \citep[e.g.,][]{bate2003,martel2006} produce stars. After the gas is removed by accretion and the stellar winds of the most massive O~stars, pure N-body simulations \citep[e.g.,][]{ecology4,kroupa1999,ecology7} can be used to study the subsequent evolution of star clusters and the binary population.

\obs{} are well suited for studying the primordial binary population. They are young, with ages ranging from $\sim 2$~Myr \citep[Ori~OB1b, Cyg~OB2; see][]{brown1999,hanson2003} to $\sim 50$~Myr \citep[$\alpha$~Persei, Cep~OB6; see][]{dezeeuw1999}. Among \obs{} younger than $\la 20$~Myr only a handful of the most massive systems have changed due to stellar evolution. The effects of dynamical evolution are expected to be limited due to their young age and low stellar density ($< 0.1 \msun\,{\rm pc}^{-3}$). Moreover, OB associations cover the full range of stellar masses \citep[e.g.][]{brown2001}, in contrast to the T~associations, their low-mass counterparts.

In this paper we focus on \object{Scorpius~Centaurus} (\sco{}), the nearest young \ob{}, and thus a prime candidate for studying the binary population. The proximity of \sco{} ($118-145$~pc) facilitates observations, and the young age ($5-20$~Myr) ensures that dynamical evolution has not significantly altered the primordial binary population since the moment of gas removal. 
The membership and stellar content of the association were established by \cite{dezeeuw1999} using \hipp{} parallaxes and proper motions, and its binary population is relatively well-studied.

Due to selection effects, it is not possible to observe the binary population in \sco{} directly. The dataset is hampered by the selection of the targets and instrumental limits on the observable range in semi-major axis, period, eccentricity and mass ratio. The fact that the observed binary population is biased makes it difficult to draw conclusions about the true binary population. However, by using the method of simulating observations of modeled stellar populations \citep[e.g.][]{kouwenhoventhesis}, it is possible to put constraints on the binary population. We accurately model the selection effects of the six major binarity surveys of \sco{}, and compare simulated observations with the true observations, to determine the properties of the current binary population in \sco{}. 

The organization of this paper is as follows.
In \S~\ref{section: true_method} we briefly describe the method and terminology that we use to recover the true binary population.
In \S~\ref{section: scoob2properties} we discuss the \sco{} association and membership issues.
In \S~\ref{section: true_observations} we describe the available datasets with visual, spectroscopic, and astrometric binaries in \sco{}, and outline our models for the respective selection effects.
In \S~\ref{section: true_pairingfunction}--\ref{section: true_eccentricitydistribution} we recover the mass ratio distribution and the semi-major axis distribution, and we constrain the eccentricity distribution for the binary population in \sco{}, respectively. Conclusions on the intrinsic binary fraction of \sco{} are drawn in \S~\ref{section: true_binaryfraction}. 
The possible differences between the current binary population and primordial binary population in \sco{} are discussed in \S~\ref{section: primordialbinarypopulation}.
In \S~\ref{section: true_discussion} we compare our results with those of others, and we discuss the validity of our assumptions. 
Finally, we summarize our main results in \S~\ref{section: summary}.

% ====================================================================
% ====================================================================
% ====================================================================
% ==INTRODUCTION======================================================
% ====================================================================
% ====================================================================
% ====================================================================

\section{Method and terminology} \label{section: true_method}

We recover the binary population in \sco{} from observations using the method of simulating observations of modeled stellar populations. This method is extensively described in \cite{kouwenhoventhesis}, and is briefly summarized below.

With increasing computer power, it has become possible to create sophisticated models of star clusters and \obs{}. One can compare these simulated associations with the observations of real associations in order to constrain the properties of the intrinsic binary population. However, this cannot be done directly, as the interpretation of the observational dataset is hampered by selection effects. Only a small (and biased) subset of the binary population is known. With the method of simulating observations of simulations (S.O.S.) one characterizes the selection effects, and applies these to the simulated association. The simulated observations that are then obtained can be compared directly with the real observations \citep[see, e.g.,][]{kouwenhoventhesis}.

In order to recover the binary population in \sco{} we simulate \ob{} models with different properties. We compare each model with observational data, by simulating observations for each major binarity survey. With this comparison we identify which association model is consistent with the observations, and thus constrain the binary population in \sco{}.
\cite{kouwenhoventhesis} shows that this is a safe method to derive the binary population. As long as the parameter space (of the binary population) is fully searched, and as long as the selection effects are well-modeled, this method allows recovery of the intrinsic binary population, as well as the uncertainties on each derived property. Furthermore, unlike the S.O.S. method used in this paper, the traditional method of correcting for selection effects (using a ``correction factor'') may lead to erroneous or unphysical results.

In this paper we make several assumptions when recovering the binary population in \sco{}. 
In our model we consider only single stars and binary systems; no higher order multiples are assumed to be present. In \S~\ref{section: triples} we will briefly return to the consequences of this assumption. We assume the distributions of the different observed parameters to be independent of each other:
\begin{equation} \label{equation: true_independence}
  \begin{tabular}{l}
    $f_{\rm BP}(M_1,M_2,a,e,i,\omega,\Omega,\mathcal{M})$  \\
    $\quad = \, f_{M_1,M_2}(M_1,M_2)\,f_a(a)\,f_e(e)\,f_i(i)\,f_\omega(\omega)\,f_\Omega(\Omega)\,f_\mathcal{M}(\mathcal{M})$ \,,  \\
    %$\quad \mbox{or}\ f_{M_1,M_2}(M_1,M_2)\,f_P(P)\,f_e(e)\,f_i(i)\,f_\omega(\omega)\,f_\Omega(\Omega)\,f_\mathcal{M}(\mathcal{M})$ \,, 
    \end{tabular}
\end{equation}
where $M_1$ and $M_2$ are the primary and companion mass, $a$ the semi-major axis, $e$ the eccentricity, $i$ the inclination, $\omega$ the argument of periastron, $\Omega$ the position angle of the line of nodes, and $\mathcal{M}$ the mean anomaly at some instant of time. Alternatively, one can replace $f_a(a)$ by the orbital period distribution $f_P(P)$.

In our models the overall binary fraction for the association can be described with a single number, independent of the primary mass $\binfracm \equiv \binfrac$. Observations of other stellar populations have suggested that the binary fraction may depend on primary mass or system mass, and that binary fraction tends to increase with increasing primary mass \citep[e.g.][]{preibisch1999newa,sterzik2004}, though a quantitative description for $\binfracm$ is still unavailable. In our models we therefore adopt a binary fraction independent of primary mass, so as to keep our description for the binary population in \sco{} as simple as possible. Note that \cite{kouwenhoventhesis} has shown that selection effects may introduce a trend between binary fraction and primary mass in the observations, even though such an intrinsic trend may not be present in reality.
However, our assumption does not influence the results significantly, as in this paper we only study the population of binary stars with an intermediate mass primary, mostly of spectral type B or A. As this corresponds to a small mass range, we neglect the possible correlation between binary fraction and primary mass (see \S~\ref{section: bf_vs_m1} for a further discussion).

For the same reason, we assume the semi-major axis $a$ and eccentricity $e$ to be independent of primary mass.
The independence of the semi-major axis $a$ with respect to the eccentricity $e$ may be a good approximation, as observations suggest that these parameters are only mildly correlated for solar-type stars in the solar neighbourhood \citep[e.g.,][]{duquennoy1991,heacox1997}; see also \S~\ref{section: true_eccentricitydistribution}. Note, however, that even in the case that this dependence is absent in the {\em intrinsic} population, a correlation may still be present in the observations due to selection effects.
We assume that the inclination $i$, the argument of periastron $\omega$, the position angle of the ascending node $\Omega$, and the mean anomaly ${\mathcal M}$ at some instant of time are independent of each other and of all other parameters. 
Finally, we assume that the binary systems have a random orientation in space (which is not necessarily implied by the previous assumptions). Even in the unlikely case that binary systems do not have a random orientation, the results do not change measurably \citep[see][]{kouwenhoventhesis}. Note that the primary and companion mass distributions are {\em never} independent, $f_{M_1,M_2}(M_1,M_2) \neq f_{M_1}(M_1)f_{M_2}(M_2)$, as by definition $M_1 \geq M_2$. 

For reasons of simplicity, we ignore the interaction between close binary stars; our models do not include Roche Lobe overflow or common envelope evolution. Low-mass contact binaries, such as cataclysmic variables, WUMa binaries, and symbiotic stars generally appear on a timescale which is significantly longer than the age of \sco{}. The higher-mass contact binaries, such as high-mass X-ray binaries, close binaries with mass reversal, and double pulsars could be present, or may have escaped the association as runaways. Due to the youth of \sco{}, a few of the closest binaries may have evolved into such objects. The non-inclusion of this close binaries, however, is unlikely to affect our conclusions on the {\em primordial} binary population, as we adopt a lower limit to the period of 12~hours (\S~\ref{section: recover_amin}). If these binaries are present, our inferred binary fraction (\S~\ref{section: true_binaryfraction}) may be slightly underestimated.

Throughout this paper we denote the (intrinsic) probability density function of a binary parameter $x$ as $f_x(x)$ and its cumulative distribution as $F_x(x)$. The corresponding {\em observed} distributions for a binarity survey are denoted as $\tilde{f}_x(x)$ and $\tilde{F}_x(x)$, respectively.

% ============================================================================
% ============================================================================
% ============================================================================
% ============================================================================

\section{The \sco{} association } \label{section: scoob2properties}

% ============================================================================
% ============================================================================
% ============================================================================
% ============================================================================

\begin{table*}
  \begin{tabular}{lc cc cc cccc ccc}
    \hline \hline
    Subgroup & $D$ & $R$ & Age & $A_V$ & $N_\star$ & $S_\star$ & $B_\star$ & $T_\star$ & $>3$ & $F_{\rm M,\star}$ & $F_{\rm NS,\star}$ & $F_{\rm C,\star}$ \\
          & pc  & pc  & Myr & mag \\
    \hline 
    \object{US}  &  145$^1$ & $\sim 20^5$  & 5$-$6$^{2,3}$ & 0.47$^5$  & 120$^1$ & 64  & 44 & 8  & 3 & 0.46 & 0.67 & 0.61 \\
    \object{UCL} &  140$^1$ & $\sim 35^5$  & 15$-$22$^4$   & 0.06$^5$  & 221$^1$ & 132 & 65 & 19 & 4 & 0.40 & 0.61 & 0.52\\
    \object{LCC} &  118$^1$ & $\sim 35^5$  & 17$-$23$^4$   & 0.05$^5$  & 180$^1$ & 112 & 57 & 9  & 1 & 0.37 & 0.56 & 0.44 \\
    \hline 
    Model& 130     & 20           & 5  & 0.00 & 9\,000 & varying & varying & 0  & 0 & varying & varying & varying \\
    \hline 
    \hline 
  \end{tabular}
  \caption{Properties of the subgroups \object{Upper Scorpius} (\object{US}), \object{Upper Centaurus Lupus} (\object{UCL}), and \object{Lower Centaurus Crux} (\object{LCC}) of \sco{}, and of our model for \sco{}. Columns 2--4 list for each subgroup its distance, effective radius, and age. Column~5 lists the median interstellar extinction towards each subgroup. Column~6 lists the number of confirmed \hipp{} members of each subgroup, and is followed by the {\em observed} number of singles, binaries, triples, and higher-order systems among the confirmed members in columns 7--10, taken from \cite{kouwenhoven2007a}. Finally, columns 11--13 list the observed multiplicity fraction, non-single star fraction, and companion star fraction among the confirmed members \citep[see][for a definition of these fractions]{kouwenhoven2005}. Note that the latter quantities are lower limits due to the presence of unresolved binary and multiple systems. In the bottom row we list the properties of our \sco{} model. The number of systems $N=S+B$ (i.e., singles and binaries) used in our model includes substellar objects with masses down to $0.02 \msun$. References: (1) \cite{dezeeuw1999}, (2) \cite{degeus1989}, (3) \cite{preibisch2002}, (4) \cite{mamajek2002}, (5) \cite{debruijne1999}. 
 \label{table: subgroups}}
\end{table*}

\sco{} is currently the best studied \ob{}. It consists of three subgroups: \object{Upper Scorpius} (\object{US}), \object{Upper Centaurus Lupus} (\object{UCL}) and \object{Lower Centaurus Crux} (\object{LCC}) \citep[e.g.,][]{blaauw1964A,dezeeuw1999}. These three subgroups are likely the result of triggered star formation \citep[e.g.,][]{blaauw1991,preibisch1999,preibisch2006}, and in turn may have triggered star formation in the $\rho$~Ophiuchus region. Several properties of the three subgroups of \sco{} are listed in Table~\ref{table: subgroups}.

\cite{preibisch2002} performed an extensive study of the single star population of the US subgroup of \sco{}. They combine their observations of PMS-stars with those of \cite{preibisch1999} and \cite{dezeeuw1999} and derive an empirical mass distribution in the mass range $0.1 \msun \leq M \leq 20 \msun$ (Equation~\ref{equation: preibischimf}). \cite{lodieu2006} on the other hand studied the low-mass and substellar population of \sco{} and find a best-fitting value $\alpha=-0.6\pm 0.1$ of the mass distribution $f_M(M) \propto M^\alpha$ in the mass range $0.01-0.3 \msun$. The results of both studies overlap in the region $0.1-0.3 \msun$. In this region \cite{preibisch2002} find a slope $\alpha=-0.9\pm 0.2$ of the mass distribution, while \cite{lodieu2006} find $\alpha=-0.6 \pm 0.1$. The slight difference between the measured slopes is likely statistical. It is clear, however, that the mass distribution for \sco{} has a break at a certain value $M_\beta$ in (or near) the mass range $0.1-0.3 \msun$. For this reason we model the mass distribution $f_M(M)$ in \sco{} as follows:
\begin{equation} \label{equation: preibischimf}
  f_M(M) \equiv \frac{dN(M)}{dM} \propto \left\{
  \begin{array}{llll}
    M^{-0.6\pm0.1}  &  \quad  M_{\rm min} & \leq M & < M_\beta \\
    M^{-0.9\pm0.2}  &  \quad  M_\beta    & \leq M & < 0.6  \msun\\
    M^{-2.8\pm0.5}  &  \quad  0.6 \msun   & \leq M & < 2  \msun  \\
    M^{-2.6\pm0.5}  &  \quad  2   \msun   & \leq M & < 20 \msun \\
  \end{array}
  \right. \,,
\end{equation}
where $M_{\rm min} \la 0.01\msun$ and $0.1\msun \la M_\beta \la 0.3\msun$. Note that our adopted prescription for $f_M(M)$ roughly corresponds to the mass distribution derived by \cite{kroupa2001}, while it is slightly steeper than Salpeter ($\alpha=-2.35$) in the intermediate-mass regime.

In our analysis we focus on deriving the properties of the intermediate mass binary population, as ample observations of these are available. Due to a lack of systematic surveys for binarity among low-mass stars in \sco{} we cannot constrain these. For these reasons, the form of the mass distribution $f_M(M)$ for $M \la 1.5\msun$ is irrelevant, {\em unless} both the primary and the companion are directly drawn from $f_M(M)$. In \S~\ref{section: true_pairingfunction} we show that the observations exclude the latter possibility, given any reasonable value of  $M_{\rm min}$ and $M_\beta$. In \S~\ref{section: true_pairingfunction} we will also show that binary systems in \sco{} are well described with a primary mass distribution $f_M(M)$ and a mass ratio distribution $f_q(q)$, so that the exact values of $M_{\rm min}$ and $M_\beta$ are irrelevant.

\subsection{The model for \sco{}}

We create association models using the STARLAB simulation package \citep[see, e.g.,][]{ecology4}. The properties of the stellar and binary population are projected onto the space of observables using an extension of the STARLAB package. We adopt a Plummer model \citep{plummer1911} with a projected half-mass radius of $20$~pc, and assume virial equilibrium. Note that, as in this paper we do not evolve the models over time, the latter assumptions do not affect our results.

In our model for \sco{} we adopt a distance of 130~pc (the median distance of the confirmed members of \sco{}) and an age of 5~Myr. Although the subgroups UCL and LCC are older than US, the systematic error introduced by our choice of the age is small. In our models we slightly overestimate the luminosity of stars in the UCL and LCC subgroups, but this affects only the stars close to the detection limit (see \S~\ref{section: photometry}), and does not affect the properties of our simulated observations significantly. The error in the age neither affects the interpretation of the observed mass ratio distribution, as each observed mass and mass ratio is derived from the absolute magnitude of the stars, assuming the correct age for the subgroup, and using the \hipp{} parallax for each star individually. The distribution of these observed mass ratios are then compared with those of the model.

We adopt the extended Preibisch mass distribution in Equation~\ref{equation: preibischimf} for our model of \sco{}. We make the assumptions that (1) the mass distribution for the subgroups UCL and LCC is identical to that of US, (2) we adopt a minimum mass $M_{\rm min}=0.02\msun$ (i.e., we do not consider planetary-mass objects in our mass distribution), and (3) we adopt $M_\beta = 0.1\msun$ for the mass distribution. As in our study we focus on intermediate-mass binaries, assumptions (2) and (3) are only of importance if {\em both} components are directly drawn from the mass distribution. In \S~\ref{section: true_pairingfunction} we discuss this issue and we will show that the exact values of $M_{\rm min}$ and $M_\beta$ are irrelevant for our study.

\cite{preibisch2002} estimate that the US subgroup contains approximately 2525 single/primary stars in the mass range $0.1-20 \msun$.
 With the extension to lower mass in Equation~\ref{equation: preibischimf} the number of singles/primaries is higher, as we also include the very low mass stars and brown dwarfs. For a minimum mass $M_{\rm min}=0.01\msun$ and a value $M_\beta=0.1\msun$, and assuming that the UCL and LCC subgroups have an equal number of singles/primaries, the total number of singles/primaries in \sco{} is approximately $9\,200$. For a value $M_\beta \approx 0.1\msun$ in Equation~\ref{equation: preibischimf} the number of singles/primaries is approximately $8\,000$. Free-floating planets ($M_1 \la 0.02\msun$) are not included in the above statistics. We will therefore adopt $N=S+B=9\,000$ systems ($M_1 > 0.08\msun$) in our simulations, where $S$ is the number of single stars, and $B$ the number of binary systems.

\subsection{Photometry} \label{section: photometry}

We obtain the magnitude of each simulated star in the optical and near-infrared bands using the isochrones described in \cite{kouwenhoven2005}. These isochrones consist of models from \cite{chabrier2000} for $0.02 \msun \leq M < 1 \msun$, \cite{palla1999} for $1 \msun \leq M < 2 \msun$, and  \cite{girardi2002} for $M > 2 \msun$.  We adopt the isochrone corresponding to an age of 5~Myr and solar metallicity. 
By adopting 5~Myr isochrones we overestimate the brightness of 20~Myr old stars by $\sim 0.05$~mag in $JHK_S$ for stars with $M \ga 1 \msun$ and by $\sim 0.5$~mag in $JHK_S$ for stars with $M \la 1 \msun$. 
The error introduced by the metallicity ($\sim 0.05$~mag in $JHK_S$) is negligible for our purposes: see \cite{kouwenhoven2007a} for a more detailed description of these matters. The \hipp{} magnitude $H_p$ for each star is derived from its $V$ magnitude and $V-I$ colour, using the tabulated values listed in the \hipp{} Catalogue (ESA 1997, Vol.~1, \S~14.2). For each star we convert the absolute magnitude into the apparent magnitude using the \hipp{} parallax of each star. We do not include interstellar extinction in our models. \sco{} is practically cleared of gas. The median visual extinction for the member stars of the three subgroups is $A_{V,{\rm US}}=0.5$~mag, $A_{V,{\rm UCL}}=0.06$~mag, and, $A_{V,{\rm LCC}}=0.05$~mag, respectively \citep{debruijne1999}, which translate to values of $A_{K_S,{\rm US}} \approx0.05$~mag, $A_{K_S,{\rm UCL}} \approx0.006$~mag, and, $A_{K_S,{\rm LCC}} \approx0.006$~mag in the near-infrared \citep{mathis1990}. For the purpose of our study the interstellar extinction can thus be neglected, in particular for the study of the near-infrared surveys of \cite{shatsky2002}, \cite{kouwenhoven2005}, and \cite{kouwenhoven2007a}.

% ====================================================================
% ====================================================================
% ====================================================================

\subsection{\sco{} membership} \label{section: membership}

\begin{figure}[!bt]
  \centering
  \includegraphics[width=0.5\textwidth,height=!]{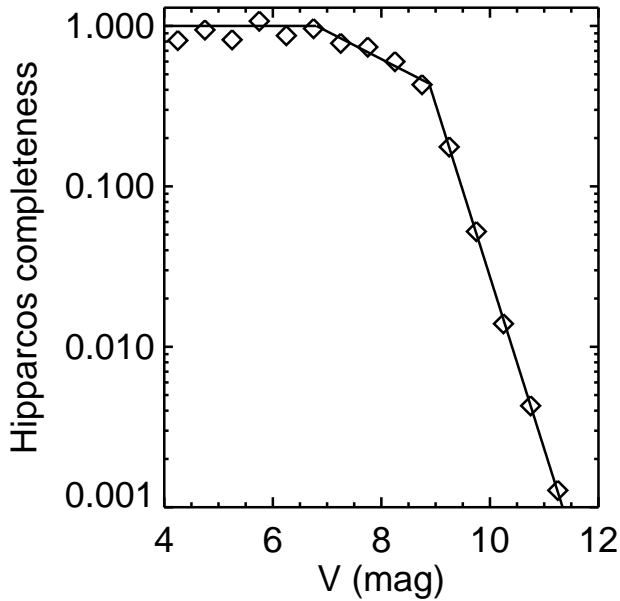}
  \caption{The completeness of the \hipp{} catalogue in the \sco{} region, as a function of $V$ magnitude. The diamonds represent the ratio between the number of stars in the \hipp{} catalogue and the number of stars in the {\em TYCHO-2} catalogue, in each $V$ magnitude bin. The comparison above is made for the \sco{} region, and is similar for each of the three subgroups of \sco{}. The solid line represents the model for the completeness adopted in this paper (Equation~\ref{equation: hipparcos_fraction}).  
\label{figure: hipparcos_fraction} }
\end{figure}

De~Zeeuw et al. (1999) have published a census of the stellar content and membership of nearby ($\la 1$~kpc) \obs{}. They present a list of 521~members of the \sco{} association, based on the \hipp{} position, proper motion, and parallax of each star. Of these members, 120~are in the US subgroup, 221~in UCL, and 180~in LCC. Due to the \hipp{} completeness limit, most of the confirmed members are bright  ($V \la 8$~mag) and mostly of spectral type B, A, and~F. 

In the analysis of the observational data (\S~\ref{section: true_observations}) we consider only the confirmed members of \sco{} \citep[i.e., those identified by][]{dezeeuw1999}, all of which are in the \hipp{} catalogue. Among the stars observed by \hipp{} it is unlikely that a \sco{} member star is not identified as such. On the other hand, it is possible that non-members are falsely classified as members of \sco{}; the so-called interlopers. The fraction of interlopers among the ``confirmed'' \sco{} members stars is estimated to be $\sim 6\%$ for B~stars, $\sim 13\%$ for A~stars, and $\sim 22\%$ for F~and G~stars \citep[see Tables~A2 and~C1 in][]{dezeeuw1999}. The interlopers among B and A~stars are likely Gould Belt stars, which have a distance and age comparable to that of the nearby \obs{}. In our analysis we assume that all confirmed members in the list of \cite{dezeeuw1999} are truly member stars, and do not attempt to correct for the presence of interlopers.

The \hipp{} completeness limit is studied in detail by \cite{soderhjelm2000}. His prescription for the completeness is based on all entries in the \hipp{} catalogue. However, many \obs{} were studied in detail by \hipp{}, based on candidate membership lists. Due to the \hipp{} crowding limit of 3~stars per square degree, only a selected subset of the candidate members of \sco{} was observed \citep[see][for details]{dezeeuw1999}, which significantly complicates the modeling of the \hipp{} completeness.
We therefore calibrate the completeness of \hipp{} in the \sco{} region by comparing the number of \hipp{} entries with the number of stars of a given magnitude in the same region. We use the {\em TYCHO-2} catalogue for this comparison. The {\em TYCHO-2} catalogue is complete to much fainter stars than \hipp{}. In Figure~\ref{figure: hipparcos_fraction} we show the proportion $P$ of stars that is in the \hipp{} catalogue, relative to the number of stars in the {\em TYCHO-2} catalogue, as a function of $V$ magnitude. We model the proportion $P$ as a function of $V$ with three line segments:
\begin{equation} \label{equation: hipparcos_fraction}
  \log P(V) = \left\{
  \begin{array}{cll}
    0                & \mbox{for}\quad                   & V \leq 6.80~\mbox{mag} \\
    1.18 - 0.17\, V & \mbox{for}\quad 6.80~\mbox{mag} < & V \leq 8.88~\mbox{mag} \\
    9.18 - 1.07\, V & \mbox{for}\quad 8.88~\mbox{mag} < & V                      \\
  \end{array}
  \right. \,.
\end{equation}
The {\em TYCHO-2} catalogue is 99\% complete down to $V=11$~mag and 90\% complete down to $V=11.5$. The completeness of the \hipp{} catalogue for $V\ga 11$~mag is therefore not accurately described by Equation~\ref{equation: hipparcos_fraction}. However, this does not affect our results, as the surveys under study only include the brightest members of \sco{}.  In this model we ignore the fact that the \hipp{} completeness also depends on spectral type.

Apart from the large membership study of \cite{dezeeuw1999}, several others have been performed. Several recent studies have focused on the membership of low-mass objects, in particular of brown dwarfs \citep[e.g.][]{kraus2005,slesnick2006,lodieu2006}. These studies often focus on a small group of suspected members in a specific region of \sco{}, and the membership is mostly based on the (less accurate) photometric method. As no census on membership of low-mass stars and brown dwarfs is currently available, we do not include these in our analysis.

% ====================================================================
% ====================================================================
% ====================================================================
% ==INTRODUCTION======================================================
% ====================================================================
% ====================================================================
% ====================================================================

\section{Observations of binary systems in \sco{}} \label{section: true_observations}

\begin{table}[!bt] 
  \begin{tabular}{ll}
    \hline 
    \hline 
    Reference & Detection method \\
    \hline
    \cite{alencar2003}     & Spectroscopic \\
    \cite{andersen1993}    & Combination \\
    \cite{balega1994}      & Visual \\
    \cite{barbier1994}     & Spectroscopic\\
    \cite{batten1997}      & Spectroscopic \\ %\cite{batten1989} \\
    \cite{buscombe1962}    & Spectroscopic \\
    \cite{chen2006}        & Visual \\
    \cite{couteau1995}     & Combination\\
    The Double Star Library & Combination\\
    \cite{duflot1995}      &Spectroscopic\\
    \cite{hartkopf2001}    &Visual\\ %\cite{hartkopf1999} \\
    \cite{jilinski2006}    & Spectroscopic \\
    \cite{jordi1997}       &Eclipsing\\
    The \hipp{} and {\em TYCHO} Catalogues &Astrometric\\ 
    \cite{kraicheva1989}   &Spectroscopic\\ %\cite{kraicheva1980} \\
    \cite{kouwenhoven2005} & Visual \\
    \cite{kouwenhoven2007a}& Visual \\
    \cite{lindroos1985}    & Visual \\
    \cite{malkov1993}      &Combination \\
    \cite{mason1995}       &Visual \\
    \cite{mcalister1993}   &Visual\\
    Miscellaneous, e.g. SIMBAD &Combination\\
    \cite{miura1992}       &Visual\\
    \cite{nitschelm2004}   & Spectroscopic \\
    \cite{pedoussaut1996}  &Spectroscopic\\ %\cite{pedoussaut1985} \\
    \cite{shatsky2002}     &Visual\\
    \cite{sowell1993}      &Visual\\
    \cite{svechnikov1984}  &Combination\\
    \cite{tokovininmsc}    &Combination\\
    \cite{wds1997}         &Combination\\
    \hline 
    \hline 
  \end{tabular}
  \caption{References to literature data with spectroscopic, astrometric, eclipsing, and visual binaries among the \hipp{} members of \sco{}. The data for a number of binary systems in \sco{} is taken from several catalogues. This table is similar to the one presented in \cite{kouwenhoven2005}, but is updated with recent discoveries.
    \label{table: referencelist}}
\end{table}

\begin{figure*}[!bt]
  \centering
  \includegraphics[width=\textwidth,height=!]{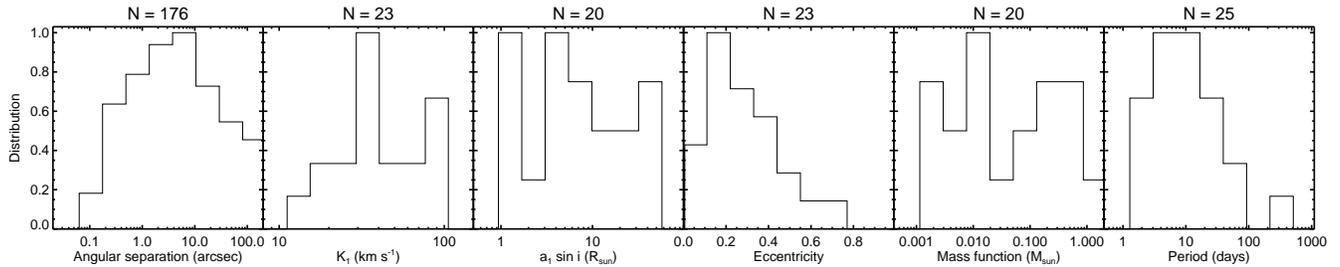}
  \caption{Properties of the {\em observed} binary population in \sco{}. Only the 521~confirmed members of \sco{} are considered. The top-left panel shows the angular separation distribution for visual binaries, at the moment of observation. The other panels show the distribution over radial velocity amplitude $K_1$, the projected semi-major axis $a_1 \sin i$, the eccentricity $e$, the mass function $\mathcal{F}(M)$, and the period $P$, derived for the orbits of the spectroscopic binaries (SB1 and SB2), and for \object{HIP78918}, the only astrometric binary in \sco{} with an orbital solution. The measurements shown in this figure include those of multiple systems. Above each panel we indicate the number of companions for which the corresponding orbital element is available. Spectroscopic and astrometric binaries without an orbital solution are not included. \label{figure: observed_distributions} }
\end{figure*}

A large fraction of the \sco{} member stars is known to be part of a binary or multiple system. In Table~\ref{table: subgroups} we have provided an overview of the {\em observed} binary fraction in the association, for which we included all known binary and multiple systems in \sco{}. In total there are 266~known companions among the 521~confirmed members of \sco{}, most of which are intermediate mass stars. The multiplicity fraction in \sco{} is {\em at least} 40\% among these stars, assuming that all proposed companions are indeed physical companions. The references for these binary and multiple systems are listed in Table~\ref{table: referencelist}. 
Figure~\ref{figure: observed_distributions} shows several observed parameter distributions. These {\em observed} distributions are not representative of the {\em intrinsic} distributions, as selection effects prohibit the detection of a significant fraction of the companion stars. Furthermore, it is possible that several of the reported companions are spurious, including for example bright background stars that are projected close to a \sco{} member star.

The known binary systems in \sco{} were discovered by different observers, using various techniques and instruments. As each of these observing runs is characterized by specific selection effects, it is difficult to study each of these in detail. We therefore focus primarily on a subset of the surveys: those of \cite{kouwenhoven2005,kouwenhoven2007a,shatsky2002,levato1987,brownverschueren}, and those detected by \hipp{} \citep{esa1997}. We refer to these papers and the corresponding datasets as KO5, KO6, SHT, LEV, BRV, and HIP, hereafter. An overview of the number of observed targets and detected binary systems in each dataset is presented in Table~\ref{table: observations_overview}. 
Combined, these datasets contain a large fraction of the known binary and multiple systems in \sco{}. The selection effects for each of these datasets can be modeled, making it possible to use the method of simulated observations. In the following sections we describe these five datasets, and discuss our model for the selection effects.

In our approach we follow the modeling of the selection effects for visual, spectroscopic, and astrometric binarity surveys, which is extensively discussed in \cite{kouwenhoventhesis}. A summary of the modeled selection effects for each of the surveys is given in Table~\ref{table: sos_biases}. 
With the {\em sample bias} we refer to the process
of selecting the targets of interest. A selected sample for a
survey usually consists of a
group of stars with specific properties, such as the solar-type stars in the
solar neighbourhood, or the B~stars in an association. In the case of a binary
survey among the members of an OB~association, the observer may erroneously
include a bright background star, assuming that it is an association member. We
consider this part of the sample bias, although we do not discuss this aspect in this paper.
When the selected targets are surveyed for binarity, observational constraints are
responsible for the {\em instrument bias}. For
example, the minimum and maximum detectable angular separation of binary stars
is determined by the properties of the telescope and the detector. We include in
the instrument bias the selection effects imposed by the telescope-instrument
combination and atmospheric conditions. We additionally include the bias that
results from the difficulties of identifying companions. For example, faint
companions at a large separation of their primary may not always be identified
as such, due to the confusion with background stars.

\begin{table*}
  %\begin{tabular}{p{0.7cm}p{4.1cm}l rr rr}
  \begin{tabular}{lll rr rr}
    \hline
    \hline
    Abbreviation & Reference & Dataset & $N_{\rm orig}$ & $B_{\rm orig}$ & $N_{\rm used}$ & $B_{\rm used}$ \\
    \hline
    KO5    & \cite{kouwenhoven2005}   & Visual         & 199 & 74 & 199 & 60 \\
    KO6    & \cite{kouwenhoven2007a}  & Visual         & 22  & 29 & 22  & 18 \\
    SHT    & \cite{shatsky2002}       & Visual         & 115 & 25 & 87  & 23 \\
    LEV    & \cite{levato1987}        & Spectroscopic  & 81  & 61  & 53  & 39 \\
    BRV    & \cite{brownverschueren}  & Spectroscopic  & 156 & 91  & 71  & 47 \\
    HIP    & \cite{hipparcos}         & Astrometric/Visual & 521 & 125 & 521 & 125 \\
    \hline
    \hline
  \end{tabular}
  \caption{An overview of the datasets used to derive the properties of the binary population in \sco{}. Columns~$1-3$ list the dataset acronym, the reference, and the type of binary studied in the dataset. Columns~4 and~5 list the number of targets in the original dataset, and the number of companions found for these targets. Columns~6 and~7 list the number of targets and companions used in our analysis. This dataset is smaller than the original dataset, as we do not include the non-members of \sco{} in our analysis and at most one companion per targeted star in the case of a multiple systems. The datasets partially overlap, which is taken into account when these are combined in the following sections. We list in this table the total number of spectroscopic binaries, including the radial velocity variables (RVVs; irrespective of their true nature), SB1s, and SB2s. For the \hipp{} observations we list the number of entries in the categories (X), (O), (G), (C), and (S), among the confirmed members of \sco{}. \label{table: observations_overview}}
\end{table*}

\begin{table}
  \begin{tabular}{ll}
    \hline
    \hline
    \multicolumn{2}{l}{KO5 --- \citep{kouwenhoven2005} --- Visual binaries} \\
    \hline
    Observer's choice         & A and late-B members of \sco{} \\
                              & (incl. \hipp{} completeness) \\
    Brightness constraint     & $5.3~\mbox{mag} \leq V_1 \leq 9.5~\mbox{mag}$, $M_1 \geq 1.4 \msun$ \\
    Separation constraint     & Equation~\ref{equation: adonis_separationconstraint} \\
    Contrast constraint       & Equation~\ref{equation: adonis_constrastconstraint} \\
    Confusion constraint      & $K_{S,2} \leq 12$~mag \\    
    \hline
    \multicolumn{2}{l}{KO6 --- \citep{kouwenhoven2007a} --- Visual binaries} \\
    \hline
    Observer's choice         & A selection (11\%) of the KO5 sample  \\
    Brightness constraint     & $5.3~\mbox{mag} \leq V_1 \leq 9.5~\mbox{mag}$, $M_1 \geq 1.4 \msun$ \\
    Separation constraint     & Equation~\ref{equation: naco_separationconstraint} \\
    Contrast constraint       & Equation~\ref{equation: naco_constrastconstraint}  \\
    Confusion constraint      & Not applicable \\
    \hline
    \multicolumn{2}{l}{SHT --- \cite{shatsky2002}  --- Visual binaries}\\
    \hline
    Observer's choice         & B members of \sco{} \\
                              & (incl. \hipp{} completeness) \\
    Brightness constraint     & $V_1 \leq 7.0~\mbox{mag}$, $M_1 \geq 3.5 \msun$  \\
    Separation constraint     & Non-coronographic mode: Equation~\ref{equation: tokovinin_separationconstraint_noncoro} \\
    \quad \quad (idem)        & Coronographic mode:  Equation~\ref{equation: tokovinin_separationconstraint_coro} \\
    Contrast constraint       & Non-coronographic mode: Equation~\ref{equation: tokovinin_contrastconstraint_noncoro}  \\
    \quad \quad (idem)        & Coronographic mode:  Equation~\ref{equation: tokovinin_contrastconstraint_coro}  \\
    Confusion constraint      & $K_{S,2} \leq 12$~mag, $J_2 \leq 13$~mag, \\
                              & and $J_2-K_{S,2} < 1.7$~mag   \\
    \hline
    \multicolumn{2}{l}{LEV --- \cite{levato1987}  --- Spectroscopic binaries}\\
    \hline
    Observer's choice         & B members of \sco{} \\
                              & (incl. \hipp{} completeness) \\
    Brightness constraint     & $V_{\rm comb} \leq 8.1$~mag, $M_1 \geq 3 \msun$  \\
    Contrast constraint       & Not applicable \\
    Amplitude constraint      & \quad Spectroscopic bias model SB-W \\
    Temporal constraint       & \quad with $T=2.74$~year, $\Delta T = 0.38$~year, \\
    Aliasing constraint       & \quad and $\rverr = 3.1$~km\,s$^{-1}$. \\
    Sampling constraint       & Not applied  \\
    \hline
    \multicolumn{2}{l}{BRV --- \cite{brownverschueren} --- Spectroscopic binaries}\\
    \hline
    Observer's choice         & B members of \sco{} \\
                              & (incl. \hipp{} completeness) \\
    Brightness constraint     & $V_1 \leq 7.0~\mbox{mag}$, $M_1 \geq 3.5\msun$  \\
    Contrast constraint       & Not applicable \\
    Amplitude constraint      & \quad Spectroscopic bias model SB-W, \\
    Temporal constraint       & \quad with $T=2.25$~year, $\Delta T = 0.75$~year, \\
    Aliasing constraint       & \quad and $\rverr = 1.4$~km\,s$^{-1}$. \\
    Sampling constraint       & Not applied  \\
    \hline
    \multicolumn{2}{l}{HIP --- \hipp{} mission --- Astrometric binaries}\\
    \hline
    Brightness constraint     & \hipp{} completeness \\
    Amplitude constraint      & \quad Classification into the \\
    Temporal constraint       & \quad categories (X), (C), (O), or (G) \\
    Aliasing constraint       & \quad depending on the observables \\
    Sampling constraint       & \quad of each binary system (see Table~\ref{table: true_hipparcosbiases}) \\
    \hline
    \hline
  \end{tabular}
  \caption{An overview of the models for the selection effects used to generate simulated observations of simulated \obs{}, for the six major datasets discussed in Sections~\ref{section: adonisobservations} to~\ref{section: hipparcosobservations}. The {\em sample bias}, resulting from the choice of the sample alone, includes the observer's choice and the brightness constraint. All other constraints result from the properties of the telescope, detector, atmospheric conditions, and confusion with background stars, and are in this paper referred to as the {\em instrument bias}. For a detailed description of the constraints mentioned in this table we refer to \S~4.5 of \cite{kouwenhoventhesis}. \label{table: sos_biases}}
\end{table}

% ====================================================================
% ====================================================================
% ====================================================================

\subsection{KO5 --- \cite{kouwenhoven2005} observations} \label{section: adonisobservations}

\cite{kouwenhoven2005} performed a near-infrared adaptive optics binarity survey among A and late-B members of Sco OB2. Their observations were obtained with the ADONIS/SHARPII+ system on the ESO~3.6~meter telescope at La~Silla, Chile. Adaptive optics was used to obtain high spatial resolution, in order to bridge the gap between the known close spectroscopic and wide visual binaries. 
The survey was performed in the near-infrared, as in this wavelength regime the contrast between the components of a binary system with a high mass ratio is less than in the visual regime.
All targets were observed in the $K_S$ band, and several additionally in the $J$ and $H$ bands. 
KO5 selected their sample of A and late-B targets from the list of confirmed \hipp{} members that were identified by \cite{dezeeuw1999}. All targets have $6~\mbox{mag} \la V \la 9~\mbox{mag}$, which corresponds to similar limits in the $K_S$ band.

With their observations KO5 are sensitive to companions as faint as $K_S \approx 15.5$~mag, corresponding to the brightness of a massive planet in \sco{}. Due to the large probability of finding faint background stars in the field of view, KO5 classify all secondaries with $K_S > 12$~mag as background stars, and those with $K_S \leq 12$~mag as candidate companion stars. The $K_S = 12$~criterion separates companion stars and background stars in a statistical manner, and is based on the background star study of SHT. A member of \sco{} with $K_S=12$~mag has a mass close to the hydrogen-burning limit. The follow-up study of KO6 with VLT/NACO (see \S~\ref{section: nacoobservations}) has shown that the $K_S=12$~criterion correctly classifies secondaries as companions in $80-85\%$ of the cases. 
With their survey KO5 find 151~secondaries around the 199~target stars. Out of these 151~secondaries, 74 are candidate companions ($K_S \leq 12$~mag), and 77~are background stars ($K_S > 12$~mag).  
KO5 find that the mass ratio distribution $f_q(q)$ for late-B and A~type stars in \sco{} is consistent with $f_q(q) \propto q^{-0.33}$, and exclude random pairing between primary and companion.

% ====================================================================
% ====================================================================
% ====================================================================

\subsubsection{Treatment of the KO5 dataset} \label{section: treatment_adonis}

All 199~targets in the KO5 dataset are confirmed members of \sco{}, and are therefore included in our analysis. We use in our analysis a subset of the companions identified in KO5.
Several targets in the ADONIS survey have more than one candidate or confirmed companion. In this paper we do not study triples and higher-order multiples; we consider at most one companion per target star. For each of these candidate multiple systems we include the (candidate) companion that is most likely a physical companion. 
For \object{HIP52357} we include the companion with $(\rho,K_S) = (0.53'',7.65~\mbox{mag})$, as it is brighter and closer to the target star than the candidate companion with $(\rho,K_S) = (10.04'',11.45~\mbox{mag})$.
For the same reason, we do not include the wide and faint candidate companion of \object{HIP61796} with $(\rho,K_S) = (12.38'',11.86~\mbox{mag})$ in our analysis. KO5 find two bright and close companions of \object{HIP76001}, with $(\rho,K_S) = (0.25'',7.80~\mbox{mag})$ and $(\rho,K_S) = (1.48'',8.20~\mbox{mag})$, respectively. Although \object{HIP76001} is likely a physical triple, we choose to retain only the innermost candidate companion. 
KO5 find a bright secondary separated $1.8''$ from \object{HIP63204}. With their follow-up study, KO6 find a close companion at $\rho=0.15''$. KO6 show that this close companion is physical, while the secondary at $1.8''$ is optical; we do not consider the latter secondary in our analysis.

\object{HIP68532} and \object{HIP69113} are both confirmed triple systems, each with a primary and a ``double companion''. For both \object{HIP68532} and \object{HIP69113}, the two stars in the ``double companion'' have a similar separation and position angle with respect to the primary, and a similar magnitude. In physical terms, the double companions of \object{HIP68532} and \object{HIP69113} could have originated from a more massive companion that fragmented into a binary. We therefore model the double companions of these stars as single companions, taking the average $\rho$ and $\varphi$, the combined $K_S$ magnitude, and the total mass of each double companion.

For the comparison with the simulated observations the targets \object{HIP77315} and \object{HIP77317} are both considered as individual, single stars. The star \object{HIP77317} is known to be a companion of \object{HIP77315} at $\rho=37.37''$, and is for that reason listed as such in KO5. This binary system is far too wide to be detected with the observing strategy of KO5; both stars are therefore treated as individual stars.
With the ADONIS survey KO5 find three candidate companions of \object{HIP81972}. Of these three, only the secondary at separation $5.04''$ is a confirmed companion in the follow-up study of \cite{kouwenhoven2007a}. As \object{HIP81972} is near the Galactic plane, the other two secondaries are likely background stars, and are thus not included in the dataset. 

KO5 separated the secondaries into candidate companions and background stars using the $K_S$~magnitude of each secondary. The follow-up study of KO6, using multi-colour analysis, has shown that several of these candidate companions are background stars. We do not consider in our analysis these secondaries, indicated with \object{HIP53701}-1 ($K_S=8.9$ mag), \object{HIP60851}-1 ($K_S=11.5$ mag), \object{HIP60851}-2 ($K_S=11.3$ mag), \object{HIP80142}-1 ($K_S=9.51$ mag), and \object{HIP80474}-1 ($K_S=10.8$ mag) in KO6.

The resulting KO5 dataset that we use for our analysis contains data for 199~targets with a total of 60 companion stars. For each of these targets and their companions we use the measurements given in KO5, unless more recent (and more accurate) measurements for these stars are presented in the follow-up study of KO6.
For each of the targets \object{HIP63204}, \object{HIP73937}, and \object{HIP79771} a new close companion is resolved by KO6, which was unresolved in the observations of KO5. For these three targets we use the properties of the primary star as provided by KO6.

The mass of each primary and companion is derived from the near-infrared magnitude. If available, the mass of each star is taken from KO6, who use the $JHK_S$ magnitude to derive the mass. In all other cases the mass is taken from KO5, who derive the mass from the $K_S$ magnitude only. The more recently determined masses allow us to better constrain the properties of the binary population in \sco{}. Finally, the mass ratio $q = M_2/M_1$ is calculated for each binary system. In the Appendix (Table~\ref{table: data_adonis}) we list the properties of the binaries used for comparison with simulated observations. 

% ====================================================================
% ====================================================================
% ====================================================================

\subsubsection{Modeling the observational bias of KO5}

We model the sample bias in KO5 as follows. The authors selected the A~and late-B members of \sco{}. As these members were identified as such in the \hipp{} membership study of \cite{dezeeuw1999} we first impose the \hipp{} completeness (see \S~\ref{section: membership}) on the simulated association. Based on the properties of the target list of KO5, we model the observer's choice and brightness constraint by removing all targets (i.e. singles and primaries) with $V < 5.3$~mag, all targets with $V > 9.5$~mag, and all targets with $M < 1.4 \msun$ from the sample. 

\begin{figure}[!bt]
  \includegraphics[width=0.5\textwidth,height=!]{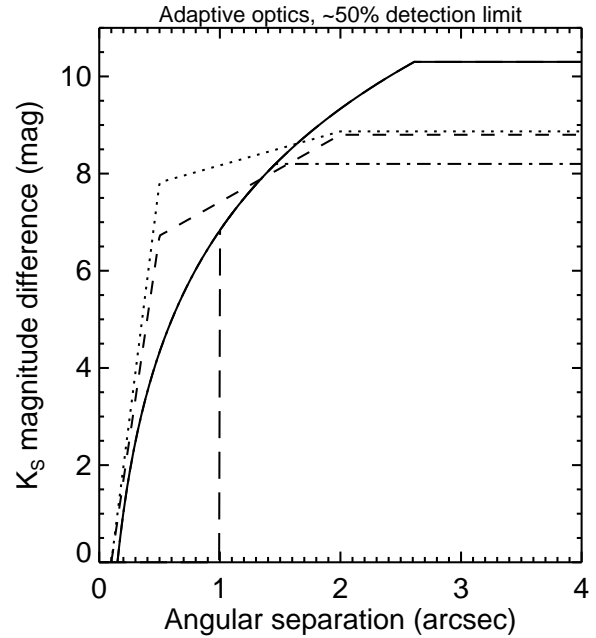}
  \caption{The 50\% detection limit $\Delta K_S$ as a function of angular separation $\rho$, for the KO5 observations (short-dashed curve), the KO6 observations (dotted curve), the non-coronographic SHT observations (dash-dotted curve), the coronographic SHT observations (long-dashed curve), and the combined SHT observations (solid curve). For the KO5 and KO6 observations the curves represent those for average Strehl ratios of 30\% and 24\%, respectively. 
\label{figure: probability_detlim_ka} }
\end{figure}

We model the detection limit of the KO5 observations using the analysis presented in KO6, who describe these in detail. We study the 50\% detection limit (in terms of the magnitude {\em difference} $\Delta K_S \equiv K_2-K_1$) and find its dependence on angular separation and Strehl ratio (SR). We parametrize the dependence of the detection limit $\Delta K_{\rm S,det}(\rho)$ in magnitudes on Strehl ratio (SR) as 
\begin{equation} \label{equation: adonis_constrastconstraint}
  \begin{array}{l}
    \Delta K_{\rm S,det}(\rho) = \\
    \quad \left\{
    \begin{array}{lrl}
      0                                         &  \quad                        \rho &< \rho_{\rm lim,A}\\
      (22.0 - 3.75\, s({\rm SR}))\ (\rho-0.1'') &  \quad \rho_{\rm lim,A}  \leq \rho &< 0.5''\\
      8.8 + s({\rm SR})\, (\rho-2'')            &  \quad 0.5''             \leq \rho &< 2''\\
      8.8                                       &  \quad 2''               \leq \rho & \\
    \end{array}
    \right. \,,
  \end{array}
\end{equation}
where $\rho_{\rm lim,A} = 0.2''$ is the angular resolution of the KO5 observations.
Following the properties of the KO5 observations, we model $ s({\rm SR})$ with
\begin{equation} \label{equation: adonis_constrastconstraint_slope}
  s({\rm SR})= 2.54 - 3.85 \times {\rm SR} \,. 
\end{equation}
As an example we plot the detection limit $\Delta K_{\rm S,det}(\rho)$ in Figure~\ref{figure: probability_detlim_ka} for observations with SR~=~30\%.
We simulate the distribution over SR by drawing for each target the SR randomly from the observed distribution $\tilde{f}_{\rm SR}({\rm SR})$, which is approximated with
\begin{equation} \label{equation: adonis_constrastconstraint_strehl}
  \tilde{f}_{\rm SR}({\rm SR}) \propto \exp \left( - \frac{({\rm SR}-\mu_{\rm SR})^2}{2\sigma_{\rm SR}^2} \right) \quad \quad 5\% < {\rm SR} < 50\% \,,
\end{equation}
where $\mu_{\rm SR}=30\%$ and $\sigma_{\rm SR}=5\%$.

KO5 considered only the secondaries with $K_S \leq 12$~mag as physical companions. The follow-up study of KO6 has shown that this $K_S=12$ criterion indeed correctly classifies most of the companions and background stars. We therefore identify in our simulated observations only the companions with $K_S \leq 12$~mag as true companions.

\begin{figure}[!bt]
  \centering
  \includegraphics[width=0.5\textwidth,height=!]{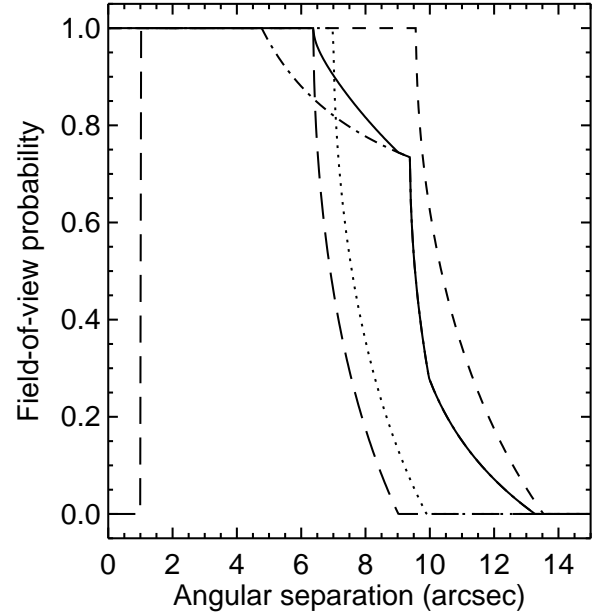}
  \caption{In the three imaging surveys for binarity discussed in this paper (KO5, KO6, and SHT) the field-of-view is non-circular. Whether a secondary is in the field of view, depends therefore not only on its separation $\rho$, but also on its position angle $\varphi$. This figure shows the probability that a secondary is in the field-of-view, as a function of $\rho$, assuming random orientation of the binary systems, for the KO5 observations (short-dashed curve), the KO6 observations (dotted curve), the non-coronographic SHT observations (dash-dotted curve), the coronographic SHT observations (long-dashed curve), and the combined SHT observations (solid curve). Whether a secondary is detected or not, depends additionally on its brightness and on the brightness difference with the primary (see Figure~\ref{figure: probability_detlim_ka}). \label{figure: probability_rho_vs_pa} }
\end{figure}

Each measurement is assigned a detection probability $D_A(\rho)$ as a function of angular separation $\rho$. This detection probability refers solely to whether or not a companion is projected into the field of view. As the field of view is not circular, the detection probability of a companion is a function of angular separation. For ADONIS we have a square field of view sized $12.76'' \times 12.76''$. KO5 observed each target four times, each time with the target in another quadrant of the field of view, so that the effective field of view is $L_{\rm A}=\frac{3}{2} \cdot 12.76''=19.14''$. The probability $D_A$ that a secondary with an angular separation $\rho$ is in the field-of-view is then given by:
\begin{equation} \label{equation: adonis_separationconstraint}
  D_A(\rho)  =
  \left\{ \begin{array}{lll}
    1                                       & {\rm for} &                         \rho < L_{\rm A}/2 \\
    1- (4/\pi) \arccos (L_{\rm A}/2\rho)    & {\rm for} & L_{\rm A}/2        \leq \rho < L_{\rm A}/\sqrt{2} \\
    0                                       & {\rm for} & L_{\rm A}/\sqrt{2} \leq \rho
  \end{array} \right. \,,
\end{equation}
which is visualized in Figure~\ref{figure: probability_rho_vs_pa}.

% ====================================================================
% ====================================================================
% ====================================================================

\subsection{KO6 --- \cite{kouwenhoven2007a} observations} \label{section: nacoobservations}

The results of the ADONIS binarity survey performed by KO5 raised several questions, in particular on the absence of faint secondaries in the $1''-4''$ separation range, and on the validity of the $K_S =12$ criterion that KO5 used to separate secondaries into companion stars and background stars. Although SHT and KO5 argue that the latter criterion statistically classifies the background stars correctly, the correct classification of the companion stars with $K_S \approx 12$~mag was still uncertain.
To address this issue, KO6 performed follow-up multi-colour $JHK_S$ observations of a subset of the ADONIS targets. With multi-colour observations, each secondary can be placed in the colour-magnitude diagram, and compared with the isochrone of the \sco{} subgroups. Companion stars are expected to be near the isochrone, while background stars are (generally) expected to be far from the isochrone.

 The observations described in KO6 were carried out with the adaptive optics instrument NAOS-CONICA (NACO), mounted on the ESO Very Large Telescope on Paranal, Chile. A subset of 22 (out of 199) KO5 targets were selected for follow-up observations. The subset was not randomly selected, but preference was given to faint and close background stars, to secondaries with $K_S \approx 12$~mag, and to newly discovered candidate companions.
KO6 analyzed the $JHK_S$ observations of these 22~stars observed with NACO, including the multi-colour ADONIS observations of 9~targets. With their observations KO6 found three new close companions (of \object{HIP63204}, \object{HIP73937}, and \object{HIP79771}) that were unresolved in the survey of KO5.

% ====================================================================
% ====================================================================
% ====================================================================

\subsubsection{Treatment of the KO6 dataset}

For our analysis we consider all 22~targets observed by KO6, all of which are confirmed members of \sco{}. The 9~ADONIS targets that were also studied in KO6 are not considered here, simply because they were not observed in the campaign of KO6. Around the 22 NACO targets KO6 find 62~secondaries, of which they classify 18 as confirmed companions (c), 11~as possible companions (?), and 33~as background stars (b).

In our analysis we use the data for 15~(out of 18) confirmed companions, and 5~(out of 8) candidate companions. Both \object{HIP68532} and \object{HIP69113} have a tight ``double companion''. We treat each of these as a single companion, by combining the separation and mass of the individual companions (see \S~\ref{section: treatment_adonis}). The targets \object{HIP67260}, \object{HIP79771}, and \object{HIP81949} all have three (candidate) companions, for which we only include the inner companion in our analysis. We do not include the very faint secondary of \object{HIP80142}, as this is likely a background star. For \object{HIP81972} we only include the companion \object{HIP81972}-3, which is by far the most massive companion, in our analysis.

The final KO6 dataset used in this paper consists of 22~targets with 18~companions. Note that when this dataset is compared with simulated observations, a discrepancy may be present, as the sample was composed to study candidate companions and background stars with particular properties. In Table~\ref{table: data_naco} we list the properties of the binaries used for comparison with simulated observations.

% ====================================================================
% ====================================================================
% ====================================================================

\subsubsection{Modeling the observational bias of KO6}

A subset of $22/199=11\%$ of the targets in the KO5 sample are observed with NACO by KO6. We model the KO6 sample by randomly drawing 11\% of the targets in the simulated KO5 target sample. Note that in reality, the subset was not random (see above); instead, the targets were selected based on the properties of their secondaries. The simulated KO6 observations therefore cannot be directly compared with the results of the KO6 observations. However, they can be used to find the expected number of close and/or faint companions with KO6; companions that could not be found with the KO5 survey.

We use the 50\% detection limit from the analysis presented in KO6, and parameterize it with the Strehl ratio (SR) of the observations. The 50\% detection limit as a function of $\rho$, for targets with a different brightness is derived using simulations \citep[][\S~3]{kouwenhoven2007a}.  From the observational data we derive a detection limit $ \Delta K_{\rm S,det}(\rho)$ in magnitude, consisting of four line segments:
\begin{equation} \label{equation: naco_constrastconstraint}
  \begin{array}{l}
    \Delta K_{\rm S,det}(\rho) = \\
    \quad\left\{
    \begin{array}{ll}
      0                                         &  \quad \rho < \rho_{\rm lim,N} \\
      (2.5\, B({\rm SR}) - 2.63 )\ (\rho-0.1'') &  \quad \rho_{\rm lim,N} \leq \rho < 0.5''\\
      B({\rm SR}) + 0.70\, (\rho-2'')           &  \quad 0.5'' \leq \rho < 2''\\
      B({\rm SR})                               &  \quad \rho \geq 2''\\
    \end{array}
    \right.\,,
  \end{array}
\end{equation}
where $\rho_{\rm lim,N} = 0.1''$ is the angular resolution of the KO6 observations and
\begin{equation} \label{equation: naco_constrastconstraint_slope}
  B({\rm SR}) = 6.86 + 8.37\times\mbox{SR} 
\end{equation}
is the magnitude difference of the faintest detectable source, for a given Strehl ratio SR. As an example we show the detection limit $\Delta K_{\rm S,det}(\rho)$ in Figure~\ref{figure: probability_detlim_ka} for observations with SR~=~24\%. We simulate the distribution over SR by drawing for each target the SR randomly from the observed distribution $\tilde{f}_{\rm SR}({\rm SR})$, which we approximate with Equation~\ref{equation: adonis_constrastconstraint_strehl},
with $\mu_{\rm SR}=24\%$ and $\sigma_{\rm SR}=7\%$. The field of view for the observations of KO6 is $14''\times 14''$. As the field of view is non-circular, the detection limit is a function of both angular separation $\rho$ and position angle $\varphi$. For our simulated observations, each measurement is assigned a detection probability $D_{\rm N}(\rho)$ as a function of angular separation $\rho$, given by
\begin{equation} \label{equation: naco_separationconstraint}
  D_{\rm N}(\rho)  =
  \left\{ \begin{array}{lll}
    1                                       & \quad \rho < L_{\rm N}/2 \\
    1- (4/\pi) \arccos (L_{\rm N}/2\rho)    & \quad L_{\rm N}/2 < \rho < L_{\rm N}/\sqrt{2} \\
    0                                       & \quad L_{\rm N}/\sqrt{2} < \rho
  \end{array} \right. \,,
\end{equation}
where $L_{\rm N}=14''$ is the linear size of the field-of-view.

% ====================================================================
% ====================================================================
% ====================================================================

\subsection{SHT --- \cite{shatsky2002} observations} \label{section: tokovininobservations}

\cite{shatsky2002} performed an imaging survey for binarity among 115~B~type stars in the \sco{} region. Their observations were carried out in 2000 with the near-infrared adaptive optics instrument ADONIS at the ESO 3.6~meter telescope on La Silla, Chile. 
Their sample is based on the study of \cite{brownverschueren}; see \S~\ref{section: brownverschuerenobservations}. Among the 115 B-type stars surveyed by SHT,  87~are confirmed members of \sco{} according to \cite{dezeeuw1999}.  Among the total sample of these 115~stars SHT find 96~secondaries in the angular separation range $0.3''-6.4''$, of which they identify 10~as new physical companions. 
The authors conclude that the mass ratio distribution $f_q(q)$ for B-type stars in \sco{} is consistent with $f_q(q) \propto q^{-0.5}$, and that random pairing can be excluded.

% ====================================================================
% ====================================================================
% ====================================================================

\subsubsection{Treatment of the SHT dataset}

Near the 87 confirmed members of \sco{} targeted by SHT, 80~secondaries are found, of which 61~likely optical and 19~likely physical companions. Of this set of 19~physical companions, we use a subset of 17~for our analysis. The target \object{HD132200} is probably a physical triple system. As we consider in this paper only single and binary systems, we do not include the widest and faintest component of \object{HD132200}, and retain the component with $\rho=0.128''$ and $K_S=5.46$~mag. The secondary \object{HD133937P} is incorrectly reported in SHT. For this secondary $\rho=0.57''$ and $J-K_S=2.06$~mag (N.~Shatsky \& A.~Tokovinin, private communication). Due to its large $J-K_S$ value, and as $J>13$~mag, this secondary is likely a background star. We therefore do not consider \object{HD133937P} in our analysis.

Several targets were not included in the observed sample of SHT. These targets were known to have close companions and thus not suitable for wavefront sensing. These unobserved targets were included in the analysis of SHT though. Seven of these (\object{HIP53701}, \object{HIP57851}, \object{HIP62322}, \object{HIP64425}, \object{HIP74117}, \object{HIP76371}, and \object{HIP77840}) are confirmed members of \sco{}. We include these non-observed targets in our analysis, either as single or as a binary system, depending on whether their companions would have been detected with the SHT observing strategy. Technically, the non-inclusion of a set of stars falls under the ``observer's choice''. The latter constraint is difficult to model, as it would involve modeling of pre-SHT observations of close binaries, as well as the determination whether or not such a binary is suitable for wavefront sensing. We choose, however, to manually add these stars to the list of observed targets, as the properties of these stars and their companions are well-understood (making detailed models of the observer's choice redundant). Note that non-inclusion of these unobserved stars introduces a small bias, as these stars would have been surveyed by SHT if their companions were unknown at that time. 
The member stars \object{HIP57851}, \object{HIP62322}, \object{HIP74117}, \object{HIP76371}, and \object{HIP77840} were reported as visually resolved (C)-binaries in the \hipp{} catalogue. We use the angular separation and magnitude of these components as given in the catalogue, and include the stars in the sample. \object{HIP64425} is a known triple system \citep{tokovininmsc} for which we use the massive inner binary in our analysis. We treat the non-observed star \object{HIP53701} as a single star, as KO6 have shown that its secondary is a background star.

For the stars observed by SHT, we derive the mass of target and companion star from the $K_S$ magnitude, using the evolutionary models described in Section~\ref{section: scoob2properties}. For the stars that are analysed by SHT, but not observed by these authors (see above), we derive the mass using the $V$ band magnitude and \hipp{} $H_p$ magnitude. For each star we adopt the distance given by the \hipp{} parallax and the age of the subgroup of which the target is a member (Table~\ref{table: subgroups}).

The final dataset from the SHT survey that we use in our analysis of the binary population in \sco{} comprises 87~targets with 23~physical companions. The properties of these 23~companions are listed in Table~\ref{table: data_tokovinin}.

% ====================================================================
% ====================================================================
% ====================================================================

\subsubsection{Modeling the observational bias of SHT}

We require that all targets are confirmed \hipp{} members of \sco{}, and so first impose the \hipp{} detection limit on the simulated association. We model the brightness constraint of the SHT observations by adopting a minimum mass of $3.5 \msun$, and a minimum brightness of $V=7$~mag for the targets.

SHT show the typical detection limit of their observations in their Figure~3. The detection limit is obviously different for the observations with and without the coronograph. The observations with coronograph are deeper, and the observations without the coronograph provide a larger range in angular separation. A companion star is detected if it is observed {\em either} in the coronographic mode {\em or} in the non-coronographic mode.

In the non-coronographic observations, each companion is assigned a detection probability $D_{\rm NC}(\rho)$ as a function of its separation $\rho$. In the non-coronographic observations, SHT observed each target twice in the non-coronographic mode, both times with the target in a quadrant of the detector. Due to the square shape of the detector, and due to the observing strategy, the position angle is of importance to whether a companion at separation $\rho$ is in the field-of-view. We model this dependence by assigning a probability $D_{\rm NC}(\rho)$ that a companion is in the field of view, depending on $\rho$. For a square field-of-view of a detector with linear size $L$, and a separation $K$ between the two observations (along the diagonal of the field-of-view), the probability is given by
\begin{equation} \label{equation: tokovinin_separationconstraint_noncoro}
  \begin{array}{l}
    D_{\rm NC}(\rho)  = \\
    \quad \left\{ 
    \begin{array}{lll}
      1                                                                      & \rho < r_1 \\ 
      \tfrac{1}{2} + \frac{2}{\pi} \arcsin \left(\frac{r_1}{\rho}\right)     & r_1    < r_2/\sqrt{2}  \\    
      \tfrac{1}{2} + \frac{2}{\pi} \arcsin \left(\frac{r_1}{\rho}\right) - \frac{4}{\pi} \arccos \left(\frac{r_2}{\rho}\right)     & r_2\sqrt{2}   < \rho < r_3 \\
      \tfrac{1}{2} - \frac{2}{\pi} \arccos \left(\frac{r_2}{\rho}\right)     & r_3 < \rho < r_2 \\
      
      0                                                                      &  r_2  < \rho
    \end{array} 
    \right.,
  \end{array} 
\end{equation}
where 
\begin{equation}
r_1=\tfrac{1}{2}L\sqrt{2}  - \tfrac{1}{2} K  \,, \quad 
r_2=\tfrac{1}{2}L\sqrt{2} + \tfrac{1}{2} K \,, \quad 
r_3 = \sqrt{\tfrac{1}{2}(r_1^2+r_2^2)} \,.
\end{equation}
For the non-coronographic observations of SHT, the linear dimension of the detector is $L=12.76''$, and the translation along the diagonal of the field of view is $K=8.5''$ (see Figure~1 in SHT, for details). 
We model the detection limit of the observations {\em without} the coronograph with
\begin{equation} \label{equation: tokovinin_contrastconstraint_noncoro}
  \Delta K_{\rm S,det} = \left\{
  \begin{array}{ll}
    0                     & \quad \mbox{for\ } \rho < \rho_{\rm lim,S} \\
    8.32 \log \rho + 6.83 & \quad \mbox{for\ } \rho_{\rm lim,S} \leq \rho < 1.46'' \\
    8.2                   & \quad \mbox{for\ } \rho \geq 1.46''\\
  \end{array}
  \right. \,,
\end{equation}
where $\rho_{\rm lim,S} = 0.1''$ is the angular resolution of the SHT observations (based on Figure~3 in SHT).

SHT additionally observe each target using the coronograph. They do not perform their coronographic observations in mosaic-mode; only one pointing is used. Each measurement is therefore assigned a detection probability $D_{\rm C}(\rho)$ as a function of angular separation $\rho$, given by
\begin{equation} \label{equation: tokovinin_separationconstraint_coro}
  D_{\rm C}(\rho)  =
  \left\{ \begin{array}{lll}
    0                               & {\rm for} & \rho < d_{\rm C} \\ 
    1                               & {\rm for} & d_{\rm C} < \rho < L/2 \\
    1- \frac{4}{\pi} \arccos \left(\frac{L}{2\rho}\right)    & {\rm for} & L/2 < \rho < L/\sqrt{2} \\
    0                               & {\rm for} & L/\sqrt{2} < \rho
  \end{array} \right. \,,
\end{equation}
where $L=12.76''$ and $d_{\rm C} = 1''$ is the radius of the coronograph. 
Based on Figure~3 in SHT, we model the detection limit of the observations {\em with} the coronograph with
\begin{equation} \label{equation: tokovinin_contrastconstraint_coro}
  \Delta K_{\rm S,det} = \left\{
  \begin{array}{ll}
    0                     & \quad \mbox{for\ } \rho < d_{\rm C} \\
    8.32 \log \rho + 6.83 & \quad \mbox{for\ } d_{\rm C} \leq \rho < 2.62'' \\
    10.3                  & \quad \mbox{for\ } \rho \geq 2.62 \\
  \end{array}
  \right. \,.
\end{equation}

Finally, we combine the simulated observations in coronographic and non-coronographic mode. We consider a binary system as detected, if it is observed in at least one of the two modes.

SHT additionally studied the background star population in the \sco{} region. Due to the large number of background stars, it is likely that a very faint or red secondary is a background star. 
SHT classify a secondary as a background star if $J>13$~mag, if $K_S>12$~mag, or if $J-K_S > 1.7$~mag (unless the secondary is a known companion). In our model for the selection effects, we adopt these limits in magnitude and colour when obtaining the simulated observations.

% ====================================================================
% ====================================================================
% ====================================================================

\subsection{LEV --- \cite{levato1987} observations} \label{section: levatoobservations}

\cite{levato1987} performed a large radial velocity survey for binarity among early-type stars in the \sco{} region. They performed their observations in May 1974 with the 0.9~meter and 1.5~meter CTIO telescopes, and in 1976 with the 2.1~meter telescope at KPNO. 
Their sample consists of 81~candidate members of \sco{}, and is based on that of \cite{slettebak1968} who composed a list of suspected \sco{} members for a study on stellar rotation. All except 4 of the 82~targets of \cite{slettebak1968}, and 3~additional targets were observed by LEV. The spectral type of the observed targets ranges from B0\,V to A0\,V. The targets in the sample have $2.5~\mbox{mag} < V < 8.1~\mbox{mag}$. 

On average, each star is observed over an interval of $\langle T \rangle = 2.74$~year, with a spread of $\sigma_T = 0.68$~year. Each target is observed $5-12$ times, with an average observing interval $\langle \Delta T \rangle = 0.38$~year and a corresponding spread of $\sigma_{\Delta T} = 0.14$~year. 
For each target LEV list the internal error in the radial velocity measurements. Averaged over all targets, this error is $\langle \rverr\rangle = 3.1$~km\,s$^{-1}$, with a spread of 1.0~km\,s$^{-1}$; approximately 90\% of the targets have $\rverr > 2$~km\,s$^{-1}$.

In their Table~3 LEV list their conclusions on binarity. Of the 53~confirmed members of \sco{} that they observed, 14~have a constant radial velocity (within the measurement errors), 23 have a variable radial velocity (RVV), 8~are SB1, and 8~are SB2. Given these observations, the spectroscopic binary fraction is {\em at least} $(8+8)/53 = 30\%$, in the case that all RVVs are spurious. If all reported RVV targets are indeed binaries, the observed spectroscopic binary fraction is $(8+8+23)/53=74\%$. The observed spectroscopic binary fraction is a lower limit for $\binfrac$, as binaries that are unresolved in the survey of LEV (e.g. visual binaries) are not included in these statistics.

% ====================================================================
% ====================================================================
% ====================================================================

\subsubsection{Treatment of the LEV dataset}

For the comparison between the observational data and the simulated observations, we only consider the 53~confirmed members \citep[according to][]{dezeeuw1999} of \sco{} that LEV observed.
In their Tables~3 and~4, LEV include the star \object{HIP76945} (\object{HD140008}), a confirmed member of the UCL subgroup. LEV did not observe this SB2, but take the orbital elements from \cite{thackeray1965}. Our simulations indicate that the radial velocity variability of a binary with properties such as those of \object{HIP76945} would practically always be detected in a survey similar to that of LEV. It is unclear, however, whether LEV would have been able to derive the orbital elements for this binary, i.e., if they would have detected it as an SB1 or SB2. We include \object{HIP76945} as an SB2 in the dataset, as \cite{thackeray1965} were able to derive the orbital elements several decades before the study of LEV.
For a subset of the targets LEV derive the orbital elements. In their Table~4, LEV list the elements of 22~targets, of which 16~are confirmed members of \sco{}. 
In Table~\ref{table: data_levato} we list the properties of these 16~SB1 and SB2 systems from the LEV dataset that are confirmed members of \sco{}. We also list the 23~radial velocity variables (RVVs), for which the orbital elements are unavailable. The LEV dataset consists of 53~targets, of which $16+23= 39$~are detected as binary systems.

% ====================================================================
% ====================================================================
% ====================================================================

\subsubsection{Modeling the observational bias of LEV} \label{section: levatomodeling}

In this paper we consider only the confirmed \hipp{} members of \sco{}, i.e., first impose the \hipp{} detection limit on the association. We model the choice of the sample of LEV by removing all binary systems with a combined magnitude fainter than $V=8.1$~mag from the simulated observations. 

We model the instrument bias of LEV using windowed sampling \citep[SB-W; see][]{kouwenhoventhesis}. Briefly summarized; in order to compare the model predictions with the observations, we simulate the detection of the spectroscopic binaries in our models as follows. We obtain radial velocity measurements of all binary systems in the simulated association, at regular intervals $\Delta T$ for a time-span $T$ (windowed sampling). We assume a value for the measurement error $\rverr$, which is constant over the time of observations. If the radial velocity measurements show a spread significantly larger than the error, the binarity is detected.
For each single star and binary system we test the hypothesis that the observed velocity measurements $\{v_i\}$ result from a constant velocity. We calculate the $\chi^2$ of the set of $N_v$ radial velocity measurements:
\begin{equation}
\chi^2 = \sum_i \frac{(v_i-\overline{v})^2}{\rverr^2}\,,
\end{equation}
where $\overline{v}$ is the mean of the measurements $\{v_i\}$. We then calculate the probability $p$ that $\chi^2$ is drawn from the $\chi^2$-distribution:
\begin{equation}
  p = 1 - \Gamma \left( \tfrac{1}{2}\nu,\tfrac{1}{2}\chi^2 \right)\,,
\end{equation}
where $\nu = N_v-1$ is the number of degrees of freedom. High values ($p\approx 1$) indicate that our hypothesis (that the radial velocity is constant) is true, and that the measurements are likely the result of statistical noise. Values of $p$ close to zero indicate that the observed variations in the radial velocity are real. We classify objects with radial velocity sets with $p \leq 0.0027$ (corresponding to the $3\sigma$ confidence level) as binary systems, while the other targets are marked as single stars.
In our modeling of spectroscopic binaries we thus only determine whether a binary is detected or not; we do not discriminate between spectroscopic binaries of type SB1, SB2 or RVV. Modeling the latter difference is sophisticated and depends on a significant number of parameters. For example, the Nyquist theorem requires that at least two measurements should be obtained per orbital period. Additional constraints are imposed by the properties of the binary system: the spectral type of the star (in particular the number of spectral lines), the brightness of the system, and the values of the radial velocity amplitude $K_1$ (relative to $\epsilon_{\rm RV}$), the eccentricity $e$ and the argument of periastron $\omega$. It is not trivial to model these.

In our model for the LEV observations we use windowed sampling, adopting an observing run of $T=2.74$~year, an observing interval $\Delta T = 0.38$~year, and a radial velocity accuracy of $\rverr = 3.1$~km\,s$^{-1}$. The latter assumption is a simplification, as the value of $\rverr$ is slightly different for each observation in the LEV dataset (with a spread of $\sim 1$~km\,s$^{-1}$). A star is more easily detected if $\rverr < 3.1$~km\,s$^{-1}$, and less easy if $\rverr$ is larger. Our simulations show, however, that our assumption of a constant $\rverr$ introduces an error significantly smaller than the error introduced by low-number statistics, justifying our assumption.

% ====================================================================
% ====================================================================
% ====================================================================

\subsection{BRV --- \cite{brownverschueren} observations} \label{section: brownverschuerenobservations}

\cite{brownverschueren} studied stellar rotation among members of the \sco{} association. The observations were carried out between 1991 and 1993 using the ECHELEC spectrograph at the ESO 1.52~meter telescope on La~Silla, Chile. The sample of BRV contains the pre-\hipp{} candidate and established members of \sco{}, based on the studies of \cite{blaauw1964A}, \cite{bertiau1958}, and \cite{degeus1989}. 
The observations and data reduction procedure are described in detail in \cite{verschueren1997}, and the results on duplicity are described in \cite{verschueren1996} and \cite{brownverschueren}. 
Their sample consists of 156~targets in the \sco{} region, mostly of spectral type~B.
They find that $\sim 60\%$ of the binary systems exhibit a significant radial velocity variation. After combination of their data with those of LEV and those of the Bright Star Catalogue \citep{hoffleit1982,hoffleit1983}, they obtain a binary fraction of 74\%.

\subsubsection{Treatment of the BRV dataset}

Among the 156~observed targets there are 71~confirmed members of \sco{} (18~in US, 30~in UCL, and 23~in LCC). Among these 71~targets, 7~are SB1, 10~are SB2, 30~are RVV, and 12~have a constant radial velocity (CON). For 12~targets, insufficient measurements are available to make a statement about the radial velocity variation.

Two out of the 30~RVV binaries are known to exhibit radial velocity variation due to line profile variability. \object{HD120324} is a non-radial pulsator and \object{HD136298} is a $\beta$~Cephei variable. For both stars, this is likely the reason  that they are classified as RVV. These stars are therefore not considered as binary systems in our analysis. 

The final BRV dataset used in our analysis consists of 71~confirmed members. Of these targets, 12~are spectroscopically single, 7~are SB1, 10~are SB2, 28~are RVV, and 12~have insufficient data to determine whether the radial velocity is variable. The binary fraction is thus {\em at least} $(7+10)/71 \approx 24\%$, if {\em none} of the RVV and CON targets are binary, and $(71-12)/71 = 83\%$ if {\em all} RVV and CON targets are binary. Among the target stars with sufficient data to make a statement on duplicity (i.e., without the 12~CON targets included), the observed binary fraction is $17/59=28\%$ at least and $45/59=76\%$ at most.

\subsubsection{Modeling the observational bias of BRV}

We model the choice of the BRV sample in a way similar as we did for the LEV dataset. Each target is observed three times over an interval of $T=2.25$~year, so that $\Delta T = 0.75$~year. Following the reduction of the original data \citep{verschueren1996}, we classify each target with a radial velocity variation larger than $3\rverr = 4.2$~km\,s$^{-1}$ as a RVV  (see \S~\ref{section: levatomodeling}).

% ====================================================================
% ====================================================================
% ====================================================================

\subsection{HIP --- \hipp{} observations} \label{section: hipparcosobservations}

\begin{table*}
  %\begin{tabular}{p{0.8cm}r ccccc p{1.45cm}p{1.45cm}p{1.45cm}}
  \begin{tabular}{lr ccccc lll}
    \hline
    \hline
    Subgroup& $N_\star$&(X)& (O) &(G) & (C)& (S)             & $\tilde{F}_{\rm M,XOG}$ & $\tilde{F}_{\rm M,XOGS}$ & $\tilde{F}_{\rm M,C}$ \\
    \hline
    US      & 120   & 1 & 0   & 4  & 15 & 8  (1)  & $4.2 \pm 1.5\%$      &  $10.0\pm 2.1\%$      & $ 12.5\pm 3.0\%$    \\
    UCL     & 221   & 0 & 1   & 9  & 36 & 7  (0)  & $4.5 \pm 1.5\%$      &  $\ 7.7\pm 1.9\%$      & $ 16.3\pm 3.0\%$   \\
    LCC     & 180   & 2 & 0   & 6  & 28 & 11  (2)  & $4.4 \pm 1.6\%$      &  $\ 9.4\pm 2.4\%$      & $ 15.6\pm 3.0\%$    \\
    \hline
    \sco{} & 521 & 3 & 1   & 19 & 79 & 26  (3)  & $4.4 \pm 1.0\%$      &  $\ 8.8\pm 1.4\%$      & $ 15.2\pm 1.8\%$   \\
    \hline
    \hline
  \end{tabular}
  \caption{Candidate and confirmed astrometric binaries in the \hipp{} catalogue. For each subgroup we list the number $N_\star$ of known members, the number of stochastic (X), orbital (O), acceleration (G), component (C), and suspected (S) binaries in the \hipp{} catalogue. For each (S) binary we list between brackets how many of these are also (X)-flagged. The last three columns list the ``astrometric binary fraction'' --- including the (X), (O), (G) binaries --- without the (S) binaries and with the (S) binaries included, and the \hipp{} ``visual'' binary fraction, for the (C) binaries only. (V) binaries are not present in \sco{}.
 \label{table: hipparcos_observations} }
\end{table*}

\begin{table*}
  %\begin{tabular}{llp{1cm}cl}
  \begin{tabular}{lllcl}
    \hline
    Constraints on $\rho$ and $\Delta H_p$ & Period constraints & Solution & Symbol & Elements \\
    \hline
    $2 \leq \langle \rho \rangle \leq 100$~mas  or $\Delta H_p > 4$   & $P \leq 0.1$~year      & Stochastic  & (X) & no \\
    $2 \leq \langle \rho \rangle \leq 100$~mas  or $\Delta H_p > 4$   & $0.1 < P \leq 10$~year & Orbital     & (O) & yes \\
    $2 \leq \rho \leq 100$~mas  or $\Delta H_p > 4$                   & $5 < P \leq 30$~year   & Acceleration& (G) & no \\
    $0.1 \leq \rho \leq 100$~arcsec and $\Delta H_p \leq 4$              & $P > 30$~year       & Resolved    & (C) & no \\
    \multicolumn{2}{l}{Not modeled}                                                            & Suspected   & (S) & no \\
    \multicolumn{2}{l}{Not modeled}                                                            & VIM         & (V) & no \\
    \hline
  \end{tabular}
  \caption{A model for the instrument bias of the \hipp{} catalogue, based on the analysis of \cite{lindgren1997}. The binary systems satisfying the above constraints are resolved with \hipp{} in our models. For the comparison between the observations and the simulated observations, we consider two sets of \hipp{} binaries: the visual binaries and the astrometric binaries. No orbital motion is detected for the (C) binaries; these are visually resolved and therefore technically visual binaries. The \hipp{} astrometric binaries contain the targets with (X), (O), (G), and optionally (S) entries. No difference between the latter categories is made for the comparison with the astrometric binaries. Binary systems that do not satisfy the constraints listed in this table remain undetected in our simulated observations for \hipp{}. We do not model the (V)-binaries (variability-induced movers; VIMs) and (S)-binaries (suspected non-single stars). Note that in our model we {\em overpredict} the number of binaries in categories (X), (O), and (S), as not all binaries with the properties above are detected by \hipp{} as such. \label{table: true_hipparcosbiases} }
\end{table*}

In the \hipp{} catalogue, the (possible) binary nature of a target is indicated in field H59 with the flags (X), (O), (G), (C), or (V), and in field H61 with the flag (S). For the targets with a (C) flag both stars in the binary system are resolved, but no orbital motion is detected. These systems are considered as visual (or optical) binaries. \hipp{} entries with an (X)-flag have a stochastic solution. These stars exhibit an apparent motion significantly larger than the statistical uncertainties, although no double star solution could be found. For entries with an (O)-flag, at least one of the orbital elements could be derived from the apparent motion. Entries with a (G)-flag show a significant acceleration in the apparent motion, but no solution could be found. These are likely long-period binaries. The (V)-flagged entries are variability-induced movers. For this group of binaries, the photocenter exhibits apparent motion due to variability of one of the components. Finally, (S)-flagged entries are suspected non-single stars.  These targets are effectively single as observed by \hipp{} \cite[][\S~2.1]{esa1997}, although no convincing single-star astrometric model could be fit to the observations. Several of the (S) binaries are also (X)-flagged, indicating that these are likely non-single.

Among the 521~confirmed members of \sco{}, 46~are candidate or confirmed astrometric binaries --- i.e., those in the categories (X), (O), (G), and (S) --- in the \hipp{} catalogue: 12~in US, 17~in UCL, and 17~in LCC. An additional 79~\sco{} members are classified as (C)-binaries; these are visually resolved binaries. Table~\ref{table: hipparcos_observations} lists for the three subgroups of \sco{} the number of entries in each of the \hipp{} categories. 
As for the binaries in the category (C) no orbital motion is detected, we will consider this group as visual binaries. (V)-binaries are not present among the confirmed \sco{} members. \object{HIP78918} is the only member of \sco{} with an orbital solution (O). The three (X)-flagged members of \sco{} are also flagged as suspected non-single stars (S).

% ====================================================================
% ====================================================================
% ====================================================================

\subsubsection{Treatment of the \hipp{} dataset}

In our analysis we consider each target in the categories (X), (O), (G), and (C) in Table~\ref{table: hipparcos_observations} as a binary system. Binarity among the stars in the (S) category (the ``suspected non-single'' targets) is rather uncertain. We therefore compare our results with the \hipp{} data, with and without the suspected (S) binaries included. 
Note that not all targets in the category (X) are necessarily binary systems. For example, the flag (X) of \object{HIP80763} (\object{$\alpha$~Sco}) may be due to the extended nature of the star, which is surrounded by a dust-shell \citep{cruzalebes1998}. This may induce photocentric motion that is not related to binarity.

% ====================================================================
% ====================================================================
% ====================================================================

\subsubsection{Modeling the observational bias of \hipp{}} \label{section: modeling_hipparcos}

The observer's choice and sample bias for the \hipp{} member list of \sco{} are discussed in Section~\ref{section: membership}; our adopted model to describe these biases is given in Equation~\ref{equation: hipparcos_fraction}. The instrument bias for \hipp{} was described in detail in \cite{lindgren1997} and is summarized in Table~\ref{table: true_hipparcosbiases} \citep[see][for further details]{kouwenhoventhesis}. 

The binaries in category (C) are considered as visual binaries, as both components of such binaries are resolved, while no astrometric motion is detected. The detection of the (C)-binaries is modeled using the prescription in Table~\ref{table: true_hipparcosbiases} and is then compared with the observations.

The binaries in the categories (X), (O), and (G) are considered as astrometric binaries, and are modeled using the prescriptions in Table~\ref{table: true_hipparcosbiases}. As we do not model the \hipp{} observations in detail, we are unable to accurately predict in which of these categories each \hipp{} target falls. The simple model that we adopt for the \hipp{} biases results in an overlap between the properties of the stars in these categories, and furthermore, we overpredict the number of stars observed in these categories. \cite{lindgren1997} have analyzed the properties of the binary systems in each of the categories. But this does not mean that each binary system with these properties is observed as such. In our model we make the latter assumption, resulting in an overestimation of the number of binaries detected by \hipp{}, and an overlap between the modeled categories (O) and (G). We therefore combine the number of detected (X), (O), and (G) binaries in the simulated observations, and use the resulting astrometric binary fraction $\tilde{F}_{\rm M,XOG}$ as an {\em upper limit} for the comparison with the observations.

We do not include stellar variability in our model, and are therefore unable to model the (V)-flagged binaries (field H59). We do not model the (S)-flagged binaries (field H61) either, as the determination whether a target is classified as such in the \hipp{} catalogue is based on an internal and external comparison performed by the two \hipp{} data reduction teams.

% ====================================================================
% ====================================================================
% ====================================================================
% ==INTRODUCTION======================================================
% ====================================================================
% ====================================================================
% ====================================================================

\section{Recovering the pairing function and mass ratio distribution} \label{section: true_pairingfunction}

In this and the following sections we discuss the determination of the binary population in \sco{} from observations. We adopt the strategy described in \cite{kouwenhoventhesis}. First we derive the pairing function and mass ratio distribution in \S~\ref{section: true_pairingfunction}. We recover the semi-major axis, the period distribution and the binary fraction in \S~\ref{section: true_smadistribution}, and the eccentricity distribution in \S~\ref{section: true_eccentricitydistribution}. Finally, we present a discussion on the derived binary fraction and its associated error in \S~\ref{section: true_binaryfraction}.

\begin{figure*}[!bt]
  \centering
  \begin{tabular}{lll}
    \includegraphics[width=0.31\textwidth,height=!]{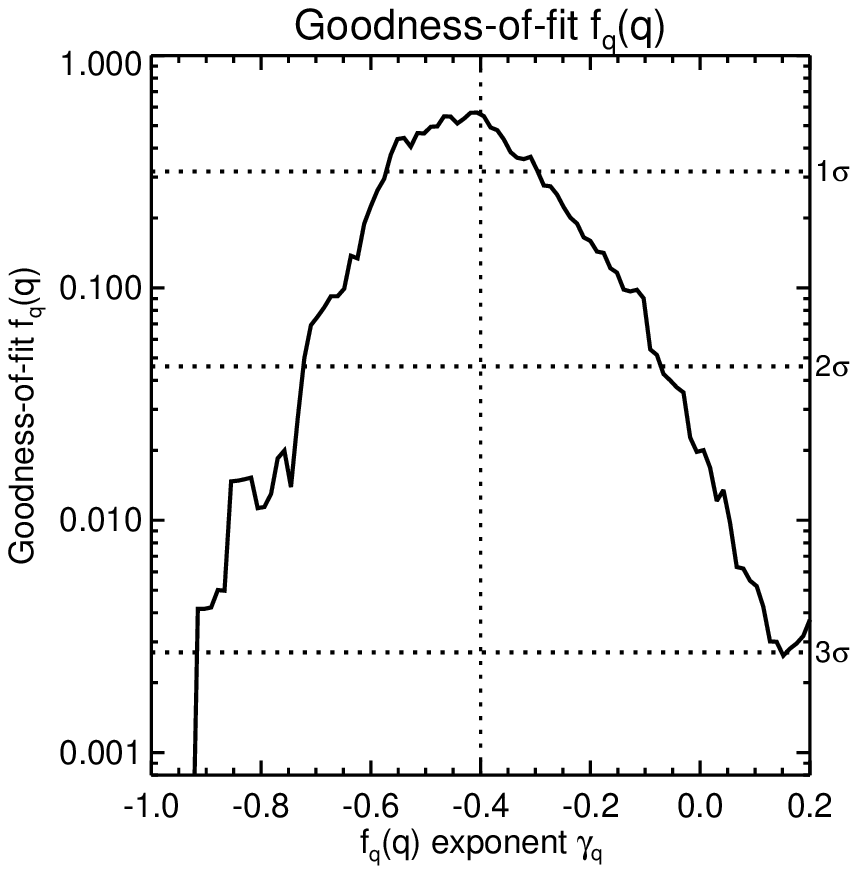} &
    \includegraphics[width=0.31\textwidth,height=!]{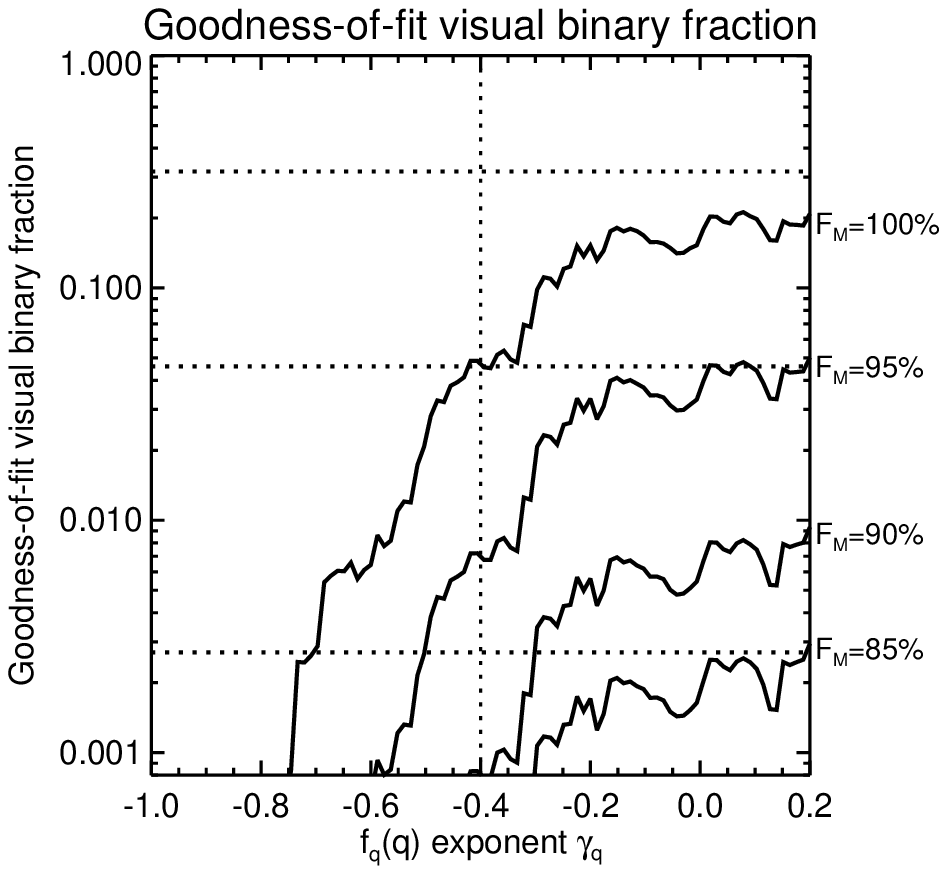} &
    \includegraphics[width=0.31\textwidth,height=!]{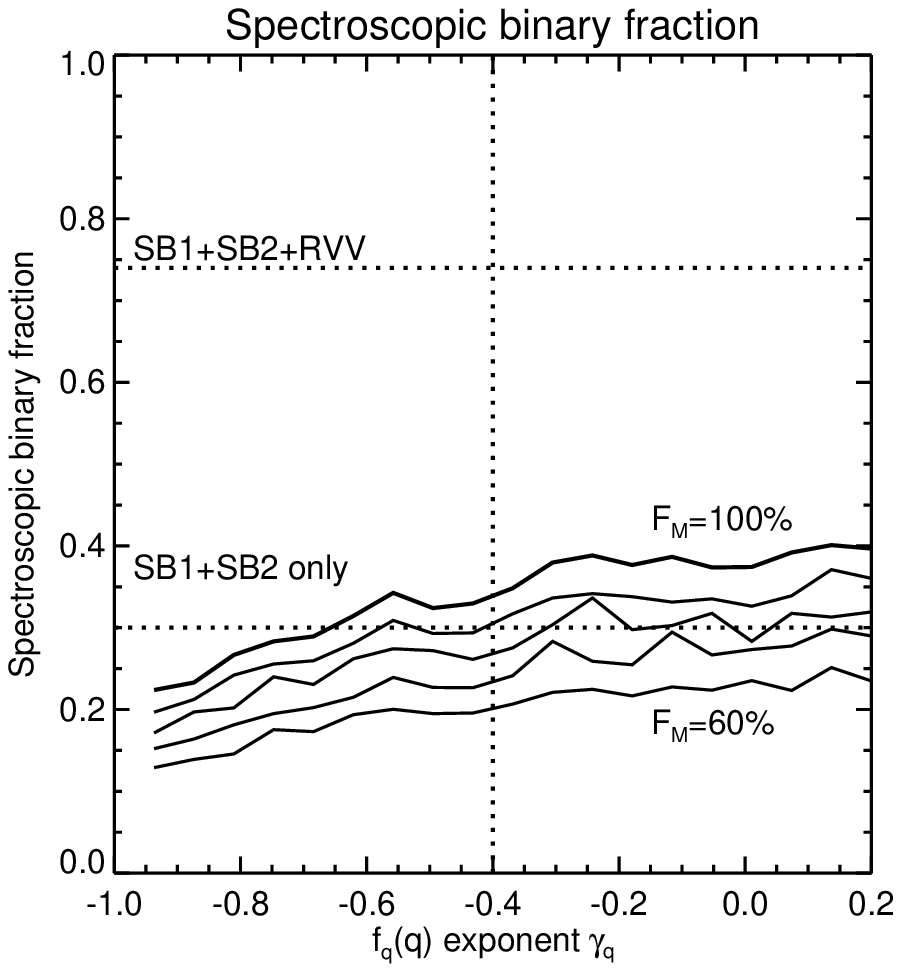} \\
  \end{tabular}
  \caption{How well can the intermediate-mass binary population in \sco{} be described by a mass ratio distribution of the form $f_q(q) \propto q^{\gamma_q}$? 
    {\em Left}: the goodness-of-fit for the comparison between the observed mass ratio distribution (from the KO5 and SHT datasets) and that predicted by models with different values of $\gamma_q$. A large value for the goodness-of-fit means that the model predictions are consistent with the observations. The $1\sigma$, $2\sigma$ and $3\sigma$ confidence limits for rejection of the model are indicated with the horizontal dotted lines.
    {\em Middle}: the goodness-of-fit for the comparison between the observed visual binary fraction (from the KO5 and SHT datasets) and that predicted by the models with an intrinsic binary fraction of 85\% (bottom curve), 90\%, 95\% and 100\% (top curve). 
    {\em Right:} the predicted (SB1, SB2 and RVV) spectroscopic binary fraction (for the LEV dataset) as a function of $\gamma_q$. The five curves indicate the results for a model binary fraction of 60\% (bottom curve), 70\%, 80\%, 90\% and 100\% (top curve). The bottom horizontal dotted line indicates the minimum observed spectroscopic binary fraction, where only the SB1s and SB2s are included, assuming that the radial velocity variations in all RVVs are caused by line-profile variability rather than binarity. The top horizontal line indicates the maximum spectroscopic binary fraction, assuming that {\em all} RVVs are indeed binaries. 
    Taken together, the three panels indicate that the mass ratio distribution for intermediate mass stars in \sco{} can be described by $f_q(q) \propto q^{\gamma_q}$ with exponent $\gamma_a \approx -0.4$ (vertical dotted line in each panel). An intrinsic binary fraction close to 100\% is required in order to produce the large visual and spectroscopic binary fractions that are observed. 
    See \S~\ref{section: true_pairingfunction} for a further discussion of this figure. 
    \label{figure: bestfit_opik_q} }
\end{figure*}

\cite{kouwenhoventhesis} discusses five possible ways of pairing the components of a binary system. These include random pairing (RP), primary-constrained random pairing (PCRP), and three variants of primary-constrained pairing (PCP-I, PCP-II, and PCP-III). For the models with random pairing the primary mass $M_1$ and companion mass $M_2$ are both drawn from the mass distribution. For PCRP both masses are drawn from the mass distribution, with the additional constraint that the companion mass is smaller than that of the primary. In the three PCP pairing models the primary mass is drawn from the mass distribution, and the companion mass is derived from the mass ratio $q \equiv M_2/M_1$ which is drawn from a distribution $f_q(q)$. The difference between the three PCP models lies in the treatment of the very low mass companions that are generated. Each of these five pairing functions result in a different binary population. 
\cite{kouwenhoventhesis} has shown that for binary systems with intermediate mass primaries the pairing functions PCP-I, PCP-II, PCP-III practically give the same results. As we focus on intermediate mass binaries in this paper, we will therefore consider three pairing functions in our analysis: RP, PCRP, and PCP (where PCP represents either PCP-I, PCP-II, or PCP-III).

As shown by \cite{kouwenhoventhesis} the mass ratio distribution $f_q(q)$ resulting from pairing function RP or PCRP depends strongly on the mass distribution $f_M(M)$, in particular on the shape of the mass distribution in the brown dwarf regime. The mass distribution for \sco{} is fairly well constrained (Equation~\ref{equation: preibischimf}), although the exact values of the parameters $M_{\rm min}$ and $M_\beta$ are as yet unknown. However, the mass distribution can still be used to rule out pairing functions RP and PCRP for \sco{}, by considering the extreme values for  $M_{\rm min}$ and $M_\beta$.

\cite{shatsky2002} and \cite{kouwenhoven2005} have already shown that their observed mass ratio distribution is inconsistent with random pairing (RP). Below we show that both pairing functions RP and PCRP can be excluded based on their observations. The two free parameters in the mass distribution of \sco{} (Equation~\ref{equation: preibischimf}) are in the range $M_{\rm min} \la 0.01\msun$ and $0.1 \msun \la M_\beta \la 0.3\msun$. The shape of the resulting mass ratio distribution for these pairing functions depends on the exact values of $M_{\rm min}$ and $M_\beta$. Below we will show that even for the ``most favourable'' values of $M_{\rm min}$ and $M_\beta$ both RP and PCRP can be excluded.

The mass distribution in Equation~\ref{equation: preibischimf} results in a large number of low-mass stars. The probability to obtain a binary consisting of two intermediate or high-mass stars is small \citep[e.g.][]{kouwenhoventhesis}. We use this property of pairing functions RP and PCRP to show that these are inconsistent with the observations. Let $Q$ denote the ratio between the number of binaries with mass ratio $q \geq 0.8$ and the number of targets $N_{\rm targets}$:
\begin{equation}
  Q 
  = \frac{\mbox{\# binaries with $q>0.8$}}{N_{\rm targets}} \,.
\end{equation}
The value of $Q$ increases with increasing $M_{\rm min}$ and increasing $M_\beta$ due the smaller probability of drawing a low-mass object from the mass distribution. Furthermore, $Q$ is proportional to the intrinsic binary fraction $\binfrac$. For pairing functions RP and PCRP, the largest value of $Q$ is therefore reached if $M_{\rm min} \approx 0.01\msun$, $M_\beta \approx 0.3\msun$, and $\binfrac=100\%$. 

We simulate two associations with the latter properties (one with pairing function RP and one with PCRP) and extract the sample of KO5 using the constraints listed in Table~\ref{table: sos_biases}. The resulting {\em intrinsic} values of $Q$ for the KO5 sample are $Q_{\rm RP} \approx Q_{\rm PCRP} \approx 0.004\%$. These values are upper limits because of the adopted values for $M_{\rm min}$, $M_\beta$, and $\binfrac$. Also, the corresponding {\em observed} values $\tilde{Q}$ of the simulated observations are significantly smaller due to the non-detection of very close and wide binaries, binaries with a high mass ratio, etc.

Among the 199~targets in the KO5 sample, 10~binary systems with $q \geq 0.8$ are detected. The observed fraction of binaries with $q>0.8$ is therefore $\tilde{Q}_{\rm KO5} = 5.0 \pm 1.6\%$. Due to selection effects KO5 have certainly missed several binaries with $q>0.8$ at separations smaller than the spatial resolution, or outside the field of view, so that the true value is $Q_{\rm KO5} > \tilde{Q}_{\rm KO5} = 5.0 \pm 1.6\%$. As $Q_{\rm RP} \ll \tilde{Q}_{\rm KO5}$ and $Q_{\rm PCRP} \ll \tilde{Q}_{\rm KO5}$, both pairing functions RP and PCRP can thus be excluded with strong confidence.

The observations are consistent with pairing function PCP, which is characterized by a mass ratio distribution $f_q(q)$. The unknown slope $\alpha$ and the value $M_\beta$ for the mass distribution (Equation~\ref{equation: preibischimf}) are now irrelevant, as the distribution over companion masses among intermediate mass stars is independent of $\alpha$. The mass ratio distribution and binary fraction for high-mass and intermediate mass targets is the same for the three pairing functions \citep{kouwenhoventhesis}. As we do not have detailed information on binarity among low-mass stars in \sco{}, we cannot discriminate between PCP-I, PCP-II, and PCP-III. A detailed membership study for low-mass stars in \sco{}, followed by a detailed binary study, is necessary to establish the difference. \cite{kouwenhoventhesis} also shows that for the three PCP pairing functions the mass ratio distribution for binaries with a high-mass primary is approximately equal to the generating mass ratio distribution $f_q(q)$ of an association. 

The mass ratio distribution among intermediate mass stars in \sco{} was obtained by both KO5 and SHT. SHT find a mass ratio distribution of the form $f_q(q) \propto q^{\gamma_q}$ with $\gamma_q=-0.5$ in their B~star survey, and KO5 find $\gamma_q=-0.33$ in their A and late-B star survey. 
Subsequently, we adopt a mass ratio distribution of the form $f_q(q) \propto q^{\gamma_q}$ and study for which value of $\gamma_q$ the simulated observations correspond best to the observations of KO5 and SHT. We compare the observed mass ratio distribution $\fqobs$ with that of the simulated observations using the Kolmogorov-Smirnov (KS) test, and we compare the observed binary fraction $\binfracobs$ with the predictions using the Pearson $\chi^2$ test. In both cases, we test the hypothesis that the observations and simulated observations are realizations of the same underlying association model. 
All models have a semi-major axis distribution of the form $f_a(a) \propto a^{-1}$ with $0.5\rsun < a < 5\times 10^6\rsun$ (see \S~\ref{section: recover_amin}, \S~\ref{section: recover_amax}, and \S~\ref{section: fa_vs_fq} for a motivation) and a thermal eccentricity distribution $f_e(e) = 2e$ with $0\leq e < 1$ (see \S~\ref{section: true_eccentricitydistribution} for a motivation). We perform our simulations with different intrinsic binary fractions, ranging from $\binfrac=60\%$ to $100\%$. 

Figure~\ref{figure: bestfit_opik_q} shows the results of this comparison for models with a varying value of $\gamma_q$ and a varying intrinsic binary fraction $\binfrac$. 
The left-hand panel shows the probability associated with the KS comparison between the observed $\fqobs$ and that of the simulated observations. A small value for the goodness-of-fit means that the model can be excluded with high confidence. The three horizontal dotted lines indicate the $1\sigma$, $2\sigma$ and $3\sigma$ confidence limits for rejection of the model, respectively. Models outsize the range $-0.6\la\gamma_q\la-0.3$ can be rejected with $1\sigma$ confidence. The best-fitting models have $\gamma_q\approx -0.4$ (vertical dotted line), which is the value we adopt for \sco{}.  However, $\gamma_q$ is also constrained using the additional information provided by the observed visual and spectroscopic binary fraction.

The middle panel of Figure~\ref{figure: bestfit_opik_q} indicates how well the observed visual binary fraction for \sco{} compares with the predictions of the models with different values of $\gamma_q$ and $\binfrac$. Models with a small value of $\gamma_q$ underpredict $\binfracobs$. Most binaries have a small mass ratio in this case, making it more difficult to detect the companion star, and hence resulting in a lower visual binary fraction. Models with a $\binfrac<95\%$ can be excluded with $2\sigma$ confidence or more, for any value of $\gamma_q < 0.2$ (under the condition that our assumptions hold). 

The right-hand panel of Figure~\ref{figure: bestfit_opik_q} shows the predicted spectroscopic binary fraction as a function of $\gamma_q$ and $\binfrac$. Note that, unlike the middle panel of Figure~\ref{figure: bestfit_opik_q}, we do not show the goodness-of-fit between the observed and predicted spectroscopic binary fraction, as we only have lower and upper limits of the former (see below).
More spectroscopic binaries are detected for models with a large $\gamma_q$, as the radial velocity amplitude increases with higher companion mass. The horizontal dotted lines in this panel indicate the limits for the observed spectroscopic binary fraction: $30\% \la \binfracobs \la 74\%$ (with likely $\binfracobs \approx 74\%$). For models with $\gamma_q \ga -0.6$, the models with an intrinsic binary fraction $\binfrac \approx 100\%$ are most consistent with the observations. Note that the spectroscopic binary fraction as predicted by the models is rather small as compared to the observed spectroscopic binary fraction, even for an intrinsic binary fraction $\binfrac=100\%$. This apparent underprediction is also present in the following sections, where we study different semi-major axis and period distribution, and may result from the presence of triple and higher-order systems among the members of \sco{}. In Section~\ref{section: triples} we will return to this issue. 

A good model for \sco{} should correctly predict $\fqobs$, the visual binary fraction and spectroscopic binary fraction. The distribution $\frhoobs$ suggests that $-0.6\la\gamma_q\la-0.3$, with a best-fit for $\gamma_q\approx -0.4$, while the observed binary fractions are best fitted by models with a large value of $\gamma_q$. A combination of these data is not trivial, as it involves a choice for the relative weights given to each of the goodness-of-fits for each quantity. For example, if $\frhoobs$ and the visual binary fraction are given equal weight, the combined best-fitting value is $\gamma_q\approx -0.35$, while if $\frhoobs$ is a ten times heavier weight, the combined best-fitting value is $\gamma_q\approx-0.41$. Inclusion of the spectroscopic binary fraction is not trivial, as we only have lower and upper limits for this quantity. An inspection of the three panels suggets that the mass ratio distribution can be well described by the expression $f_q(q) \propto q^{\gamma_q}$ with $\gamma_q \approx -0.4$. This value is bracketed by, and consistent with the values derived by KO5 and SHT. The comparison further indicates that the intrinsic binary fraction is larger than $\approx95\%$, with $1\sigma$ confidence.

% ============================================================================
% ============================================================================
% ============================================================================
% ============================================================================

\section{Recovering the semi-major axis distribution, the period distribution and the binary fraction} \label{section: true_smadistribution}

The orbital size distribution of a binary population can be quantified in several ways. Observers of visual binaries often express their results with a semi-major axis distribution $f_a(a)$, derived from the observed angular separation distribution. Observers of spectroscopic binary systems on the other hand, often express their results with an orbital period distribution $f_P(P)$, as the period of a spectroscopic binary can often be measured directly. Authors of theoretical/simulation papers mostly describe the orbital size distribution using the orbital energies (e.g. in units of $kT$). Over the past several decades, observational studies have brought forward two widely accepted distributions: a flat distribution in logarithmic semi-major axis, and a log-normal period distribution. In the this section we compare these distributions with the observations. We wish to stress that the discussion of $f_a(a)$ and $f_P(P)$ should {\em nowhere} be taken to imply that a power-law distribution in $a$ would result in a log-normal distribution in $P$.

The flat distribution in $\log a$, commonly known as \"{O}pik's law, has been derived for a wide range of stellar populations \citep[e.g.,][]{opik1924,vanalbada1968,vereshchagin1988,poveda2004}, and is equivalent to
\begin{equation} \label{equation: true_opikslaw}
f_a(a) \propto a^{\gamma_a} 
\quad \quad \amin \leq a \leq \amax \,,
\end{equation}
with $\gamma_a=-1$. For a set of binaries of total mass $M_T$, this results in an orbital period distribution 
\begin{equation}
f_P(P) \propto M_T^{(\gamma_a+1)/3} \, P^{(2\gamma_a-1)/3} \,, 
\end{equation}
which, in the case of $\gamma_a=-1$, is a flat distribution in $\log P$ \citep[see Section~4.D in][]{kouwenhoventhesis}. Throughout this paper we will consider the distribution $f_a(a)$ and study for which values of $\gamma_a$ the models are consistent with the observations. 

\cite{duquennoy1991} studied binarity among solar-type stars in the solar neighbourhood  and find a log-normal period distribution: 
\begin{equation} \label{equation: true_duquennoyperiods}
f_{\rm DM}(P) \propto \exp \left\{ - \frac{(\log P - \mu_P )^2 }{ 2 \sigma_P^2 }  \right\} 
\quad \quad \pmin \leq P \leq \pmax \,,
\end{equation}
here $\mu_P \equiv \overline{\log P} = 4.8$, $\sigma_P \equiv \sigma_{\log P} = 2.3$, and $P$ is in days. The latter distribution is often used as the standard reference for the orbital size distribution of a binary population. This log-normal period distribution results in an approximately log-normal semi-major axis distribution, the shape of which is mildly dependent on the distribution over binary mass $M_T$. For a set of binaries of total mass $M_T$, the resulting semi-major axis distribution is exactly described by a log-normal distribution with mean and width 
\begin{equation}
\overline{\log a} = \tfrac{2}{3}\overline{\log P} - \tfrac{1}{3}\log\left(\tfrac{4\pi^2}{G M_T}\right)
\quad  \mbox{and} \quad
\sigma_{\log a}=\tfrac{2}{3}\sigma_{\log P} \, 
\end{equation}
respectively, where $a$ is in astronomical units \citep[see Section~4.D in][]{kouwenhoventhesis}. Throughout this paper we will consider the general form for the log-normal period distribution $f_{P; \mu,\sigma}(P)$, and study for which combination of $\mu$ and $\sigma$ the models are consistent with the observations.

\begin{table*}
  \begin{tabular}{ll l cc cc ll}
    \hline
    \hline
    Member   & Spectral   & $P$    &  \multicolumn{2}{c}{$M_1\,(\msun)$}    & \multicolumn{2}{c}{$a\,(\rsun)$} & Subgroup & Reference \\
    HIP      & type      & day    &   ($q\approx0$) & ($q=1$)     & ($q\approx0$) & ($q=1$) \\ 
    \hline
    74449 & B3\,IV& 0.90 & 5.2 & 4.0 & 6.8  & 7.8  & UCL  & \cite{buscombe1962} \\
    77911 & B9\,V & 1.26 & 3.5 & 2.2 & 7.5  & 8.0  & US   & \cite{levato1987} \\
    78265 & B1\,V & 1.57 & 10.9& 8.3 & 12.6 & 14.5 & US   & \cite{levato1987} \\
    74950 & B9\,IV& 1.85 & 4.2 & 3.2 & 10.2 & 11.8 & UCL  & \cite{andersen1993} \\
    77858 & B5\,V & 1.92 & 4.5 & 3.4 & 10.7 & 12.3 & US   & \cite{levato1987} \\
    \hline
    \hline
  \end{tabular}
  \caption{The five known binaries in \sco{} with an orbital period less than two days. Columns 1--3 list the primary star, the primary spectral type, and the measured orbital period. Columns 4 and 5 list the primary mass estimate as derived from the $V$-band magnitude, under the assumption that the mass ratio is $q\approx0$ and $q=1$, respectively. Columns~6 and~7 list extremes for the semi-major axis of the binary, as derived using Kepler's third law, under the assumption of a mass ratio $q\approx0$ (6th column), and $q=1$ (7th column). Finally, columns 8 and 9 list the subgroup of which the binary is a member and the reference for the orbital period. It is possible that a small number of closer, yet undiscovered binaries exist in \sco{}. \label{table: recovering_amin}}
\end{table*}

In our binarity dataset for \sco{}, most information on the orbital size distribution is provided by the observed angular separation distribution $\frhoobs$, the observed visual binary fraction and the observed spectroscopic binary fraction. The latter two parameters are additionally linearly dependent on the intrinsic binary fraction $\binfrac$ of \sco{}, so we derive $\binfrac$ simultaneously.

For the comparison of the observed angular separation distribution $\frhoobs$ we use the angular separation measurements of the combined observations of the KO5 and SHT datasets (see \S~\ref{section: adonisobservations} and \S~\ref{section: tokovininobservations}). The observed (visual) binary fraction $\binfracobs$ is $30\pm4\%$ for the KO5 dataset and $26\pm4\%$ for the SHT dataset. The visual binary fraction for the combined KO5/SHT dataset is $\binfracobs=31\pm4\%$ (taking into account the 23~targets that appear in both datasets, and their companions).  We will use the latter value for comparison with the simulated observations. 
For the comparison between the observed spectroscopic binary fraction and the predictions we use the observations of LEV. The LEV sample (see \S~\ref{section: levatoobservations}) consists of 53~confirmed members of \sco{}. Among these targets there are 8~SB1s, 8~SB2s, 23~RVVs, and 14~targets with a constant radial velocity (within the measurement errors). For the 23~RVVs it is unknown whether these are truly binary systems, as the radial velocity variations may also result from line profile variability. The true value for the observed spectroscopic binary fraction in the LEV dataset is thus in the range between $30\pm6\%$ and $74\pm6\%$, depending on how many of the RVVs are true binary systems. Note that it is unlikely that the majority of the observed RVVs are spurious binaries; the true spectroscopic binary fraction is likely close to 74\%.

In this section we recover the orbital size distribution of binaries in \sco{}, which can be described with either $f_a(a)$ or $f_P(P)$. We simultaneously determine the intrinsic binary fraction $\binfrac$. We determine the lower limits $\amin$ and $\pmin$ in \S~\ref{section: recover_amin} and the upper limits $\amax$ and $\pmax$ in \S~\ref{section: recover_amax}.
 The best-fitting (power-law) semi-major axis distribution is discussed in \S~\ref{section: bestfit_fa_powerlaw}, and the best-fitting (log-normal) period distribution in \S~\ref{section: bestfit_fp_gaussian}. A further discussion on the validity of the {\em independent} derivation of the orbital size distribution and $f_q(q)$ is presented in \S~\ref{section: fa_vs_fq}. Finally, we summarize our conclusions on the orbital size distribution and $\binfrac$ in \S~\ref{section: recovery_ap_summary}. 

% ============================================================================
% ============================================================================
% ============================================================================
% ============================================================================

\begin{table}
%  \begin{tabular}{p{1.2cm}c p{0.4cm}c p{0.6cm}p{1.2cm}l}
  \begin{tabular}{lccclp{1.2cm}c}
    \hline
    \hline
    Member & $\rho$ & \multicolumn{1}{c}{$\pi$} & $a_{\rm est}$    & Group  &  SpT & Ref \\
    HIP       & arcsec & \multicolumn{1}{c}{mas}   & $10^6\rsun$ \\
    \hline
64004 & 25.1 & 7.92 & 0.68 & LCC & B1.5V    &   1   \\
71860 & 27.6 & 5.95 & 0.99 & UCL & B1.5III    & 2     \\
69113 & 28.6 & 4.57 & 1.34 & UCL & B9V    &     1 \\
75647 & 30.0 & 7.79 & 0.82 & UCL & B5V    &     3  \\
69749 & 30.2 & 4.07 & 1.59 & UCL & B9IV    &    4  \\
60320 & 32.4 & 9.71 & 0.71 & LCC & Am    &      3 \\
69618 & 33.9 & 6.71 & 1.08 & UCL & B4Vne    &   3   \\
77315 & 34.7 & 7.64 & 0.97 & UCL & A0V    &     3  \\
63003 & 34.8 & 8.64 & 0.86 & LCC & B2IV-V    &  1    \\
72192 & 35.3 & 7.72 & 0.98 & UCL & A0V    &     1  \\
%69618 & 36.0 & 6.71 & 1.15 & UCL & B4Vne    &      \\
78104 & 38.3 & 7.97 & 1.03 & US & B2IV/V    &   3    \\
72984 & 39.0 & 5.93 & 1.41 & UCL & A0/A1V    &  3     \\
79374 & 41.4 & 7.47 & 1.19 & US & B2IV    &     5 \\
83693 & 43.3 & 7.73 & 1.20 & UCL & A2IV    &    3 \\
80024 & 46.7 & 6.98 & 1.43 & US & B9II/III    & 4       \\
67472 & 48.0 & 6.19 & 1.66 & UCL & B2IV/Ve    & 3      \\
78265 & 49.2 & 7.10 & 1.48 & US & B1V+B2V    &  3    \\
64661 & 60.0 & 8.04 & 1.60 & LCC & B8V    &     1    \\
65271 & 60.0 & 9.20 & 1.40 & LCC & B3V    &     5   \\
78384 & 115.0 & 6.61 & 3.73 & UCL & B2.5IV    & 5    \\
    \hline
    \hline
  \end{tabular}
  \caption{The widest known binary systems in \sco{}. For each of these binary systems we list the angular separation, the \hipp{} parallax, an estimate for the semi-major axis $a_{\rm est} \equiv D\tan \rho$, the subgroup, and the spectral type of the primary. The last column lists the reference. Note that this list must be incomplete, as very wide binaries are difficult to detect. Furthermore, several of these binaries may be optical due to confusion with background stars, i.e., not physically bound. References: (1) \cite{lindroos1985}, (2) \cite{worley1978}, (3) \cite{wds1997}, (4) \cite{oblak1978}, (5) \cite{tokovininmsc}. \label{table: recovering_amax} }
\end{table}

\subsection{The minimum period and semi-major axis} \label{section: recover_amin}

The shortest period $\pmin$ can in principle be constrained using observations of spectroscopic binaries. Table~\ref{table: recovering_amin} lists the five known binaries in \sco{} with an orbital period less than two days. These data indicate that the minimum period $\pmin$ is less than of order one day.  Binary systems with an orbital period shorter than one day may be present in \sco{}, but this orbital period is close to the physical minimum period, leading to Roche Lobe overflow. Only a very small fraction of systems is expected to have such a short orbital period, as stars in such binaries would have been in physical contact during their contraction phase.

For each short-period binary in  Table~\ref{table: recovering_amin} we have obtained an estimate for the semi-major axis, using Kepler's third law. We derive the absolute $V$-band magnitude from the observed $V$-band magnitude using the improved \hipp{} parallax and interstellar extinction for each star \citep{debruijne1999}. We estimate a lower and upper mass limit for the primary star, assuming that the mass ratio is either $q\approx0$ or $q=1$. 
As the mass ratio of each of these binaries is unknown, we show the corresponding inferred semi-major axis for the extremes $q\approx0$ and $q=1$. As the five short-period binary systems are all spectroscopic binaries, the value of the mass ratio is likely $q\ga 0.5$. From the semi-major axis estimates in  Table~\ref{table: recovering_amin} we thus conclude that $\amin$ must be of order $10\rsun$ or smaller.

As a result of selection effects, the observations cannot be used to constrain $\pmin$ and $\amin$ any further. However, physical arguments can be used to obtain estimates for the lower limits for $\pmin$ and $\amin$. 
The value of $\amin$ cannot be significantly smaller than the semi-major axis at which Roche lobe overflow occurs for one of the components of a binary system. One of the components of such a tight binary may fill its Roche Lobe if the binary separation $a$ is less than about 2--3 times the radius of that star \citep[see, e.g.,][]{eggleton1983,hilditch2001}. Assuming zero eccentricity, the estimates for $\amin$ are of order 12, 4.0, 2.5 and $1.5\rsun$ for binaries with a primary of spectral type B0\,V, A0\,V, F0\,V and G0\,V, respectively. Adopting a typical mass ratio of $q\approx0.3$, these correspond to minimum orbital periods of $\pmin=25$, 11, 8, and 6 hours, respectively. In reality the latter values are larger due to the presence of eccentric binary systems in \sco{}. A combination of these estimates indicates that $\amin \ga 2\rsun$ and $\pmin \ga 12$~hours. Binary systems with shorter orbital periods are known (see \S~\ref{section: true_method}). Although several members of \sco{} may have evolved into such a tight systems (with $P\la 12$~hours), our conclusions on the {\em primordial} binary population will not be affected by this assumption. If these binaries are present, our inferred binary fraction (\S~\ref{section: true_binaryfraction}) may be mildly underestimated.

Using observations and physical limitations, we have constrained the values $2\rsun \la \amin \la 10\rsun$ and $0.5~\mbox{day} \la \pmin \la 1$~day. Note that these are rather rough constraints. In Section~\ref{section: true_binaryfraction} we will return to this issue when deriving the binary fraction of \sco{}.

% ============================================================================
% ============================================================================
% ============================================================================
% ============================================================================

\subsection{The maximum semi-major axis and period} \label{section: recover_amax}

It is difficult to characterize the properties of the widest orbits from observations due to confusion with background stars. However, observations of the widest binaries can be used to derive a lower limit for the maximum semi-major axis $\amax$. Table~\ref{table: recovering_amax} lists the known binaries in \sco{} with angular separation larger than $25''$. Under the assumption that these wide binary systems are all physically bound, we additionally list a first-order estimate for the semi-major axis $a_{\rm est} \equiv D\tan \rho$, where $D$ is the distance to the binary, and $\rho$ the angular separation. Table~\ref{table: recovering_amax} shows that the maximum semi-major axis is of order $2\times 10^6\rsun$ or larger. Using Kepler's third law we find that the orbital period of the binary systems listed in Table~\ref{table: recovering_amax} is in the range between $\sim 0.150$~Myr and $\sim 0.25$~Myr (assuming a system mass of $5 \msun$ for each binary listed in Table~\ref{table: recovering_amax}). 

A theoretical limit to the maximum semi-major axis $\amax$ is obtained from the argument of tidal disruption of
binary systems. In the Galactic field the maximum observed semi-major axis is of
the order of $\sim 0.1$~pc \citep[$5\times 10^6$~R$_\odot$][]{bahcall1985,close1990,chaname2004}. The analysis of \cite{close1990}
shows that $\sim 3\%$ of the Galactic disk binaries has a separation larger than 0.01~pc
($0.5\times 10^6$~R$_\odot$). Binaries in the Galactic halo could be as wide as 1~pc due to the lower ambient stellar density \citep{chaname2004}. The results of the wide binary searches and dynamical simulations above indicate that binary systems with a semi-major axis larger than a certain value $a_{\rm tidal}$ are unstable in the Galactic tidal field and are ionized quickly, and that $a_{\rm tidal} \approx 0.2$~pc ($=9\times 10^6\rsun$) .

OB~associations are expanding groups and are likely unbound \citep{blaauw1964A,brown1999}, so that they will dissolve in the field star population within a few tens of Myr. If \sco{} is indeed an expanding association, the association must have been denser in the past. The upper limit for the semi-major axis may therefore be smaller than $a_{\rm tidal}$. For binary systems with a total mass of $5 \msun$, this tidal limit corresponds to $\sim 4$~Myr. Note that if such a binary system would exist in Upper Scorpius, it would have completed only one revolution since its birth. For lower-mass binaries, the period corresponding to $a=0.2$~pc would be significantly larger.

The goal of our study is to find the primordial binary population, so that theories on star formation can be constrained with observations. For our purpose the exact value of $\amax$ is not a parameter of crucial importance. Binary systems with a semi-major axis of order $a=0.2$~pc have an orbital period larger than $\sim 4$~Myr. In the context of star formation, the components of these wide binaries may possibly be considered as single, as both stars may have formed practically independent from each other.

Combining the information above, we have constrained $2 \times 10^6\rsun \la \amax \la 8.9 \times 10^6\rsun$, or alternatively $0.15~\mbox{Myr} \la \pmax \la 4~\mbox{Myr}$. In Section~\ref{section: true_binaryfraction} we will return to this issue when deriving the binary fraction of \sco{}.

% ============================================================================
% ============================================================================
% ============================================================================
% ============================================================================

\subsection{A power-law semi-major axis distribution?} \label{section: bestfit_fa_powerlaw}

\begin{figure*}[!bt]
  \centering
  \begin{tabular}{ccc}
    \includegraphics[width=0.31\textwidth,height=!]{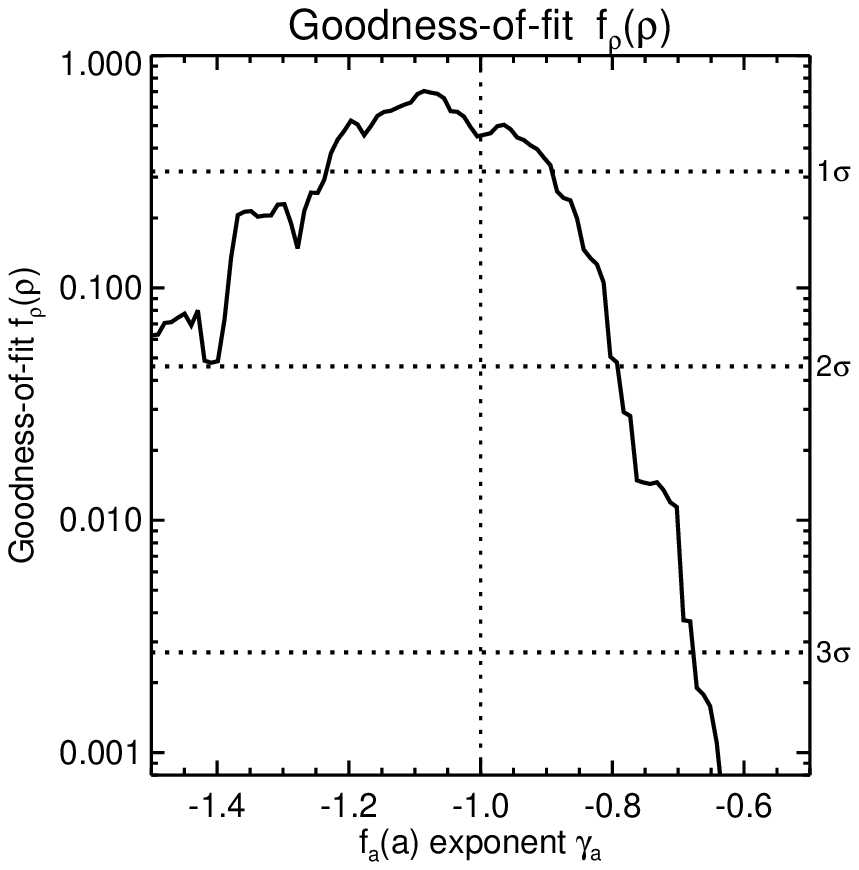} &
    \includegraphics[width=0.31\textwidth,height=!]{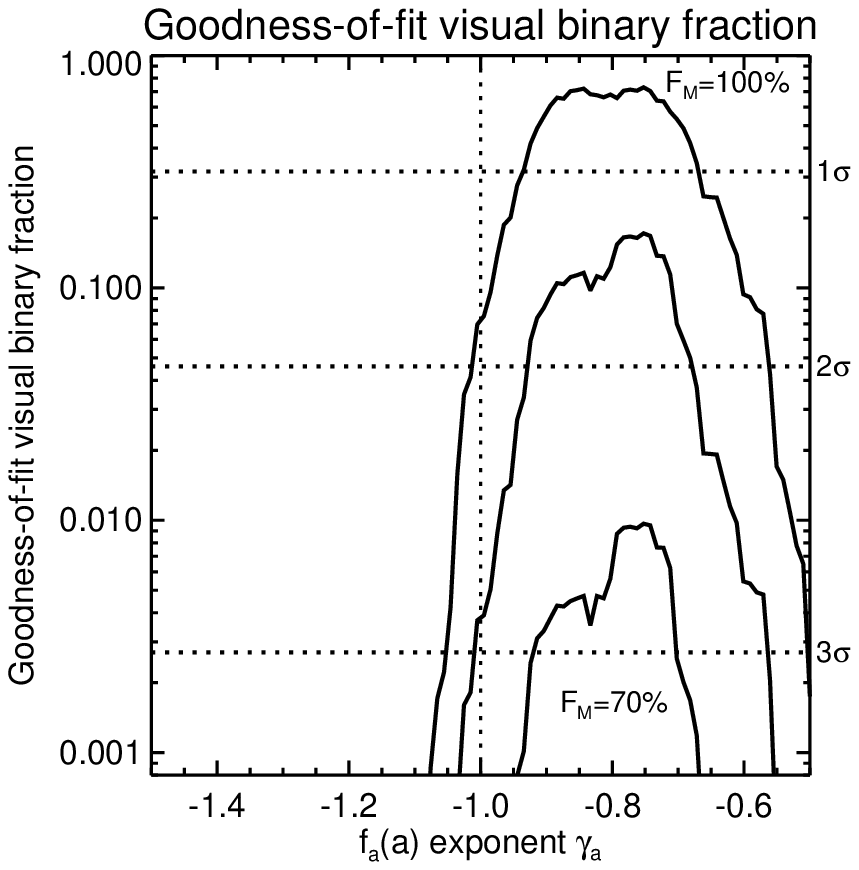} &
    \includegraphics[width=0.31\textwidth,height=!]{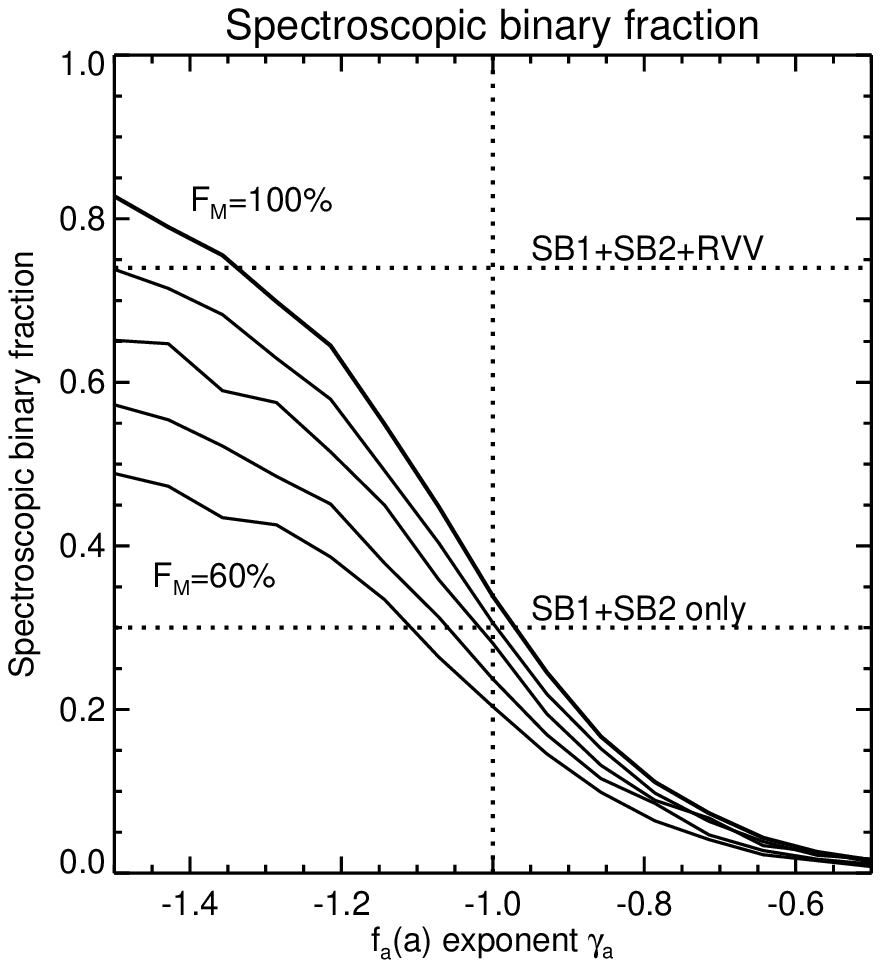} \\
  \end{tabular}
  \caption{How well can the intermediate-mass binary population in \sco{} be described by a semi-major axis distribution of the form $f_a(a) \propto a^{\gamma_a}$? 
    {\em Left}: the goodness-of-fit for the comparison between the observed angular separation distribution (from the KO5 and SHT datasets) and that predicted by the models with different values of $\gamma_a$. A large value for the goodness-of-fit means that the model predictions are consistent with the observations. The $1\sigma$, $2\sigma$ and $3\sigma$ confidence limits for rejection of the model are indicated with the horizontal dotted lines.
    {\em Middle}: the goodness-of-fit for the comparison between the observed visual binary fraction (from the KO5 and SHT datasets) and that predicted by the models with an intrinsic binary fraction of 70\% (bottom curve), 80\%, 90\% and 100\% (top curve). 
    {\em Right:} the predicted (SB1, SB2 and RVV) spectroscopic binary fraction (for the LEV dataset) as a function of $\gamma_a$. The five curves indicate an intrinsic binary fraction of 60\% (bottom curve), 70\%, 80\%, 90\% and 100\% (top curve). The bottom horizontal line indicates the lower limit for the observed spectroscopic binary fraction, where only the SB1s and SB2s are included, assuming that the radial velocity variations in {\em all} RVVs are caused by line-profile variability rather than binarity. The top horizontal line indicates the upper limit for the spectroscopic binary fraction, assuming that all RVVs are indeed binaries. 
    A combination of the results in these three panels indicates that the semi-major axis distribution for intermediate mass stars in \sco{} can be described by $f_a(a) \propto a^{\gamma_a}$ with exponent $\gamma_a \approx -1.0$ (vertical dotted line in each panel), commonly known as \"{O}pik's law. An intrinsic binary fraction close to 100\% is required in order to produce the large number of visual and spectroscopic binaries detected in \sco{}.
    \label{figure: bestfit_opik_r} }
\end{figure*}

\begin{figure}[!bt]
  \includegraphics[width=0.5\textwidth,height=!]{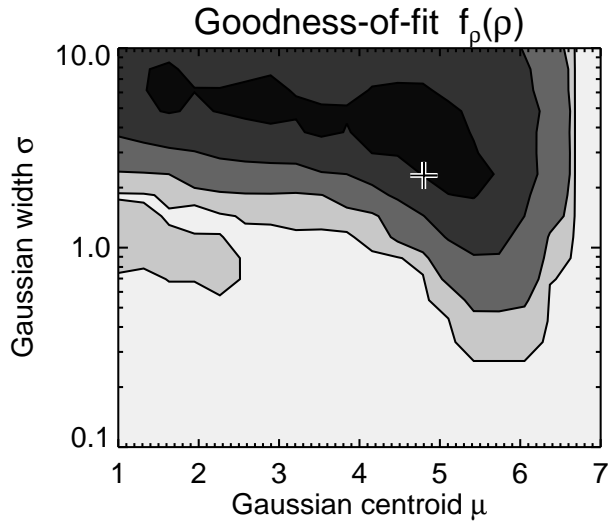} \\
  \caption{How well does a model with a log-normal period distribution $f_{P;\mu,\sigma}(P)$ with centroid $\mu$ and width $\sigma$ reproduce the observed angular separation distribution $\frhoobs$? The darkest colors in this figure indicate the best-fitting models. From dark to light, the contours indicate the $\tfrac{1}{3}\sigma$, $1\sigma$, $2\sigma$, and $3\sigma$ confidence limits for rejection of the model, respectively. The cross at $(\mu=4.8,\sigma=2.3)$ indicates the distribution derived by \cite{duquennoy1991} for solar-type stars in the solar neighbourhood. The comparison with the observed visual and spectroscopic binary fraction is shown in Figure~\ref{figure: bestfit_dm_r2}.  \label{figure: bestfit_dm_r1} }
\end{figure}

Now that the lower and upper limits for the semi-major axis and period distributions are constrained, it is possible to evaluate these distributions.
We first study the possibility that the binary population is characterized by a semi-major axis distribution of the form $f_a(a) \propto a^{\gamma_a}$, and determine for which value of $\gamma_a$ the simulated observations correspond best to the observations. We compare the observed angular separation distribution $\tilde{f}_{\rho}(\rho)$ with that of the simulated observations using the KS test, and we compare the observed visual and spectroscopic binary fractions with the predictions using the Pearson $\chi^2$ test. In both cases, we test the hypothesis that the observations and simulations are realizations of the same underlying association model. In our models we adopt semi-major axis limits $5\rsun < a < 5\times 10^6 \rsun$, a thermal eccentricity distribution $f_e(e) = 2e$ and a mass ratio distribution of the form $f_q(q) \propto q^{\gamma_q}$ with $\gamma_q=-0.4$. Later in this paper we will discuss the validity of these assumptions.

Figure~\ref{figure: bestfit_opik_r} shows the results of this comparison for models with a varying value of $\gamma_a$. The left-hand panel shows the probability associated with the KS test when comparing the observed $\frhoobs$ and that of the simulated observations. The models with $-1.15 \la \gamma_a \la -0.90$ produce an observed angular separation distribution very similar to that of KO5/SHT, and consequently have a large goodness-of-fit value.

The middle panel shows the probability associated with the Pearson $\chi^2$ test when comparing the observed visual binary fraction $\binfracobs=31\pm4\%$ and that of the simulated observations for models as a function of $\gamma_a$ with varying intrinsic binary fraction $\binfrac$. Models with $-0.9\la \gamma_a \la -0.6$ produce a visual binary fraction similar to that of the observations, but only if the intrinsic binary fraction is close to 100\%. 

The right-hand panel shows the predicted spectroscopic binary fraction (SB1, SB2 and RVV) as a function of $\gamma_a$ and $\binfrac$. Note that, unlike the middle panel of Figure~\ref{figure: bestfit_opik_r}, we do not show the goodness-of-fit between for the observed and predicted spectroscopic binary fraction, as we only have lower and upper limits of the former (see below). The two horizontal dotted lines in the right-hand panel indicate the lower limit of $30\pm 6\%$ (SB1 and SB2) and the ``upper limit'' of $74\pm 6\%$ (SB1, SB2 and RVV) for the observed spectroscopic binary fraction $\binfracobs$. The spectroscopic binary fraction increases with decreasing $\gamma_a$, as the average orbital separation decreases. Models with $\gamma_a \la -1$ result in a spectroscopic binary fraction that is compatible with the observations. For models with $\gamma_a\approx -1$, an intrinsic binary fraction very close to 100\% is required in order to reproduce the lower limit for the observed spectroscopic binary fraction. 

Considering the results in all three panels of Figure~\ref{figure: bestfit_opik_r}, we find that the semi-major axis distribution for intermediate mass stars in \sco{} is best described by  $f_a(a) \propto a^{\gamma_a}$ with $\gamma_a \approx -1.0$. The observations are therefore reasonably consistent with models with \"{O}pik's law ($\gamma_a = -1$). Models with slightly smaller values of $\gamma_a$ correspond somewhat better to $\frhoobs$ and the spectroscopic binary fraction, while those with slightly larger values of $\gamma_a$ correspond better to the observed visual binary fraction. The observed visual and spectroscopic binary fractions are only reproduced by models with an intrinsic binary fraction close to $100\%$.

% ============================================================================
% ============================================================================
% ============================================================================
% ============================================================================

\begin{figure*}[!bt]
  \begin{tabular}{c}
    \includegraphics[width=\textwidth,height=!]{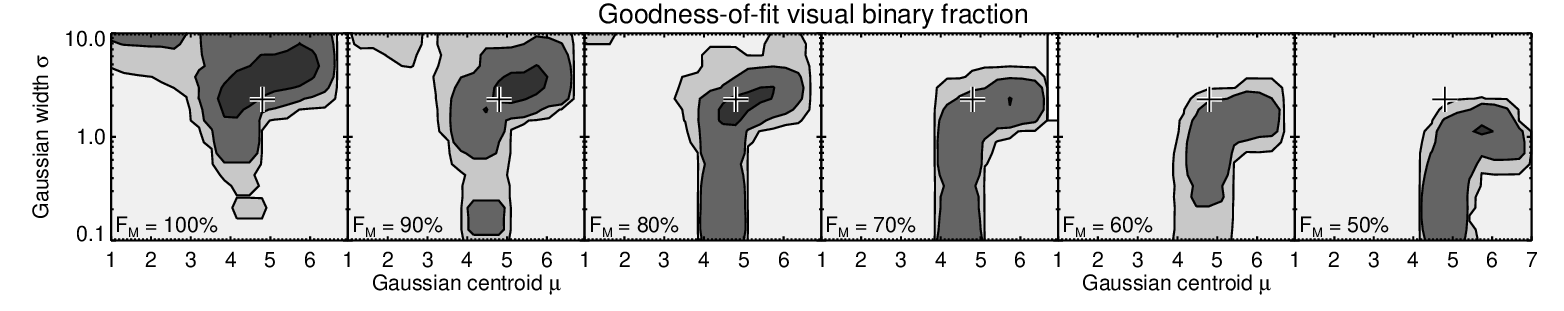} \\
    \includegraphics[width=\textwidth,height=!]{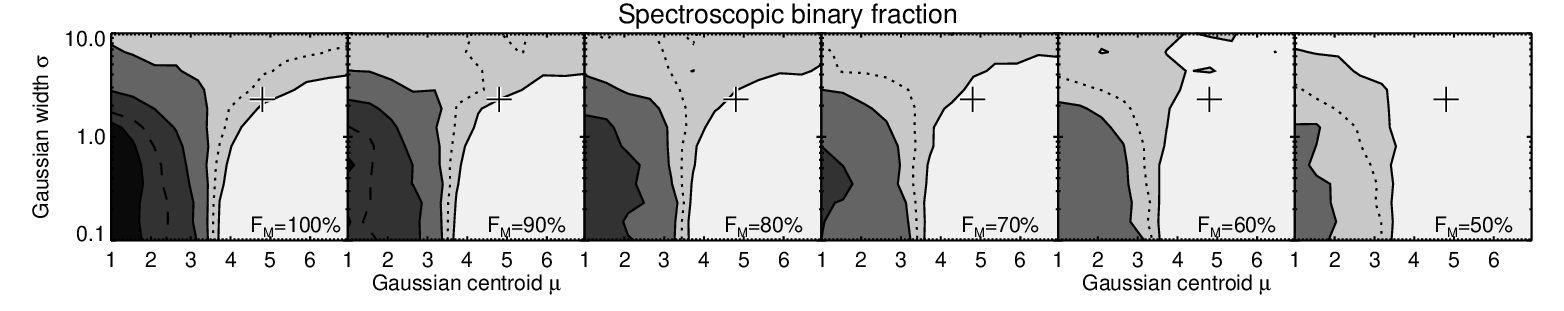} \\
  \end{tabular}
  \caption{How well does a log-normal period distribution with centroid $\mu$ and width $\sigma$ reproduce the observed visual and spectroscopic binary fractions? 
    {\em Top}: the goodness-of-fit for the comparison between the observed and predicted visual binary fraction, as a function of $\mu$ and $\sigma$. Each panel represents a set of models with a different intrinsic binary fraction ranging from $\binfrac=100\%$ (left) to $50\%$ (right). The darkest colors indicate the best-fitting models. From dark to light, the three contours indicate the $1\sigma$, $2\sigma$, and $3\sigma$ confidence limits for rejection of the model, respectively. The cross at $(\mu=4.8,\sigma=2.3)$ in each panel indicates the distribution derived by \cite{duquennoy1991} for solar-type stars in the solar neighbourhood. 
    {\em Bottom}: The predicted (SB1, SB2 and RVV) spectroscopic binary fraction as a function of $\mu$, $\sigma$ and $\binfrac$ (note that these panels do {\em not} indicate the goodness-of-fit). The spectroscopic binary fraction is indicated with the gray-shade in each panel, where the solid contours indicate the combinations of $\mu$ and $\sigma$ which predict a spectroscopic binary fraction (from black to white) of 80\%, 60\%, 40\% and 20\%, respectively. The other curves indicate the extreme constraints imposed by the observations, of $30\pm6\%$ (dashed contour) and $74\pm6\%$ (dotted contour), respectively, where the higher value is more likely to represent reality. 
    This figure, combined with the $f_\rho(\rho)$ comparison of Figure~\ref{figure: bestfit_dm_r1}, shows that {\em if} the binary population can be described with a log-normal period distribution, we require (1) a binary fraction near 100\%, and (2) a large value for $\sigma$ (which mimics \"{O}pik's law).  
    \label{figure: bestfit_dm_r2} }
\end{figure*}

\subsection{A log-normal period distribution?} \label{section: bestfit_fp_gaussian}

We also study the possibility that the binary population can be described with a log-normal period distribution (Equation~\ref{equation: true_duquennoyperiods}) for a certain combination of $\mu$ and $\sigma$. We simulate associations with period distribution $f_{P;\mu,\sigma}(P)$ for various values of $\mu$ and $\sigma$, and compare simulated observations with the real observations. Again, we consider the observed angular separation distribution $\frhoobs$, the observed visual binary fraction and the observed spectroscopic binary fraction. The best-fitting log-normal distribution is obtained by combining the results in Figures~\ref{figure: bestfit_dm_r1} and~\ref{figure: bestfit_dm_r2}, which we describe below.

In Figure~\ref{figure: bestfit_dm_r1} we compare the observed angular separation distribution $\frhoobs$ for models with varying values of $\mu$ and $\sigma$. The darkest regions in the figure indicate the best-fitting combinations of $\mu$ and $\sigma$. In general, the models with a large value for $\sigma$ fit the observations well. These models have a very broad period distribution, resulting in a $f_a(a)$ that is very similar to \"{O}pik's law, which was shown to be consistent with the observations in \S~\ref{section: bestfit_fa_powerlaw}. Models with a period distribution similar to that found by \cite{duquennoy1991}, i.e., $\mu\approx4.8$ and $\sigma\approx2.3$ are consistent with $\frhoobs$ as well. Models with $5 \la \mu \la 6$ are consistent with the observed distribution $\frhoobs$, as for these models the peak of $\log P$ falls in the visual binary regime. This results in an angular separation distribution similar to  \"{O}pik's law.
Figure~\ref{figure: bestfit_dm_r1} shows that most models with a large value of $\sigma$ are consistent with the observed angular separation distribution. These models resemble those with \"{O}pik's law, which was shown to be consistent with the observations in the previous section. Of the models with a large $\sigma$, only those with extremely wide binary populations ($\mu \ga 7$) produce an incompatible $\frhoobs$, as only the tail of the distribution then falls in the visual regime. 

Figure~\ref{figure: bestfit_dm_r2} shows for the same set of models the results for the observed visual and spectroscopic binary fractions. The top panels show the goodness-of-fit of the visual binary fraction, for models with a different value of $\mu$, $\sigma$, and for different intrinsic binary fractions $\binfrac$ (indicated in the bottom-left corner of each panel). The darkest regions in each panel indicate the best-fitting combinations of $\mu$ and $\sigma$. The bottom panels show the observed spectroscopic binary fraction as a function of $\mu$, $\sigma$ and $\binfrac$, where the darkest regions correspond to the highest spectroscopic binary fraction. The lower limit ($30\pm6\%$) and ``upper limit'' ($74\pm6\%$) for the spectroscopic binary fraction in the LEV sample are indicated with the dotted and dashed contours, respectively. Any well-fitting model has a combination of $\mu$ and $\sigma$ within these contours, preferably near the left-hand contour, as a significant number of the RVVs detected by LEV is likely to be a true binary. In each panel of Figure~\ref{figure: bestfit_dm_r2} we indicate the values $(\mu,\sigma)=(4.8,2.3)$ found by \cite{duquennoy1991} for solar-type stars in the solar neighbourhood.

By comparing the top and bottom panels in Figure~\ref{figure: bestfit_dm_r2} we find that the intrinsic binary fraction is larger than $\approx 70\%$, whatever the values of $\mu$ and $\sigma$ are. Models with $\binfrac \approx 100\%$ are most consistent with the observations. 
The \cite{duquennoy1991} period distribution with $(\mu,\sigma)=(4.8,2.3)$ is consistent with the observed angular separation distribution and visual binary fraction, but underpredicts the number of spectroscopic binaries, even if the intrinsic binary fraction is 100\%, and all RVVs detected by LEV are spurious. Similar models with a larger value for $\sigma$ are consistent with the observations. However, these period distributions are very broad, making the binary population indistinguishable from that resulting from \"{O}pik's law.

% ============================================================================
% ============================================================================
% ============================================================================
% ============================================================================

\subsection{The ambiguity in deriving the orbital size distribution and $f_q(q)$} \label{section: fa_vs_fq}

\begin{figure*}[!bt]
  \centering
  \begin{tabular}{cc}
    \includegraphics[width=0.39\textwidth,height=!]{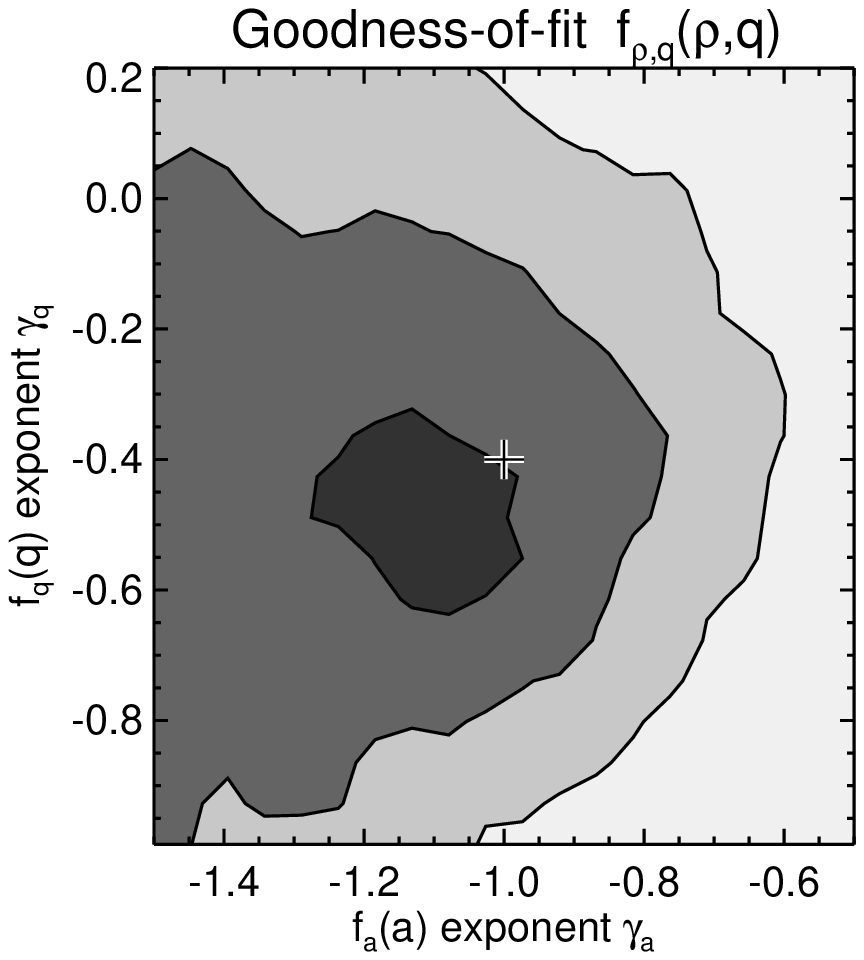} &
    \includegraphics[width=0.59\textwidth,height=!]{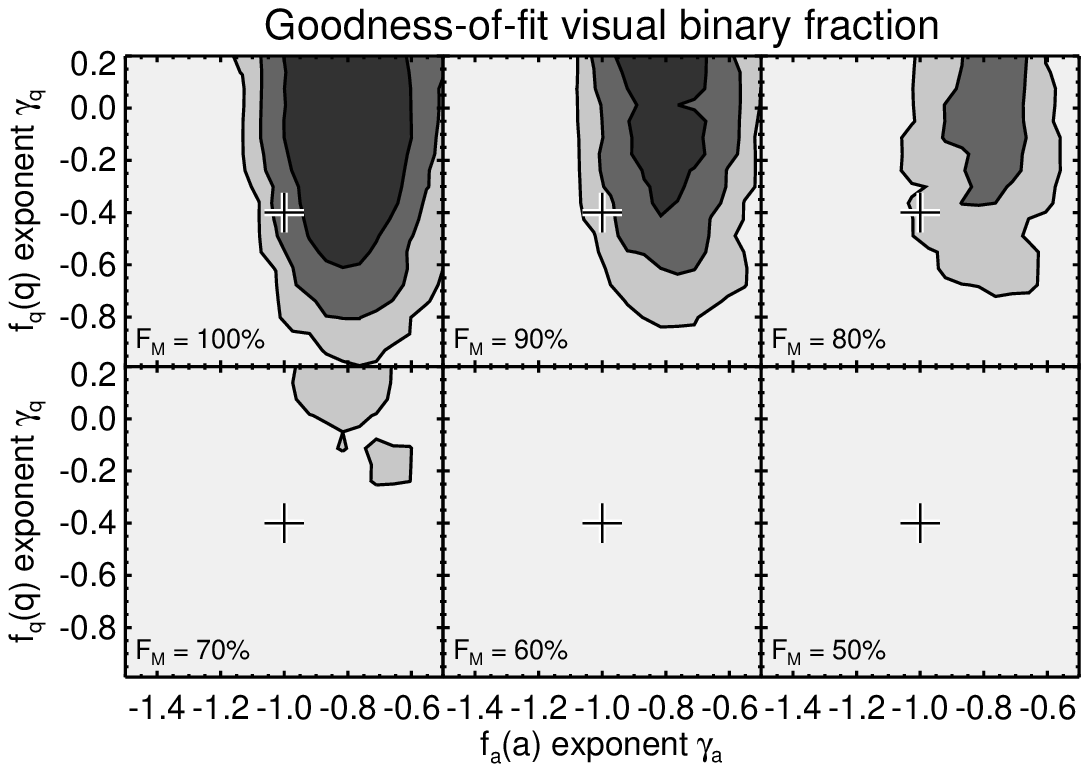} \\
  \end{tabular}
  \caption{How well do models with a semi-major axis distribution $f_a(a) \propto a^{\gamma_a}$ and a mass ratio distribution $f_q(q) \propto q^{\gamma_q}$ correspond to the observations? This figure is the multi-dimensional equivalent of Figures~\ref{figure: bestfit_opik_q} and~\ref{figure: bestfit_opik_r}, and indicates that with the {\em independent} derivation of $\gamma_q$ and $\gamma_a$ we have found a globally best-fitting solution. 
{\em Left}: the consistency between the observed two-dimensional distribution $\tilde{f}_{\rho,q}(\rho,q)$ and that of simulated observations, for models with different values of $\gamma_a$ ({\em horizontal axis}) and $\gamma_q$ ({\em vertical axis}), respectively. The darkest colors indicate the best-fitting models. From dark to light, the three contours indicate the $1\sigma$, $2\sigma$, and $3\sigma$ confidence limits for rejection of the model, respectively. The cross in each panel indicates the values of $\gamma_a$ and $\gamma_q$ that we adopt to describe the properties of \sco{}. 
{\em Right}: the consistency between the observed visual binary fraction and that of the simulated observations, for models with different values of $\gamma_a$ and $\gamma_q$. Each panel corresponds to a set of models with a different binary fraction, which is indicated in the bottom-left corner. The darkest colors indicate the best-fitting models. From dark to light, the three contours indicate the $1\sigma$, $2\sigma$, and $3\sigma$ confidence limits for rejection of the model, respectively. 
Models with a binary fraction smaller than $\approx 100\%$ are inconsistent with the observations, as they underpredict the observed number of visual binaries. The cross in each panel indicates the best-fitting values of $\gamma_a \approx -1.0$ and $\gamma_q \approx -0.4$ for \sco{}. \label{figure: bestfit_opik_rq} }
\end{figure*}

\begin{figure*}[!bt]
  \centering
  \begin{tabular}{c}
    \includegraphics[width=1\textwidth,height=!]{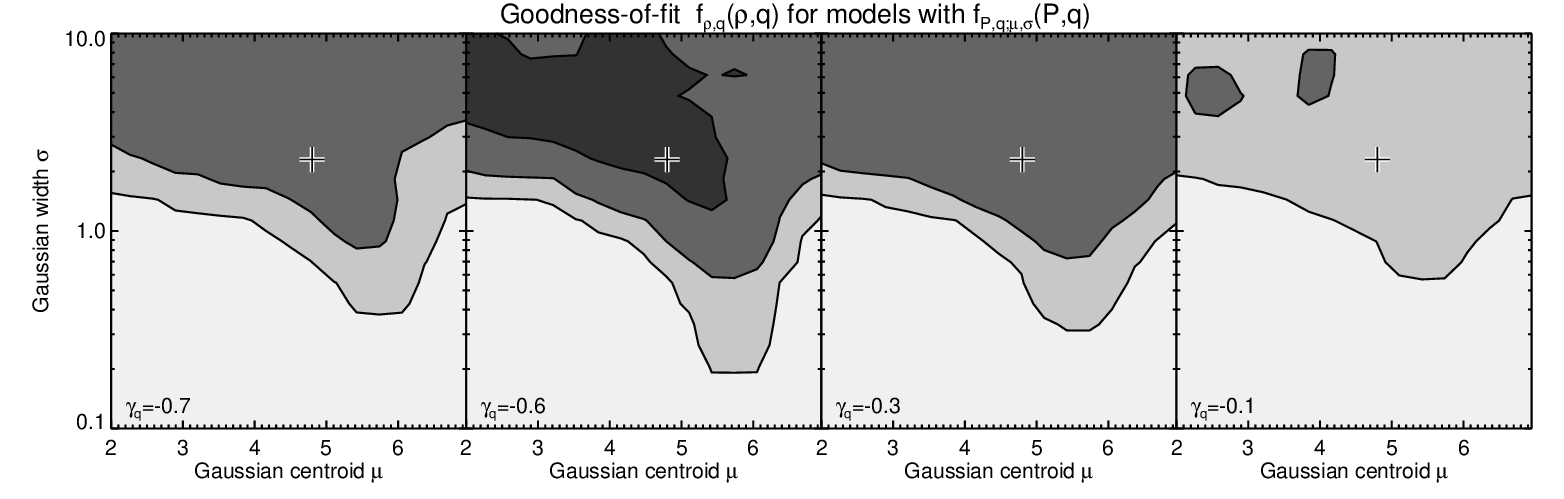}\\
    \includegraphics[width=1\textwidth,height=!]{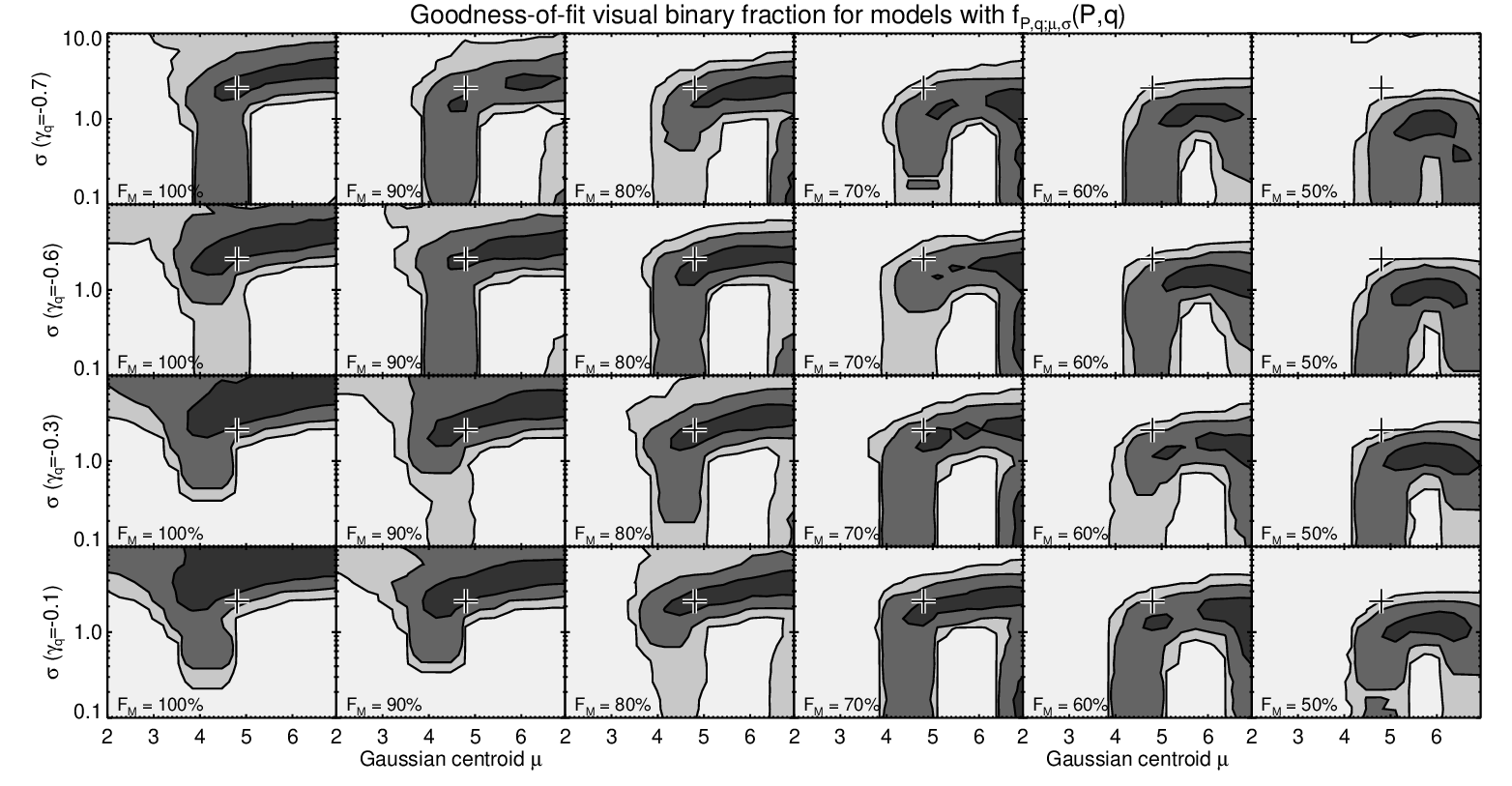}\\
  \end{tabular}
  \caption{How well do models with a log-normal period distribution $f_{P;\mu,\sigma}(P)$ and a mass ratio distribution $f_q(q) \propto q^{\gamma_q}$ correspond to the observations?
{\em Top:} a comparison for the observed two-dimensional distribution $\tilde{f}_{\rho,q}(\rho,q)$ for models with different values of $\mu$, $\sigma$ and $\gamma_q$. The cross in each panel indicates the values observed by \cite{duquennoy1991}: $\mu=4.8$ and $\sigma=2.3$. The four panels correspond to models with a mass ratio distribution exponent $\gamma_q=-0.7$ ({\em left}), $-0.6$, $-0.3$ and $-0.1$ ({\em right}), respectively (cf. Figure~\ref{figure: bestfit_dm_r2}). The goodness-of-fit is best for the dark-shaded regions. Contours are plotted at $1\sigma$, $2\sigma$ and $3\sigma$ confidence regions for rejection of the model. {\em Bottom}: the comparison for the observed visual binary fraction for models with different values of $\mu$, $\sigma$ and $\gamma_q$. The binary fraction, indicated in the bottom-left corner of each panel, decreases from 100\% (left) to 50\% (right). Contours are plotted at $1\sigma$, $2\sigma$ and $3\sigma$ confidence regions for rejection of the model.  \label{figure: bestfit_dm_rq} }
\end{figure*}

In Sections~\ref{section: true_pairingfunction} and~\ref{section: true_smadistribution} we have constrained the orbital size distribution and the mass ratio distribution $f_q(q)$ independently. A risk associated with this approach is that one may find a {\em local}, rather than a {\em global} set of best-fitting solutions. In this section we demonstrate that our choice has been appropriate, and that we have indeed found the best-fitting model. 

Even if the orbital size distribution and $f_q(q)$ are intrinsically independent, there could be a correlation between $a$ (or $P$) and $q$ among the detected binaries. The latter is important for the analysis of the results of a spectroscopic binary survey \citep[e.g.,][]{kobulnicky2006}, but is of less importance for visual binaries. In a visual binary survey the orbital size is related to the angular separation $\rho$, and $q$ is related to the brightness contrast between primary and companion. As the maximum detectable magnitude difference depends on the separation between the binary components, a correlation between the orbital size and $q$ may be introduced among the closest detected companions.

In order to study whether we have indeed found the {\em global} set of best-fitting solutions, we extend the parameter space, and make comparisons for models with varying orbital size distribution {\em and} varying $f_q(q)$. We limit ourselves to mass ratio distributions of the form $f_q(q) \propto q^{\gamma_q}$. We compare the two-dimensional distribution $\tilde{f}_{\rho,q}(\rho,q)$ and the visual binary fraction $\binfracobs$ of the simulated observations of the combined KO5/SHT dataset with that of the observations. We compare the observed distribution $\tilde{f}_{\rho,q}(\rho,q)$ using the two-dimensional KS test \citep[e.g.,][]{numericalrecipies}, and we compare the observed visual binary fraction using the Pearson $\chi^2$ test. 

Figures~\ref{figure: bestfit_opik_rq} and~\ref{figure: bestfit_dm_rq} show plots corresponding to those in Figures~\ref{figure: bestfit_opik_q} and~\ref{figure: bestfit_opik_r}, but now with the additional free parameter $\gamma_q$. The results for a power-law semi-major axis distribution in Figure~\ref{figure: bestfit_opik_rq} are similar to those derived in Sections~\ref{section: true_pairingfunction} and~\ref{section: true_smadistribution}.  The best-fitting values for $\gamma_a$ and $\gamma_q$ in these diagrams are equal (within the error bars) to those of the independent derivations of $f_a(a)$ and $f_q(q)$. Note that the adopted combination $(\gamma_a\approx -1.0,\gamma_q \approx -0.4)$ does not fall in the darkest regions of Figure~\ref{figure: bestfit_opik_rq}. The comparison for $\tilde{f}_{\rho,q}(\rho,q)$ (left panel) suggests somewhat smaller values of $\gamma_a$ and $\gamma_q$, while the comparison for the visual binary fraction suggests larger values. An inspection of both comparisons suggests that our adopted combination $(\gamma_a\approx -1.0,\gamma_q \approx -0.4)$ is among the best-fitting combinations for both $\tilde{f}_{\rho,q}(\rho,q)$ and $\binfracobs$. The figure also clearly shows again that an intrinsic binary fraction close to 100\% is required. We do not attempt to further constrain the combination $(\gamma_a,\gamma_q)$, as this would involve a combination of the {\em independent} comparisons for $\tilde{f}_{\rho,q}(\rho,q)$ and $\binfracobs$, as this would require accurate knowledge of the relative weights that should be assigned to each of these comparisons.

The results for the log-normal period distribution $f_{P;\mu,\sigma}(P)$ shown in Figure~\ref{figure: bestfit_dm_rq} also indicate that our independent derivation of $f_{P;\mu,\sigma}(P)$ and $f_q(q)$ is justified. The top panels show that the models with a mass ratio distribution exponent $\gamma_q=-0.7$, $-0.6$, $-0.3$ and $-0.1$ produce a distribution  $\tilde{f}_{\rho,q}(\rho,q)$ that is less consistent than that for the models with the best-fitting solution $\gamma_q=-0.4$, for any combination of $\mu$ and $\sigma$. The bottom panels of Figure~\ref{figure: bestfit_dm_rq} indicate for the observed visual binary fraction the best-fitting combinations of $\mu$, $\sigma$, $\gamma_q$ and $\binfrac$ (darkest regions). These comparisons indicate that, if the log-normal period distribution with $(\mu,\sigma)=(4.8,2.3)$ holds, the intrinsic binary fraction should be $\binfrac \ga 70\%$. Again, we do not attempt to find the optimum combination $(\mu,\sigma,\gamma_q,\binfrac)$ by combining the results of all panels, as this would require a detailed prescription of the relative weights given to the comparisons of $\tilde{f}_{\rho,q}(\rho,q)$ and the visual binary fraction, which is not trivial. Note that in our models we adopt a binary fraction that is independent of semi-major axis and mass ratio.

% ============================================================================
% ============================================================================
% ============================================================================
% ============================================================================

\subsection{Conclusions on $f_a(a)$, $f_P(P)$ and $\binfrac$} \label{section: recovery_ap_summary}

In the sections above we have constrained the properties of the orbital size distribution among intermediate-mass binaries in \sco{}, using the visual binary surveys of KO5 and SHT and the spectroscopic binary survey of LEV. The tightest binaries have $2\rsun \la \amin \la 10\rsun$, corresponding to $0.5~\mbox{day} \la \pmin \la 1$~day. The widest orbits have $2 \times 10^6\rsun \la \amax \la 8.9 \times 10^6\rsun$, or $0.15~\mbox{Myr} \la \pmax \la 4~\mbox{Myr}$. We have compared the observations with models that have {\em either} a power-law semi-major axis distribution {\em or} a log-normal period distribution, in order to find the best-fitting orbital size distribution.

We considered a power-law semi-major axis distribution $f_a(a)$ with varying exponent $\gamma_a$. Models with a semi-major axis distribution exponent $\gamma_a \approx -1.0$ are consistent with the observed angular separation distribution and the observed visual binary fraction, but only if the binary fraction is close to 100\%. We have additionally considered models with a log-normal period distribution $f_{P;\mu,\sigma}(P)$ with varying centroid $\mu$ and width $\sigma$. The distribution derived by \cite{duquennoy1991}, with $\mu=4.8$ and $\sigma=2.3$ is shown to be consistent with the observed angular separation distribution and the observed visual binary fraction for models with an intrinsic binary fraction near 100\%. The observed spectroscopic binary fraction, however, is rather large compared to those predicted by the above models, even if all stars are in binary systems.

\"{O}pik's law could be considered as resulting from a very broad log-normal period distribution ($\sigma \rightarrow \infty$). Although \"{O}pik's law and the log-normal period distribution with $(\mu,\sigma)=(4.8,2.3)$ are consistent with the observations (i.e., they cannot be excluded), the observations indicate that the best description of the binary population in \sco{} is intermediate, with $2.3 < \sigma < \infty$.

We have constrained the mass ratio distribution and the orbital size distribution separately. Our investigation of this approach in section~\ref{section: fa_vs_fq} indicates that we have indeed found the global best-fitting model.

% ============================================================================
% ============================================================================
% ============================================================================
% ============================================================================

\section{Recovering the eccentricity distribution} \label{section: true_eccentricitydistribution}

The eccentricity distribution for binaries in \sco{} is not known. The field star population, on the other hand, has been studied in detail by \cite{duquennoy1991}, who surveyed solar-type stars in the solar neighbourhood for binarity and find the following composite eccentricity distribution. The close binaries ($P<10$~days) have a negligible eccentricity, presumably due to tidal circularization, although they may have formed in nearly circular orbits. Binaries with a period $10<P<1000$~days show a bell-shaped eccentricity distribution that peaks near $e=0.3$. Binaries with a larger orbital period tend to have a thermal eccentricity distribution $f_e(e)=2e$, though the presence of large-eccentricity binaries are inferred by incompleteness corrections. The latter distribution is expected from energy equipartition \citep{heggie1975} as a result of multiple soft encounters. A similar result was found by \cite{mathieu1994}, who studied the properties of pre-main-sequence binaries. 

Both studies indicate that the eccentricity distribution may well be a function of orbital period, which is perhaps also the case for \sco{}. The very tight binaries tend to have more or less circular orbits.
Tidal circularization may have affected these binaries in \sco{}. The tidal circularization period $P_{\rm circ}$, defined as the orbital period at which a binary orbit with a typical initial orbital eccentricity circularizes, is $P_{\rm circ}=7-10$ days for young stellar populations \citep{meibom2005}. 
The intermediate-period binaries have a somewhat flat eccentricity distribution, possibly reflecting the formation process. The widest binaries tend to have a more thermal eccentricity distribution, which may (partially) be a result of binary-binary interactions. It must be noted, however, that the eccentricity distribution for the latter two period ranges is very difficult to measure because of the large orbital periods.

It is practically impossible to derive the properties of the eccentricity distribution from observations of visual binaries \citep[e.g.,][]{kouwenhoventhesis}. On the other hand, this means that the results we derived from the observations of visual binaries (i.e. $f_q(q)$, $f_a(a)$, $f_P(P)$ and $\binfrac$) are {\em unaffected} by our choice for $f_e(e)$ in the models. In principle it is possible to constrain $f_e(e)$ using observations of spectroscopic and astrometric binaries. However, only one \sco{} binary with an astrometrically determined orbital solution is known. 

Only the measurements of the spectroscopic binaries can thus be used to constrain the eccentricity distribution for binaries in \sco{}. The properties of the eccentricity distribution derived below are only valid for the short-period ($P \la 14$~days; $a \la 40\rsun$) and intermediate-mass binaries. Caution should be taken with generalizing the results for the full population of \sco{}.
As we do not model the detection of SB1s, SB2s and RVVs in detail, we cannot compare the eccentricity distribution resulting from the simulated observations (SB1s, SB2s, and RVVs) with the observed eccentricity distribution (SB1s and SB2s only). However, it is possible to set several constraints on the intrinsic eccentricity distribution using the observations of LEV. Due to the limited number of eccentricity measurements, the complicated selection effects, and the possible correlation between $e$ and $P$, we do not attempt to fully recover $f_e(e)$. Instead, we consider three commonly adopted eccentricity distributions, and compare the resulting predictions with the observations.

In our analysis we will consider three frequently used eccentricity distributions: the flat eccentricity distribution $f_{\rm flat}(e) = 1$, the thermal eccentricity distribution $f_{2e}(e) = 2e$, and the single-valued eccentricity distribution $f_{e_0}(e) = \delta(e-e_0)$ with $0\leq e_0 < 1$. Simply by analyzing Figure~\ref{figure: observed_distributions}, we can rule out an eccentricity distribution of the form $f_{e_0}$, where all binaries have the same eccentricity $e_0$. For associations with $f_e(e)=f_{e_0}(e)$ the distribution $\tilde{f}_e(e)$ is unbiased, as all binaries have $e_O$. An error is associated with each eccentricity measurement, so that $\tilde{f}_e(e)$ is broader than $f_e(e)$. In the spectroscopic binary sample, the error in the eccentricity is of order 0.05, ruling out the best-fitting distribution with $e_0=0.27$ with $\sim 3\sigma$ confidence. 

The observed eccentricity distribution is in better agreement with the thermal distribution $f_{2e}$ and the flat distribution $f_{\rm flat}$. For these distributions the relative (intrinsic) fraction of binary systems with $e<0.5$ is:
\begin{equation}
E_{2e} = \frac{\mbox{\# binaries with $e<0.5$}}{\mbox{\# binaries}} = 25\% \,,
\end{equation}
\begin{equation}
E_{\rm flat} = \frac{\mbox{\# binaries with $e<0.5$}}{\mbox{\# binaries}} = 50\% \,.
\end{equation}
In the LEV sample, 14~of the 16~targets for which the orbital elements are determined have $e<0.5$, and only 2~have $e>0.5$. The apparent overabundance of low-eccentricity ($e < 0.5$) systems (see also Figure~\ref{figure: observed_distributions}) can partially be explained by selection effects. Highly eccentric systems spend a large fraction of their orbit near apastron, and are therefore more difficult to detect. 

In \S~\ref{section: true_smadistribution} we have shown that the observed distribution $\frhoobs$ and binary fraction $\binfracobs$ of visual binaries are consistent with both \"{O}pik's law and a log-normal period distribution with $\binfrac\approx 100\%$. For each of these distributions we predict the number of spectroscopic binaries (SB1s, SB2s and RVVs) in the LEV dataset. For models with \"{O}pik's law, our simulations indicate that LEV would be able to detect $\sim 38\%$ of the binary systems spectroscopically (i.e., as SB1, SB2 or RVV) for a flat eccentricity distribution, and $\sim36\%$ for a thermal eccentricity distribution. If the log-normal period distribution of \cite{duquennoy1991} holds, we find $\sim27\%$ for a flat eccentricity distribution, and $\sim24\%$ for a thermal eccentricity distribution, respectively. The observed spectroscopic binary fraction is thus only mildly dependent on $f_e(e)$, and decreases with an increasing average orbital eccentricity.
The spectroscopic binary fraction is also proportional to $\binfrac$.
The expected fraction of spectroscopic binary systems with $e < 0.5$ among the observed targets is thus given by:
\begin{eqnarray} \label{equation: e_values_levato1}
  \tilde{E}_{2e,{\rm DM}}    & = 24\% \times \binfrac \times E_{2e}       &\leq 6.0\% \,; \\
  \tilde{E}_{2e,\mbox{\tiny \"{O}pik}}    & = 36\% \times \binfrac \times E_{2e}       &\leq 9.0\% \,; \\\label{equation: e_values_levato2}
  \tilde{E}_{{\rm flat, DM}} & = 27\% \times \binfrac \times E_{\rm flat} &\leq 13.5\% \,; \\\label{equation: e_values_levato3}
  \tilde{E}_{\mbox{\tiny flat, \"{O}pik}} & = 38\% \times \binfrac \times E_{\rm flat} &\leq 19.0\% \,,\label{equation: e_values_levato4}
\end{eqnarray}
where the upper limits on the right-hand side are obtained for models with an intrinsic binary fraction $\binfrac=100\%$. 

For each of the 23~RVVs in the LEV sample we do not know whether it is truly a binary, and if so, we do not know its eccentricity. We therefore constrain $f_e(e)$ by considering two extreme cases for $\tilde{E}_{\rm LEV}$: (1) none of the RVVs have $e<0.5$ (i.e., they are either spectroscopically single, or they are binaries with $e>0.5$), and (2) all RVVs are binary systems with $e<0.5$. In these extreme cases, $ \tilde{E}_{\rm LEV} $ is constrained by
\begin{equation}
  %\tilde{E}_{\rm LEV} & = \frac{\mbox{\#~binaries~with~$e<0.5$}}{\mbox{53~targets}} & =  26\pm6\%-70\pm6\% \,.
  26\pm6\% < \tilde{E}_{\rm LEV} < 70\pm6\% \,.
\end{equation}
The observed spectroscopic binary fraction $\binfracobs$ is inconsistent with the predictions in Equations~\ref{equation: e_values_levato1}--\ref{equation: e_values_levato3}. The models with the log-normal period distribution (with $\mu=4.8$ and $\sigma=2.3$) or the thermal eccentricity distribution do not reproduce the observations; they underpredict the number of binaries in the LEV sample with $e<0.5$, even if {\em all} RVVs detected by LEV are spurious (which is very unlikely), and the model binary fraction is 100\%. The model with \"{O}pik's law, a flat eccentricity distribution, and a binary fraction of 100\% (Equation~\ref{equation: e_values_levato3}) is consistent with the observations, but only if most of the RVV candidate binaries are (spectroscopically) single stars exhibiting line profile variability. 

The analysis above indicates that the short-period, intermediate-mass binaries do not follow the thermal eccentricity distribution, but have lower eccentricities on average (see, however, \S~\ref{section: triples}). Note that the discussion on the eccentricity distribution is constrained only by a small number of short-period binaries with a massive primary. In the LEV sample, 12 out of the 16 systems with a measured eccentricity have a period $P<P_{\rm circ}$, suggesting that the derived properties of the eccentricity distribution may not be representative for the binary population of \sco{}. Accurate radial velocity and astrometric surveys among intermediate- and low-mass members of \sco{} are necessary to further characterize the eccentricity distribution.

% ============================================================================
% ============================================================================
% ============================================================================
% ============================================================================

\section{The binary fraction in \sco{}} \label{section: true_binaryfraction}

\begin{table*}
  \centering
  \begin{tabular}{|l|r|r|ccc|r|ccc|}
    \hline
    \hline
    Dataset & \multicolumn{1}{c}{Obs.}  & \multicolumn{4}{|c|}{$f_{\gamma_a}(a)$, $\gamma_a=-1$}     & \multicolumn{4}{c|}{$f_{\mu,\sigma}(P)$, $\mu=4.8$, $\sigma=2.3$}   \\
    \hline
    & \multicolumn{1}{c|}{$\binfracobs$} 
    & \multicolumn{1}{c|}{$\binfracobs$} & $\binfrac$$(1\sigma)$ & $\binfrac$$(2\sigma)$ & $\binfrac$$(3\sigma)$ 
    & \multicolumn{1}{c|}{$\binfracobs$} & $\binfrac$$(1\sigma)$ & $\binfrac$$(2\sigma)$ & $\binfrac$$(3\sigma)$ \\
    \hline  
    KO5                  & $30\pm4$      & 26     & $\approx 100$ & $91-100$ & $81-100$ & 35    & $97-100$ & $70-100$ & $61-100$ \\
    KO6                  & $<82\pm8$     & 27     & $\dots$ & $\dots$ & $\dots$         & 37    & $\dots$ & $\dots$ & $\dots$ \\
    SHT                  & $26\pm4$      & 23     & $94-100$ & $77-100$ & $63-100$      & 34    & $64-97$ & $52-100$ & $43-100$ \\
    \hline
    LEV SB1/SB2 only     & $>30\pm6$     & 36     & $>68$ & $>54$ & $>42$               & 24    & $\approx 100$ & $>81$ & $>64$ \\
    LEV SB1/SB2/RVV      & $<74\pm6$     & 36     & $\dots$ & $\dots$ & $\dots$         & 24    & $\dots$ & $\dots$ & $\dots$ \\
    BRV analysis -- RVV  & $<60\pm5$     & 39     & $\dots$ & $\dots$ & $\dots$         & 27    & $\dots$ & $\dots$ & $\dots$ \\
    BRV SB1/SB2/RVV      & $<66\pm5$     & 39     & $\dots$ & $\dots$ & $\dots$         & 27    & $\dots$ & $\dots$ & $\dots$ \\
    \hline
    HIP (X)/(O)/(G)      & $4\pm1$       & $<31$  & $>12$ & $>10$ & $>8$                & $<25$ & $>14$ & $>11$ & $>9$ \\
    HIP (X)/(O)/(G)/(S)  & $9\pm1$       & $<31$  & $>25$ & $>22$ & $>19$               & $<25$ & $>30$ & $>26$ & $>22$ \\
    HIP (C)              & $15\pm2$      & 13     & $\approx 100$ & $96-100$ & $86-100$ & 18    & $77-94$ & $68-100$ & $61-100$ \\
    \hline
    \hline
  \end{tabular}
  \caption{The observed binary fraction (in \%) and inferred intrinsic binary fraction (in \%) for the different datasets discussed in this paper. Columns~1 and~2 list the dataset and the observed binary fraction. The predicted observed binary fraction resulting from \"{O}pik's law for each dataset (adopting $\binfrac=100\%$) is listed in column~3, followed by the inferred $1\sigma$, $2\sigma$ and $3\sigma$ confidence ranges of the inferred binary fraction. The predicted observed binary fraction for the log-normal period distribution with $\mu=4.8$ and $\sigma=2.3$ (adopting $\binfrac=100\%$) and corresponding confidence ranges for the intrinsic binary fraction are listed in columns 7--10. The adopted association parameters are listed in Table~\ref{table: subgroups}. For models in columns 3--6 the semi-major axis range is $5\rsun \leq a \leq 5\times 10^6 \rsun$.  For models in columns 7--10 the period range is $0.7~\mbox{day} \leq P \leq 3\times 10^8~\mbox{day}$. For each model we assume a mass ratio distribution of the form $f_q(q) \propto q^\gamma_q$ with $\gamma_q=-0.4$ and a thermal eccentricity distribution. The comparison between observations and simulated observations indicates that the binary fraction among intermediate-mass binaries in \sco{} is close to 100\% ($\protect\ga 70\%$ with $3\sigma$ confidence).   
    \label{table: fm_inferred} }
\end{table*}

In the previous sections we have constrained the pairing function, the mass ratio distribution, the orbital size distribution and the eccentricity distribution. The analysis of each of these properties indicates that the binary fraction among intermediate mass stars in \sco{} has to be close to 100\%. 

Table~\ref{table: fm_inferred} lists the fraction of detected binaries among the targeted sample, for the different surveys. Results are listed for the observations and for the simulated observations of models with \"{O}pik's law and for those with the \cite{duquennoy1991} period distribution. For each of these models we have adopted an intrinsic binary fraction $\binfrac = 100\%$. The values listed in Table~\ref{table: fm_inferred} are upper limits, and are proportional to $\binfrac$. 

The observed visual binary fraction is consistent with \"{O}pik's law, but also with the log-normal period distribution ($\mu=4.8$, $\sigma=2.3$). A binary population with intermediate properties ($2.3 < \sigma < \infty$), however, is most consistent with the observations.
The observed spectroscopic binary fraction cannot be compared directly with the simulated observations, as the true nature of the detected radial velocity variables is unknown. The observed spectroscopic binary fraction can be limited by $30\pm6\%$ (in the unlikely case that {\em all} RVVs are spurious) and $74\pm6\%$ (if all RVVs are indeed binaries). Even if the intrinsic binary fraction is 100\%, models with \"{O}pik's law only marginally reproduce the observed spectroscopic binary fraction, while those with the \cite{duquennoy1991} period distribution underpredict the observed value. In the case that 50\% or 100\% of the RVVs are indeed binaries, the log-normal period distribution can be excluded with $4\sigma$ and $5\sigma$ confidence, respectively, {\em if} our other assumptions hold (see \S~\ref{section: triples}).

Our model for the \hipp{} selection effects for the (X), (O), and (G) binaries is not strict enough. This is because in our models we consider {\em all} binaries with the properties listed in Table~\ref{table: true_hipparcosbiases} as astrometric binaries. In reality, however, only a subset of these would have been marked as an astrometric binary by \hipp{} (see \S~\ref{section: modeling_hipparcos}). The predicted astrometric binary fraction can therefore only be used as an upper limit when a comparison with the observations is made. The predicted fraction of visual binaries in category (C) on the other hand, is well-modeled. Models with a binary fraction $\binfrac\approx100\%$ correctly predict the number of (C) binaries.

An inspection of the observed binary fractions for the different datasets (Table~\ref{table: fm_inferred}) and their associated statistical errors above indicates that the binary fraction among intermediate mass stars in \sco{} must be larger than $\approx85\%$ at the $2\sigma$ confidence level, and larger than $\approx70\%$ at the $3\sigma$ confidence level. We find that models with a binary fraction of 100\% are most consistent with the observations; models with a significantly smaller binary fraction are inconsistent with the observed binary fraction.  \cite{mason1998} found that almost all high-mass stars are multiple. Our complementary results indicate that also practically all intermediate mass stars form in a binary or multiple system.

% ============================================================================
% ============================================================================
% ============================================================================
% ============================================================================

\subsection{Dependence of the derived binary fraction on the limits of $a$ and $P$} \label{section: limits_dependence}

For the models with a power-law semi-major axis distribution $f_a(a)$ we have adopted a semi-major axis distribution $5\rsun \leq a \leq 5\times 10^6\rsun$, and for models with a log-normal period distribution $f_{P;\mu,\sigma}(P)$ we have adopted the period range $0.7~\mbox{day}\leq P \leq 3\times 10^8~\mbox{day}$. These lower and upper limits are derived from the observational data and have their associated uncertainties (see \S~\ref{section: recover_amin} and~\ref{section: recover_amax}). In this section we briefly study the effect of these uncertainties on our results.

The binary population properties derived from the angular separation distribution of visual binaries are not affected by our choice of the lower and upper limits. The results in the left-hand panels in Figures~\ref{figure: bestfit_opik_q}, \ref{figure: bestfit_opik_r}, \ref{figure: bestfit_dm_r2}, \ref{figure: bestfit_opik_rq}, the top panels in Figure~\ref{figure: bestfit_dm_rq}, and the derived distributions $f_a(a)$, $f_P(P)$ and $f_q(q)$ are insensitive to our choice for the tightest or widest orbits.
The inferred binary fraction, however, does depend on these limits.  Table~\ref{table: limitsdependence} shows the expected visual and spectroscopic binary fractions, for models with different $\amin$ and $\amax$ (adopting \"{O}pik's law), and different $\pmin$ and $\pmax$ \citep[adopting the period distribution of][]{duquennoy1991}. If, for a given model, we increase the value of $\amax$ or $\pmax$, this leads to smaller values for the simulated visual and spectroscopic binary fractions. If we increase $\amin$ or $\pmin$, this leads to smaller spectroscopic binary and a larger visual binary fraction. In other words, an overestimation of $\amax$ or $\pmax$ leads to an {\em overestimation} of the inferred intrinsic binary fraction $\binfrac$. An overestimation of $\amin$ or $\pmin$ leads to an {\em overestimation} of $\binfrac$ as derived from the observed spectroscopic binary fraction, but to an {\em underestimation} of $\binfrac$ as derived from the observed visual binary fraction.

Fortunately, the properties of the tightest and widest binaries are reasonably well constrained by observations. Table~\ref{table: limitsdependence} shows that the uncertainty in the observed visual binary fraction is approximately 2\%, as a result of the errors in $\amin$ and $\amax$. The corresponding error in the spectroscopic binary fraction ranges from about 2\% for the log-normal period distribution to approximately 5\% for \"{O}pik's law. The uncertainty in the inferred intrinsic binary fraction is slightly smaller as we combine the visual and spectroscopic results. An overestimation of $\amin$, for example, leads to a larger visual binary fraction but a smaller spectroscopic binary fraction, so that the systematic uncertainty on $\binfrac$ partially cancels out. 
The effects of the uncertainty in the lower and upper limits of $f_a(a)$ and $f_P(P)$ are therefore limited. We estimate that this uncertainty results in a systematic error of $\la 2-4\%$ in the inferred intrinsic binary fraction $\binfrac$.

\begin{table}
  \begin{tabular}{cccc}
    \hline
    \hline
    \multicolumn{2}{l}{$f_{\tiny\mbox{\"{O}pik}}(a)$ limits} & $\tilde{F}_{\rm M,KO5,SHT}$ (\%) & $\tilde{F}_{\rm M,LEV}$ (\%)  \\
    \hline
    $2\rsun$  & $2.0\times 10^6\rsun$ & 25  & 42 \\
    $2\rsun$  & $8.9\times 10^6\rsun$ & 24  & 37 \\
    $10\rsun$ & $2.0\times 10^6\rsun$ & 27  & 34 \\
    $10\rsun$ & $8.9\times 10^6\rsun$ & 25  & 30 \\
    \hline
     \multicolumn{2}{l}{$f_{\rm DM}(P)$ limits} & $\binfracobs$ VB (\%)  & $\binfracobs$ SB (\%)\\
    \hline
    $0.5$ day & $0.15$ Myr & 38  & 25 \\
    $0.5$ day & $4.00$ Myr & 35  & 22 \\
    $2.0$ day & $0.15$ Myr & 39  & 22 \\
    $2.0$ day & $4.00$ Myr & 36  & 20 \\
    \hline
    \hline
  \end{tabular}
  \caption{The simulated observed binary fraction for models with a different assumption for the tightest and widest binaries. The first and second column list the properties of the tightest and widest orbits, respectively. The third and fourth column list the visual binary fraction for the simulated combined KO5/SHT observations and the spectroscopic binary fraction for the simulated LEV observations, respectively. For each model we adopt an intrinsic binary fraction $\binfrac=100\%$ and a thermal eccentricity distribution. The values listed in columns~3 and~4 are proportional to $\binfrac$. The statistical errors on the listed binary fraction are $0.5-1\%$. The values listed in this table provide error estimates for our results in Table~\ref{table: fm_inferred}. The uncertainties in the limits of $f_a(a)$ and $f_P(P)$ result in a systematic error of $<2-4\%$ in the inferred intrinsic binary of the binary population. \label{table: limitsdependence}}
\end{table}

% ============================================================================
% ============================================================================
% ============================================================================
% ============================================================================

\section{The primordial binary population in \sco{}} \label{section: primordialbinarypopulation}

\sco{} is a young \ob{} (5--20~Myr) with a low stellar density ($\sim 0.1 \msun\,{\rm pc}^{-3}$), comparable to the stellar density in the solar neighbourhood. One therefore expects that stellar and dynamical evolution have only mildly altered the properties of the binary population. In this section we study which binary systems in \sco{} may have changed one or more of their parameters since the time of formation. We first consider whether stellar (and binary) evolution has affected the binary population, and subsequently investigate the importance of dynamical evolution.

For stellar populations with an age less than about 20~Myr the fraction of binaries of which the properties have changed due to stellar evolution is small. The hydrogen-burning time as a function of initial mass and initial rotational velocity was studied by \cite{meynet2000}. According to their model, stars more massive than $32-42\msun$ (O6\,V) in US, and stars more massive than $10-11\msun$ (B2\,V) in UCL and LCC have evolved away from the main sequence. The lower and upper limits for each subgroup correspond to stars with an initial rotational velocity of 0~\kms{} and 300~\kms{}, respectively.
The most massive star in the US subgroup is \object{Antares}, an M1\,Ib supergiant with an initial mass of $22.5\pm2.5 \msun$ \citep{preibisch2002}, indicating that we may have mildly overestimated the upper mass limit for the US subgroup. The derived mass limit for the UCL and LCC subgroups corresponds roughly to the observed range of spectral types. An extrapolation of the mass distribution (Equation~\ref{equation: preibischimf}) to infinity suggests that only 3--8 stars of spectral type~O ($M \ga 20\msun$) have formed in \sco{}.

Several of the binary systems with a primary initially more massive than the turn-off mass will have produced a compact component. These compact objects and their previous companion stars may have obtained a significant kick during the supernova event, resulting in a space velocity significantly different from that of their parent association (runaway stars).  The properties of compact binary systems including at least one star with a higher initial mass may have changed due to binary evolution. For $\ga 99.9\%$ of the binaries in \sco{}, however, both components have an initial mass (much) less than those mentioned above; stellar evolution will not have affected these binaries.

Changes of the binary population due to stellar evolution can therefore be neglected. 
The change due to dynamical evolution is more complicated to quantify. Below we calculate which binary systems are affected by dynamical evolution, at the {\em current} stellar density of the association. After that we briefly discuss the consequences of the fact that the subgroups of \sco{} may have been denser in the past.

The subgroups of \sco{} can be approximated as roughly spherical, with radii of order $R = 20-30$~pc (see Table~\ref{table: subgroups}). The structure and kinematics of \sco{} have been studied by  \cite{debruijne1999}, who finds that the one-dimensional internal velocity dispersion for each of the subgroups  is $\sigmaoned \la 1.0-1.5$~km\,s$^{-1}$, corresponding to a three-dimensional velocity dispersion $\sigmathreed \la 1.7-2.6$~km\,s$^{-1}$. The crossing time of each subgroup is thus $\tau_{\rm cross} = R/\sigmathreed \ga 10$~Myr. 
\cite{preibisch2002} find that the US subgroup contains approximately 2525~members more massive than $0.1\msun$. The total number of members $N$ in the substellar regime depends on the unknown values of $M_{\rm min}$ and $M_\beta$ in Equation~\ref{equation: preibischimf}. For reasonable value of $M_{\rm min}$ and $M_\beta$, about 44\% of the members have a mass smaller than $0.1\msun$, so that the total number of systems equals $N\approx 4500$. Adopting a mass ratio distribution of the form $f_q(q) \propto q^{-0.4}$ and a binary fraction of 100\%, the total mass of US is of order $2280\msun$. We assume that the UCL and LCC subgroups contain a similar number of members. The mass density of each subgroup is therefore of order $0.04\pm 0.02 \msun \,\mbox{pc}^{-3}$, corresponding to $0.09 \pm 0.04$~systems (or $0.17 \pm 0.09$~individual stars) per cubic parsec.

An estimate for the relaxation time $\tau_{\rm relax}$, i.e., the typical timescale at which a system of $N$ {\em single} stars has ``forgotten'' the initial conditions, is 
\begin{equation} \label{equation: relaxationtime}
  \tau_{\rm relax} \approx \frac{N}{8 \ln N} \times \tau_{\rm cross} \ga 670~\mbox{Myr} 
\end{equation}
\citep{binneytremaine}. As the cross-section for binary systems is significantly larger than that of single stars, they interact more frequently, so that in reality the timescale of dynamical evolution is shorter than given by Equation~\ref{equation: relaxationtime}. Whether a binary system experiences a strong encounter, depends not only on the properties of the association, but also on the size (more specifically, the binding energy) of the binary systems. A binary system is often classified as {\em hard} or {\em soft}, depending on whether it is likely to experience a strong encounter within a relaxation time. A binary system is called ``hard'' if its orbital energy $E_{\rm orb}$ is significantly larger than the mean kinetic energy $\langle K \rangle$ of the surrounding stars: $E_{\rm orb} > 3\langle K \rangle$ \citep{heggie1975,hills1975}; otherwise a binary system is called ``soft''. The orbital energy of a binary system is given by
\begin{equation}
E_{\rm orb} = \frac{qGM_1^2}{2a}
\end{equation}
where $q$ is the mass ratio, $G$ the gravitational constant, $M_1$ the primary mass, and $a$ the semi-major axis of the binary system. The mean kinetic energy is given by 
\begin{equation}
\langle K \rangle = \tfrac{1}{2} \langle M_T \rangle \, \sigmathreed^2 \,,
\end{equation}
where $ \langle M_T \rangle $ is the average mass of a field object, and $\sigmathreed$ its three-dimensional velocity dispersion. Assuming a binary fraction of 100\%, the median mass of a field object (i.e., another binary system in the association), is $ \langle M_T \rangle \approx 0.21\msun$. The three-dimensional velocity dispersion is $\sigmathreed \la 1.7-2.6$~km\,s$^{-1}$ \citep{debruijne1999}. The hard/soft boundary $a_{\rm hs}$ for a binary star with primary mass $M_1$ and mass ratio $q$ is thus given by $E_{\rm orb} = 3\langle K \rangle$, i.e.,
\begin{equation}
  \begin{tabular}{rl}
  $a_{\rm hs}$ & $= \dfrac{G}{3\langle M_T \rangle \sigmathreed^2}\, q M_1^2$ \\
               & $\ga 350 \ q \left(\frac{M_1}{\msun} \right)^2~\mbox{AU} = 7.5\times 10^4 \ q \left(\frac{M_1}{\msun} \right)^2\rsun $\,.
  \end{tabular}
\end{equation}
The latter value is in good agreement with the widest known binaries in \sco{} (see \S~\ref{section: recover_amax}), which all have a primary mass of order $3\msun$. The datasets studied in this paper contain mainly A and B type members of \sco{}, which have $M_1 \ga 1.4\msun$. Among the binaries more massive than $1.4\msun$, about $72\pm5\%$ are hard, while the widest $28\pm5\%$ of the binaries are soft, if \"{O}pik's law holds. If the \cite{duquennoy1991} period distribution holds, $77\pm3\%$ of these binaries are hard, and $24\pm3\%$ are soft. The quoted errors include only the contribution from the uncertainty in $\amin$ and $\amax$. The population of A and B stars must therefore be close to primordial, given the young age of the association. 
On the other hand, among the lower-mass binaries ($M_1 \approx 0.5\msun$) about half of the binaries is hard. The other half of the binaries (i.e., the soft low-mass binaries) is expected to experience a close encounter within a relaxation time. However, as shown above, the (current) relaxation time is significantly larger than the age of \sco{}, suggesting that this population is unlikely to have changed significantly.
The majority of binaries, in particular the intermediate-mass binaries studied in this paper, are unlikely to have experienced a strong encounter during the lifetime of the association, assuming that the density did not change over time. 

The density of \sco{} may have been (much) higher during the star formation process. Simulations of \cite{vandenberk2007}, however, suggest that even at higher stellar densities, the binarity population is only mildly affected by dynamical interactions. In the latter paper, simulations of small ($N =100$) and initially dense ($\sim 10^5$ systems per cubic parsec) expanding star clusters are presented. Their simulations include triple systems for which the period of the outer orbit reaches up to 1000~years. Although these simulated clusters are more than a million times denser than the current density of \sco{}, the properties of the binary population do not change significantly within 20~Myr. As about 60\% of the binary systems in \sco{} has an orbital period smaller than 1000~years, these binaries are expected not to have changed significantly since the birth of the association, even if the density of \sco{} was orders of magnitude larger at the time of formation.

The discussion above indicates that the current binary population is likely very similar to the current binary population, suggesting that {\em all} intermediate mass stars have formed in a binary or multiple system. However, due to the as yet unknown initial conditions of \sco{} (such as the initial stellar density), N-body simulations of expanding \obs{} are necessary to further constrain the primordial binary population of \sco{}. By varying the initial conditions, evolving each simulated association for 5--20~Myr, and comparing the outcome with the observations, the primordial binary population can be recovered. The latter technique is referred to as inverse dynamical population synthesis \citep[see][]{kroupa1995a,kroupa1995c}.

% ============================================================================
% ============================================================================
% ============================================================================
% ============================================================================

\section{Discussion} \label{section: true_discussion}

% ============================================================================
% ============================================================================
% ============================================================================
% ============================================================================

\subsection{Triple and higher-order systems} \label{section: triples}

In our analysis we have constrained the properties of the binary population primarily using observations of visual binaries. Our simulations have indicated that even models with an intrinsic binary fraction of $100\%$ produce a rather low spectroscopic binary fraction as compared to the observations of LEV. Throughout this paper, however, we have ignored the presence of triple and higher-order systems. These systems are known to be present in \sco{}: the observed higher-order multiplicity fraction among \hipp{} members of \sco{} is $(T+Q+\dots)/(S+B+T+Q+\dots)=8.4\%$ (see Table~\ref{table: subgroups}). This value is a lower limit due to the presence of undetected companions. The studies of \cite{tokovininsmekhov2002} and \cite{correia2006} have indicated that $20-30\%$ of the wide visual binaries have a spectroscopic subsystem. 

The presence of these triple and higher-order systems among the \hipp{} members of \sco{} could explain the apparent underabundance of spectroscopic binaries in our models at least partially, if one would only include the outer components of a multiple system in the statistics. We do not attempt to calculate the contribution of spectroscopic subsystems in this paper, as this would require detailed a-priori knowledge about the triple population. Further detailed observational studies are necessary to characterize the properties of these systems, and to derive the primordial binary-and-multiple-systems population of \sco{}.

% ============================================================================
% ============================================================================
% ============================================================================
% ============================================================================

\subsection{Comparison with \cite{heacox1995}}

\cite{heacox1995} derived the mass ratio distribution for \sco{} binaries with an intermediate-mass primary from the LEV dataset. He derived the mass ratio distribution using the observations of all 22~binaries with spectroscopic elements, resulting in a mean mass ratio $\langle q \rangle \approx 0.27$ with a standard derivation $\sigma_q \approx 0.04$. 
In our analysis we adopt a mass ratio distribution of the form $f_q(q) \propto q^{\gamma_q}$. If this distribution is adopted, the mean mass ratio is $\langle q \rangle = (\gamma_q+1)/(\gamma_q+2)$, and the standard deviation is $\sigma_q = (\gamma_q+1)/(\gamma_q+3) - \langle q \rangle^2$. Our best-fitting solution has $\gamma_q = -0.4 \pm 0.1$, so that $\langle q \rangle = 0.37 \pm 0.04$ and $\sigma_q=0.090 \pm 0.002$ (formal errors). These quantities, based on our analysis of {\em visual} binaries, are mildly larger than the values derived by \cite{heacox1995}. In the latter paper the mass ratio distribution is derived from a much smaller, but {\em independent} sample. \cite{heacox1995} derives the mass ratio distribution for {\em spectroscopic} binaries with early B-type primaries, while in our analysis we derive the distribution from observations of {\em visual} binaries with B and A-type primaries.

% ============================================================================
% ============================================================================
% ============================================================================
% ============================================================================

\subsection{Background stars}

\begin{table*}
%  \begin{tabular}{p{0.6cm} p{0.3cm}p{0.2cm} cc cc cc p{1.4cm}}
  \begin{tabular}{l cc cc cc cc l}
    \hline
    \hline
    Group            & $l$ & $b$                 & N$_{\rm KO5}$ & N$_{\rm KO5}$ & N$_{\rm KO6}$ & N$_{\rm KO6}$ & N$_{\rm SHT}$ & N$_{\rm SHT}$ & $C$ (arcsec$^{-2}$) \\
    \multicolumn{3}{l}{$K_S$ limit}        & $< 12$ mag& $<18$ mag & $<12$ mag & $<18$ mag & $<12$ mag & $<18$ mag& \\ 
    \hline
    US               & $352^\circ$ & $20^\circ$  & 0.002 & 0.19  & 0.001 & 0.10  & 0.002 & 0.15 &  $1.93 \times 10^{-9}$  \\
    UCL              & $328^\circ$ & $13^\circ$  & 0.010 & 0.87  & 0.006 & 0.47  & 0.008 & 0.67 &  $8.71 \times 10^{-9}$  \\
    LCC              & $299^\circ$ & $6^\circ$   & 0.016 & 1.35  & 0.009 & 0.73  & 0.013 & 1.05 &  $13.5 \times 10^{-9}$  \\
    GP               & $300^\circ$ & $0^\circ$   & 0.057 & 4.78  & 0.031 & 2.59  & 0.044 & 3.70 &  $48.0 \times 10^{-9}$  \\
    \hline
    \hline
  \end{tabular}
  \caption{The number of background stars expected {\em per field of view} for each of the three imaging surveys discussed in this paper, according to our model. The field of view size is 361.2~arcsec$^2$ for KO5, 196.0~arcsec$^2$ for KO6, and 280.1~arcsec$^2$ for SHT. We list the results for three different pointings: to the centers of the three subgroups US, UCL, and LCC, and for the intersection of LCC with the Galactic plane. Columns~4, 6, and 8 list the number of background stars brighter than $K_S=12$~mag {\em per field of view}. Columns~5, 7, and 9 list the number of background stars brighter than $K_S=18$~mag {\em per field of view}. The last column lists for each group the value of the normalization constant $C$ in Equation~\ref{equation: backgroundstarsformula}. \label{table: backgroundstarcounts}}
\end{table*}

In imaging surveys it is not always clear whether a secondary is a physical companion or a background star. In our models for the association this is obviously not a problem (as we know the true nature of each star in our model), but in practice it is. In this section we discuss briefly how background stars can affect the interpretation of the results of a visual binary survey.
We use the prescription derived in \cite{kouwenhoven2007a} for the number of background stars brighter than $K_S$ and with an angular separation smaller than $\rho$, as a function of $K_S$ and $\rho$:
\begin{equation} \label{equation: backgroundstarsformula}
  N(K_S,\rho) = C \cdot 10^{\gamma \cdot K_S} \cdot  A(\rho) \,,
\end{equation}
where $A(\rho)$ is the enclosed area in the field of view within a radius $\rho$, $\gamma=0.32\pm 0.01$~mag$^{-1}$, and $C$ is a constant. The $K_S$ dependency was derived by KO6 using the Besan\c{c}on model of the Galaxy \citep{besancon}, and the normalization constant $C$ is determined using the background star study of SHT. Table~\ref{table: backgroundstarcounts} lists for each of the three visual surveys (KO5, KO6, and SHT) the expected number of background stars {\em per field of view}. We list the number of background stars with $K_S < 12$~mag and with $K_S < 18$~mag for four different pointings in the \sco{} region.

\begin{figure}[!bt]
  \centering
  \includegraphics[width=0.5\textwidth,height=!]{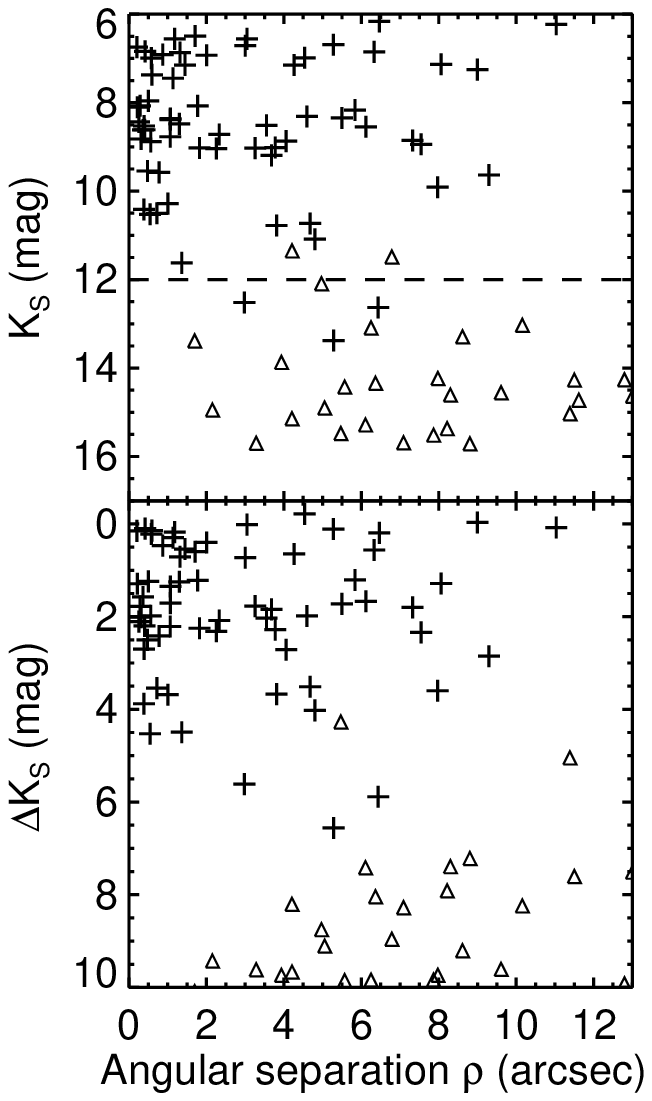}  \caption{The simulated KO5 data for a realization of the best-fitting model of \sco{} \citep[cf.][their Figure~3.8]{kouwenhoven2007a}. The plots show the distribution of physical companions (plusses) and background stars (triangles). The horizontal dashed line indicates the criterion used by KO5 (based on the analysis of SHT) to statistically separate companion stars and background stars. The simulated sample consists of 199~targets, next to which we detect 64~companion stars and 28~background stars. Although three companions have $K_S > 12$~mag, and two~background stars have $K_S <12$~mag in this example, the majority of the secondaries is correctly classified if $K_S=12$~mag is used to separate companions and background stars. 
\label{figure: detlim_plot_adonis} }
\end{figure}

For the $N$ targets in our model, we assume that $N/3$ targets are in each of the regions US and UCL, and that $N/6$ targets are in each of the regions LCC and the Galactic plane (GP). After the number of background stars $N_{\rm bg}$ is determined, each background star is assigned a $K_S$ magnitude randomly drawn from the generating distribution corresponding to Equation~\ref{equation: backgroundstarsformula}. Each of the $N_{\rm bg}$ background stars is assigned a random position in the field of one of the $N$ target stars. Finally, the angular separation and magnitude difference are calculated. For each background star we then decide whether it would be detected in the simulated observations, i.e., whether it satisfies the contrast constraint. Figure~\ref{figure: detlim_plot_adonis} shows the results for one of our models. Note how well the results in this figure resemble those of the KO5 observations in \cite{kouwenhoven2007a} in their Figure~8.

As the properties of the background star population are well described, and the parallax and age of each \sco{} member star is known, the colour and magnitude of each secondary can be used to determine whether it is a companion star or a background star. This method has been used in the analysis of the surveys of SHT, KO5, and KO6. 
SHT consider all new companions with $K_S > 12$~mag, $J > 13$~mag, or $J-K_S > 1$~mag as background stars. KO5 consider all new companions with $K_S>12$~mag as background stars. The expected number of background stars with $K_S < 12$~mag in the KO5 and SHT datasets can now be calculated. Among the 199~targets in the KO5 dataset, we expect $1-4$ background stars with $K_S < 12$~mag. In their follow-up study KO6 show that the $K_S=12$ criterion correctly classifies the companions and background stars in about 85\% of the cases. 
They performed multi-colour observations of several doubtful candidate companions, and identified six of these as possible background stars. Among the 87~confirmed \sco{} members in the SHT dataset we expect $\sim 1$ background star with  $K_S < 12$~mag.
The expected number of bright background stars is small, and many background stars have been removed by magnitude criteria and the follow-up study of KO6. The possible presence of background stars among the candidate binaries thus has a negligible effect on the conclusions of our analysis.

% ============================================================================
% ============================================================================
% ============================================================================
% ============================================================================

\subsection{Binary fraction versus primary mass} \label{section: bf_vs_m1}

Throughout this paper we have assumed that most binary parameters are mutually independent, and independent of the binary fraction as well. We have ignored the possible dependence of binary fraction, semi-major axis, eccentricity, and mass ratio on primary mass. However, due to the relatively small range of primary masses studied, these dependencies do not affect our results significantly. 

Several recent studies have focused on the very low-mass (VLM) and brown dwarf (BD) binaries in \sco{}, such as the survey for spectroscopic binaries in US and $\rho$~Oph by \cite{kurosawa2006}, and the near-infrared visual imaging survey among 12 VLM and BD members of \sco{} by \cite{kraus2005}. Surveys for binarity among the lowest mass stars and brown dwarfs find an {\em observed} binary fraction $\binfracobs \approx 20\%$ \citep{basri2006,burgasser2007}. This value is a lower limit to the {\em intrinsic} binary fraction among these targets due to the presence of undetected companions.
As the details of the binary formation mechanism are not known, we limit ourselves to a discussion of two possible relations between the binary fraction $\binfracm$ and the primary mass $M_1$: a linear and a logarithmic dependence. 

For simplicity, we adopt a binary fraction of 100\% for the most massive stars in \sco{}: $\binfrac(20\msun)=100\%$ and the observational constraint for the binary fraction in the low-mass regime: $\binfrac(0.02\msun) > \binfracobs(0.02\msun) \approx 20\%$. Adopting a linear relation between $\binfracobsm$ and $M_1$, we find that $\binfracm > \binfracobsm \approx 4 (M_1/\!\msun) + 20\%$. This prescription results in a strong dependence for high-mass stars: the binary fraction drops from $\binfrac(20\msun)=100\%$ for the earliest B~stars to $\binfrac(3\msun) > 32\%$ for the latest B~stars. If we assume, for example, that the {\em intrinsic} binary fraction for brown dwarfs is $50\%$, the binary fraction for the latest B~stars is $55\%$, and even lower for the A~stars. As our observations are dominated by targets of spectral type A ($1.5 \la M_1 \la 3\msun$), the linear prescription for $\binfracm$ can be excluded (unless the dependence is very weak). 

A logarithmic form for $\binfracm$ results in $\binfracm > 26.7\log (M_1/\!\msun) + 65.3\%$. Then stars of spectral type B have a binary fraction of $80-100\%$, and those of spectral type A have a binary fraction of $70-80\%$. Again, if we assume an {\em intrinsic} binary fraction for brown dwarfs of 50\%, the values are $85-100\%$ and $80-85\%$, respectively. The latter values are in better agreement with our observations, although they are rather low. 

A correlation between binary fraction and spectral type may be present for \sco{}, although the functional relation is presently unknown. Our derived value for $\binfrac$, however, is unlikely to be strongly affected by our choice of a binary fraction that is independent of spectral type, due to the relatively small range in mass of the binaries in our analysis. Finally, we wish to note that the derivation of the distributions $f_a(a)$, $f_P(P)$ and $f_q(q)$ are practically independent of our choice of $\binfracm$.

% ====================================================================
% ====================================================================
% ====================================================================
% ==INTRODUCTION======================================================
% ====================================================================
% ====================================================================
% ====================================================================

\section{Summary and outlook} \label{section: summary}

We have recovered the properties of the intermediate-mass binary population in the nearby \ob{} \sco{}, with the aim of finding the primordial binary population (which is present just after star formation). We have performed Monte Carlo simulations, and compared for each association model the {\em simulated observations} with the results of surveys for visual, spectroscopic, and astrometric binary systems in \sco{}. The main results of our study are the following:
\begin{itemize}
  \item[--] The current binary fraction among A and B stars in \sco{} is {\em at least} 70\% ($3\sigma$ confidence). The best agreement with the observations is obtained for models with a binary fraction of 100\%.
  \item[--] We constrain the orbital size distribution (which can be described using a semi-major axis distribution or a period distribution) using the observed angular separation distribution and the number of visual, spectroscopic, and astrometric binaries. The observations are consistent with \"{O}pik's law, i.e., $f_a(a) \propto a^{\gamma_a}$ with $\gamma_a \approx -1.0$, which corresponds to an orbital period distribution $f_P(P) \propto P^{-1}$. The log-normal period distribution found by \cite{duquennoy1991}, which corresponds to a log-normal semi-major axis distribution, is consistent with the observed angular separation distribution and visual binary fraction (if $\binfrac \approx 100\%$), but significantly underpredicts the number of spectroscopic binaries. The best-fitting models have intermediate properties, i.e., a period distribution with width  $\sigma_P > 2.3$.
  \item[--] Random pairing (RP) and primary-constrained random pairing (PCRP) from the mass distribution are excluded with high confidence. The pairing function of intermediate mass binaries in \sco{} is well-described by a mass ratio distribution of the form $f_q(q) \propto q^{\gamma_q}$ , with $\gamma_q \approx -0.4$.
  \item[--] \sco{} is a young \ob{} with a low stellar density. Stellar and binary evolution have practically only affected the binaries with O-type components. Dynamical evolution has only mildly affected the binary population. The current binary population of \sco{}, as described above, is expected to be very similar to the primordial binary population of \sco{}. 
\end{itemize}
Practically all intermediate mass stars in \sco{} are part of a binary or multiple system. Although the most massive stars in \sco{} have already evolved away from the main sequence, observations of younger populations suggest that practially all of these were binary or multiple \citep[e.g.][]{mason1998}. This implies that {\em multiplicity is a fundamental parameter in the star forming process}.

In this paper we have included the results of six major binarity surveys among \sco{} members. We have not included the smaller surveys and individual discoveries, as each of these has its specific, often poorly documented selection effects. Inclusion of these will provide a better description of the binary population in \sco{}. However, the results are likely to be similar to those described in this paper, as the six major binary surveys used contain the large majority of the known binaries in \sco{}.
Although previous observations \citep[e.g.][]{kouwenhoven2005,kouwenhoven2007a} have shown that triple and higher-order systems are present in \sco{}, we have neglected these systems here, because of the non-trivial comparison with the observations, the more complicated selection effects, and the very small number of known higher-order multiples. 

In our recovery of the current and primordial binary population in \sco{} we have made several assumptions and simplifications. Our main results are obtained using the visual binaries, and are shown to be consistent with the available spectroscopic and astrometric binaries. The modeling of the selection effects of the spectroscopic and astrometric binaries needs to be improved, so that the simulated observations can be directly compared to the observed binary parameter distributions, in order to accurately derive the eccentricity distribution and the possible correlation between the different binary parameters. 

The assumed independence of the binary parameters, as well as the properties of the low-mass binary population need to be addressed observationally. Due to selection effects a relatively small number of binaries is known among the low-mass members of \sco{}, making it difficult to derive the properties of these. 
Several studies for binarity among low-mass (candidate) members of \sco{} have been performed recently \citep[e.g.][]{bouy2006,kurosawa2006,kraus2007}. In the near future, after a further extension of the dataset, and a thorough membership study, it may be possible to characterize the binary population of \sco{} over the full range of spectral types.
This issue can be further and more accurately addressed using the results of the {\em Gaia} space mission \citep{perryman2001,turon2005}, which is a project of the European Space Agency, expected to be launched in 2011. {\em Gaia} will survey over a billion stars in our Galaxy and the Local Group, and will provide an enormous dataset of visual, eclipsing, spectroscopic, and astrometric binaries \citep{soderhjelm2005}. 
The membership and stellar content of nearby \obs{} can be accurately determined using the results of {\em Gaia}, down into the brown dwarf regime. The {\em Gaia} dataset will be homogeneous, and its selection effects can therefore be modeled in detail. The available dataset of binaries will be larger and more complete than any other binarity survey in Galactic star clusters and \obs{} thus far. 

The current binary population in \sco{} is a fossil record of the primordial binary population, as 
the young age and low stellar density of \sco{} guarantees that stellar evolution
has affected only a handful of the most massive binaries, and suggests that dynamical
evolution of the binary population has been modest. 
The latter statement needs to be verified using numerical simulations of evolving \obs{}. 
Whether the effect of dynamical
evolution has been negligible over the lifetime of \sco{} depends on
its initial conditions. If \sco{} was born as a low-density
association, similar to its present state, the binary population is expected to have changed only modestly due to dynamical evolution. On the other
hand, if the association has expanded significantly over the last
5--20~Myr, dynamical evolution may have been more prominent.

% ====================================================================
% ====================================================================
% ====================================================================
% ==INTRODUCTION======================================================
% ====================================================================
% ====================================================================
% ====================================================================

\begin{acknowledgements}
We wish to thank Hans Zinnecker, Ed van den Heuvel, Andrei Tokovinin and the referee (Anthony Moffat) for their supportive criticism, which helped to improve this paper. TK was supported by NWO under project number 614.041.006 and by PPARC/STFC under grant number PP/D002036/1. This research was supported by the Royal Netherlands Academy of Sciences (KNAW) and the Netherlands Research School for Astronomy (NOVA).
\end{acknowledgements}

\bibliographystyle{aa}
\bibliography{bibliography}

\appendix

% ============================================================================
% ============================================================================
% ============================================================================
% ============================================================================

% ====================================================================
% ====================================================================
% ====================================================================
% ==INTRODUCTION======================================================
% ====================================================================
% ====================================================================
% ====================================================================

\Online

\onecolumn

\section{Datasets used}

\setlength{\LTcapwidth}{1.0\textwidth}
\begin{longtable}{|lccc cr ccc|}
  \caption{Properties of the KO5 dataset that we use in our analysis. Note that not all candidate companions are listed here, as in our analysis we only consider single stars and binary systems. The members \object{HIP68532} and \object{HIP69113} (marked with a star) both have two companions with a similar separation, position angle, and brightness. In our analysis we consider these ``double companions'' as a single companion. \label{table: data_adonis}} \\
  \hline
  HIP & $K_{S,1}$ & $K_{S,2}$ & $\Delta K_S$ & $\rho$  & $\varphi$ & $M_1$     & $M_2$     & $q$ \\
      & mag       & mag       & mag          & arcsec  & deg       & $\msun$ & $\msun$ &     \\
  \endfirsthead
  \hline
  \multicolumn{9}{|l|}{\tablename\ \thetable{} -- continued from previous page} \\
  \hline
  HIP & $K_{S,1}$ & $K_{S,2}$ & $\Delta K_S$ & $\rho$  & $\varphi$ & $M_1$     & $M_2$     & $q$ \\
      & mag       & mag       & mag          & arcsec  & deg       & $\msun$ & $\msun$ &     \\
  \hline
  \endhead
  \hline
  \multicolumn{9}{|l|}{{Continued on next page}} \\ 
  \hline
  \endfoot
  \hline 
  \endlastfoot
  \hline
  \object{HIP50520}  &  6.23  &  6.39  &  0.16  &  2.51  &  313  &  2.12  &  1.98  &  0.93   \\
\object{HIP52357}  &  7.64  &  7.65  &  0.01  &  0.53  &  73  &  1.60  &  0.22  &   0.14  \\
\object{HIP56993}  &  7.38  &  11.88  &  4.50 &  1.68  &  23  &  1.90  &  0.180  &  0.09   \\
\object{HIP58416}  &  7.03  &  8.66  &  1.63  &  0.58  &  166  &  1.86  &  1.00  &  0.54   \\
\object{HIP59413}  &  7.46  &  8.18  &  0.72  &  3.18  &  100  &  1.62  &  1.34  &  0.83   \\
\object{HIP59502}  &  6.87  &  11.64  &  4.77  &  2.94  &  26  &  1.80  &  0.14  &  0.08   \\
\object{HIP60084}  &  7.65  &  10.10  &  2.45  &  0.46  &  330  &  1.66  &  0.62  & 0.37   \\
\object{HIP61265}  &  7.46  &  11.38  &  3.92  &  2.51  &  67  &  1.82  &  0.27  &  0.15   \\
\object{HIP61639}  &  6.94  &  7.06  &  0.12  &  1.87  &  182  &  1.82  &  1.74  &  0.96   \\
\object{HIP61796}  &  6.37  &  11.79  &  5.42  &  9.89  &  109  &  2.46  &  0.14  & 0.06   \\
\object{HIP62002}  &  7.09  &  7.65  &  0.56  &  0.38  &  69  &  1.68  &  1.20  &   0.71   \\
\object{HIP62026}  &  6.31  &  7.86  &  1.55  &  0.23  &  6  &  2.45  &  1.19  &    0.49   \\
\object{HIP62179}  &  7.20  &  7.57  &  0.37  &  0.23  &  283  &  1.84  &  1.56  &  0.85   \\
\object{HIP64515}  &  6.78  &  6.94  &  0.16  &  0.31  &  166  &  1.96  &  1.84  &  0.94   \\
\object{HIP65822}  &  6.68  &  11.08  &  4.40 &  1.82  &  304  &  2.91  &  0.38  &  0.13   \\
\object{HIP67260}  &  6.98  &  8.36  &  1.38  &  0.42  &  229  &  2.00  &  1.10  &  0.55   \\
\object{HIP67919}  &  6.59  &  9.10  &  2.51  &  0.69  &  297  &  1.97  &  0.75  &  0.38   \\
\object{HIP68080}  &  6.28  &  7.19  &  0.91  &  1.92  &  10  &  2.91  &  1.92  &   0.66   \\
\object{HIP68532}$^\star$  &  7.02  & 9.54   &  2.53  &  3.05  &  289  &  1.95  &  1.12  &   0.57   \\
\object{HIP68867}  &  7.17  &  11.61  &  4.44  &  2.16  &  285  &  2.18  &  0.24  & 0.11   \\
\object{HIP69113}$^\star$  &  6.37  & 10.29  &  3.92  &  5.34  &  65  &  3.87  &  1.49  &    0.39   \\
\object{HIP69749}  &  6.62  &  11.60  &  4.98  &  1.50  &  1  &  3.81  &  0.38  &   0.10   \\
\object{HIP70998}  &  7.06  &  10.83  &  3.77  &  1.17  &  355  &  2.54  &  0.48  & 0.19   \\
\object{HIP71724}  &  6.79  &  9.70  &  2.91  &  8.66  &  23  &  2.62  &  0.82  &   0.31   \\
\object{HIP71727}  &  6.89  &  7.80  &  0.91  &  9.14  &  245  &  2.46  &  1.64  &  0.67  \\
\object{HIP72940}  &  6.85  &  8.57  &  1.72  &  3.16  &  222  &  1.82  &  0.96  &  0.53   \\
\object{HIP72984}  &  7.05  &  8.50  &  1.45  &  4.71  &  260  &  1.90  &  1.06  &  0.56   \\
\object{HIP74066}  &  6.08  &  8.43  &  2.35  &  1.22  &  110  &  2.68  &  1.02  &  0.38   \\
\object{HIP74479}  &  6.31  &  10.83  &  4.52  &  4.65  &  154  &  3.03  &  0.38  & 0.13   \\
\object{HIP75056}  &  7.31  &  11.17  &  3.86  &  5.19  &  35  &  1.92  &  0.30  &  0.16   \\
\object{HIP75151}  &  6.65  &  8.09  &  1.44  &  5.70  &  121  &  3.19  &  1.64  &  0.51   \\
\object{HIP75915}  &  6.44  &  8.15  &  1.71  &  5.60  &  229  &  2.89  &  1.22  &  0.42   \\
\object{HIP76001}  &  7.60  &  7.80  &  0.20  &  0.25  &  3  &  1.54  &  1.36  &     0.88   \\
\object{HIP76071}  &  7.06  &  10.87  &  3.81  &  0.69  &  41  &  2.70  &  0.23  &  0.09   \\
\object{HIP77315}  &  7.24  &  7.92  &  0.68  &  0.68  &  67  &  2.08  &  1.56  &   0.75  \\
\object{HIP77911}  &  6.68  &  11.84  &  5.16  &  7.96  &  279  &  2.80  &  0.09  & 0.03   \\
\object{HIP77939}  &  6.56  &  8.09  &  1.53  &  0.52  &  119  &  3.85  &  1.82  &  0.47   \\
\object{HIP78756}  &  7.16  &  9.52  &  2.36  &  8.63  &  216  &  2.30  &  0.92  &  0.40 \\
\object{HIP78809}  &  7.51  &  10.26  &  2.75  &  1.18  &  26  &  2.03  &  0.30  &  0.15   \\
\object{HIP78847}  &  7.32  &  11.30  &  3.98  &  8.95  &  164  &  2.20  &  0.160  &0.07   \\
\object{HIP78853}  &  7.50  &  8.45  &  0.95  &  1.99  &  270  &  1.82  &  1.14  &  0.63  \\
\object{HIP78956}  &  7.57  &  9.04  &  1.47  &  1.02  &  49  &  2.40  &  1.16  &   0.48   \\
\object{HIP79124}  &  7.13  &  10.38  &  3.25  &  1.02  &  96  &  2.48  &  0.33  &  0.13   \\
\object{HIP79156}  &  7.61  &  10.77  &  3.16  &  0.89  &  59  &  2.09  &  0.27  &  0.13   \\
\object{HIP79250}  &  7.49  &  10.71  &  3.22  &  0.62  &  181  &  1.42  &  0.140  &0.10   \\
\object{HIP79530}  &  6.60  &  8.34  &  1.74  &  1.69  &  220  &  3.73  &  1.58  &  0.42   \\
\object{HIP79631}  &  7.17  &  7.61  &  0.44  &  2.94  &  128  &  1.90  &  1.58  &  0.83   \\
\object{HIP79739}  &  7.08  &  11.23  &  4.15  &  0.96  &  118  &  2.32  &  0.16  & 0.07   \\
\object{HIP79771}  &  7.10  &  10.89  &  3.79  &  3.67  &  313  &  2.14  &  0.19  & 0.09   \\
\object{HIP80238}  &  7.34  &  7.49  &  0.15  &  1.03  &  318  &  1.94  &  1.67  &  0.86   \\
\object{HIP80324}  &  7.33  &  7.52  &  0.19  &  6.23  &  152  &  1.70  &  1.54  &  0.91   \\
\object{HIP80371}  &  6.40  &  8.92  &  2.52  &  2.73  &  141  &  3.43  &  0.94  &  0.27   \\
\object{HIP80425}  &  7.40  &  8.63  &  1.23  &  0.60  &  156  &  2.08  &  1.16  &  0.56   \\
\object{HIP80461}  &  5.92  &  7.09  &  1.17  &  0.27  &  286  &  5.29  &  2.97  &  0.56   \\
\object{HIP80799}  &  7.45  &  9.80  &  2.35  &  2.94  &  205  &  1.86  &  0.34  &  0.18   \\
\object{HIP80896}  &  7.44  &  10.33  &  2.89  &  2.28  &  177  &  1.81  &  0.24  & 0.13   \\
\object{HIP81624}  &  5.80  &  7.95  &  2.15  &  1.13  &  224  &  6.53  &  2.30  &  0.35   \\
\object{HIP81972}  &  5.87  &  11.77  &  5.90 &  5.04  &  213  &  4.92  &  0.35  &  0.07   \\
\object{HIP83542}  &  5.38  &  9.90  &  4.52  &  8.86  &  196  &  1.10  &  0.91  &  0.83   \\
\object{HIP83693}  &  5.69  &  9.26  &  3.57  &  5.82  &  78  &  4.95  &  1.06  &   0.21   \\

\end{longtable}

\begin{longtable}{|lccc cr ccc|}
  \caption{Properties of the KO6 dataset that we use in our analysis. Note that not all candidate companions are listed here, as in our analysis we only consider single stars and binary systems. The members \object{HIP68532} and \object{HIP69113} (marked with a star) both have two companions with a similar separation, position angle, and brightness. In our analysis we consider these ``double companions'' as a single companion. \label{table: data_naco}} \\
  \hline
  HIP & $K_{S,1}$ & $K_{S,2}$ & $\Delta K_S$ & $\rho$  & $\varphi$ & $M_1$     & $M_2$     & $q$ \\
      & mag       & mag       & mag          & arcsec  & deg       & $\msun$ & $\msun$ &     \\
  %HIP & $K_{S,1}$ (mag) & $K_{S,2}$ (mag) & $\Delta K_S$ (mag) & $\rho$ ($''$) & $\varphi$ ($^\circ$) & $M_1$ (\msun{}) & $M_2$ (\msun{}) & $q$ \\
  \endfirsthead
  \hline
  \multicolumn{9}{|l|}{\tablename\ \thetable{} -- continued from previous page} \\
  \hline
  HIP & $K_{S,1}$ & $K_{S,2}$ & $\Delta K_S$ & $\rho$  & $\varphi$ & $M_1$     & $M_2$     & $q$ \\
      & mag       & mag       & mag          & arcsec  & deg       & $\msun$ & $\msun$ &     \\
  %HIP & $K_{S,1}$ (mag) & $K_{S,2}$ (mag) & $\Delta K_S$ (mag) & $\rho$ ($''$) & $\varphi$ ($^\circ$) & $M_1$ (\msun{}) & $M_2$ (\msun{}) & $q$ \\
  \hline
  \endhead
  \hline
  \multicolumn{9}{|l|}{{Continued on next page}} \\ 
  \hline
  \endfoot
  \hline 
  \endlastfoot
  \hline
  \object{HIP59502} & 6.87 & 11.64 & 4.77 & 2.935 & 26 & 1.80 & 0.14 & 0.08   \\
\object{HIP60851} & 6.06 & 13.69 & 7.63 & 8.159 & 231& 2.63 & 0.04 & 0.02   \\
\object{HIP61265} & 7.46 & 11.38 & 3.93 & 2.505 & 67 & 1.82 & 0.27 & 0.15   \\
\object{HIP62026} & 6.31 & 7.86  & 1.55 & 0.232 & 6  & 2.45 & 1.19 & 0.49   \\
\object{HIP63204} & 6.78 & 8.40  & 1.62 & 0.153 & 237& 2.05 & 1.06 & 0.52   \\
\object{HIP67260} & 6.98 & 8.36  & 1.38  & 0.423 & 229& 2.00 & 1.10 & 0.55  \\
\object{HIP67919} & 6.59 & 9.10  & 2.51 & 0.685 & 297& 1.97 & 0.75 & 0.38  \\
\object{HIP68532}$^\star$ & 7.02 & 9.54  & 2.53 & 3.052 & 288& 1.95 & 1.12 & 0.57  \\
\object{HIP69113}$^\star$ & 6.37 & 10.29 & 3.92  & 5.344 & 65 & 3.87 & 1.49 & 0.39   \\
\object{HIP73937} & 6.23 & 8.37  & 2.14 & 0.242 & 191& 2.94 & 1.11 & 0.38   \\
\object{HIP78968} & 7.42 & 14.26 & 6.84 & 2.776 & 322& 2.33 & 0.02 & 0.01   \\
\object{HIP79739} & 7.08 & 11.23 & 4.15 & 0.959 & 118& 2.32 & 0.16 & 0.07   \\
\object{HIP79771} & 7.10 & 11.42 & 4.33 & 0.435 & 129& 2.14 & 0.19 & 0.09   \\
\object{HIP80799} & 7.45 & 9.80  & 2.35 & 2.940 & 205& 1.86 & 0.34 & 0.18   \\
\object{HIP80896} & 7.44 & 10.33 & 2.89 & 2.278 & 177& 1.81 & 0.24 & 0.13   \\
\object{HIP81949} & 7.33 & 15.52 & 8.19 & 5.269 & 341& 2.26 & 0.02 & 0.01   \\
\object{HIP81972} & 5.87 & 11.77 & 5.90 & 5.040 & 213& 4.92 & 0.35 & 0.07   \\
\object{HIP83542} & 5.38 & 9.90  & 4.52 & 8.864 & 196& 1.10 & 0.91 & 0.83   \\

\end{longtable}

\begin{table}[!tbhp]
  \centering
  \caption{Properties of the SHT dataset that we use in our analysis. Note that not all candidate companions are listed here, as in our analysis we only consider single stars and binary systems. The six members at the bottom of the list were not explicitly observed by SHT. Due to the presence of (known) close companions these were not suitable for wavefront sensing. We have included these targets for our analysis to avoid a bias towards low binarity. \label{table: data_tokovinin}}
  \begin{tabular}{|lccc cr ccc|}
    \hline
    \hline
  HIP & $K_{S,1}$ & $K_{S,2}$ & $\Delta K_S$ & $\rho$  & $\varphi$ & $M_1$     & $M_2$     & $q$ \\
      & mag       & mag       & mag          & arcsec  & deg       & $\msun$ & $\msun$ &     \\
  %HIP & $K_{S,1}$ (mag) & $K_{S,2}$ (mag) & $\Delta K_S$ (mag) & $\rho$ ($``$)& $\varphi$ ($^\circ$)& $M_1$ (\msun{}) & $M_2$ (\msun{}) & $q$ \\
    \hline
    \object{HIP55425}  &  4.66  &  5.86  &  1.20  &  0.354  &  144  &  4.65  &  2.70  &  0.58   \\
\object{HIP56561}  &  3.17  &  6.81  &  3.64  &  0.734  &  135  &  8.30  &  2.22  &  0.27   \\
\object{HIP58884}  &  5.67  &  7.00  &  1.33  &  0.698  &  158  &  3.17  &  1.75  &  0.55   \\
\object{HIP61585}  &  3.41  &  10.94 &  7.53  &  4.853  &  198  &  6.30  &  0.19  &  0.03   \\
\object{HIP63945}  &  5.80  &  9.16  &  3.36  &  1.551  &  268  &  3.60  &  0.135 &  0.04   \\
\object{HIP65271}  &  5.12  &  7.03  &  1.91  &  0.164  &  135  &  4.25  &  1.80  &  0.42   \\
\object{HIP67472}  &  3.97  &  10.06 &  6.09  &  4.637  &  304  &  7.95  &  0.75  &  0.09   \\
\object{HIP72683}  &  5.27  &  6.84  &  1.57  &  0.099  &  86   &  4.52  &  2.17  &  0.48   \\
\object{HIP72800}  &  5.54  &  9.43  &  3.89  &  1.046  &  161  &  3.75  &  0.76  &  0.20   \\
\object{HIP73334}  &  4.09  &  5.46  &  1.37  &  0.128  &  156  &  7.83  &  5.33  &  0.68   \\
\object{HIP75264}  &  4.28  &  5.55  &  1.27  &  0.279  &  149  &  7.25  &  4.84  &  0.67  \\
\object{HIP76945}  &  5.79  &  9.47  &  3.68  &  0.507  &  133  &  3.50  &  0.80  &  0.23   \\
\object{HIP77939}  &  6.13  &  7.78  &  1.65  &  0.524  &  120  &  4.95  &  2.16  &  0.44   \\
\object{HIP78820}  &  3.86  &  6.80  &  2.94  &  0.292  &  171  &  11.20 &  2.98  &  0.27   \\
\object{HIP79374}  &  4.20  &  5.14  &  0.94  &  1.334  &  2    &  8.32  &  5.47  &  0.66   \\
\object{HIP79530}  &  6.31  &  8.07  &  1.76  &  1.693  &  220  &  3.23  &  1.35  &  0.42   \\
\object{HIP80112}  &  2.61  &  4.77  &  2.16  &  0.469  &  244  &  19.96 & 10.40  &  0.52   \\
\hline
\object{HIP57851}  &  ---    &  ---    &  ---     &  1.549  &  158  &  4.15  &  1.83  &  0.44   \\
\object{HIP62322}  &  ---    &  ---    &  ---     &  1.206  &  35   &  7.35  &  6.40  &  0.87  \\
\object{HIP64425}  &  ---    &  ---    &  ---     &  0.185  &  7    &  4.07  &  3.10  &  0.76   \\
\object{HIP74117}  &  ---    &  ---    &  ---     &  0.193  &  210  &  6.49  &  4.94  &  0.76   \\
\object{HIP76371}  &  ---    &  ---    &  ---     &  2.150  &  8    &  5.75  &  2.79  &  0.49   \\
\object{HIP77840}  &  ---    &  ---    &  ---     &  2.162  &  270  &  6.05  &  2.39  &  0.40   \\

    \hline
    \hline
  \end{tabular}
\end{table}

\begin{table}
   \centering
   \caption{The LEV dataset used in this sample, consisting of 16~binaries with orbital elements (SB1 or SB2), and 23 radial velocity variables (RVV), for which no orbital elements are available. {\em Left}: the 16~spectroscopic binaries with orbital elements among the confirmed members of \sco{}, in the LEV dataset. LEV observed 53~confirmed members of \sco{}, of which 8~SB1s, 8~SB2s, 23~RVVs, and 14~targets with a constant radial velocity. {\em Right}: the 23 RVVs. Note that several of the RVVs may not be spectroscopic binaries, as radial velocity variation may also be caused by line profile variability. \label{table: data_levato}} 
  \begin{tabular}{cc}
    \begin{tabular}{lc cr l}
      \hline
      \hline
      HIP & $P$  & $e$ & \multicolumn{1}{c}{$\omega$} & Group \\
          & days &     & \multicolumn{1}{c}{deg}      &       \\
      \hline
      \object{HIP67464}	& 2.6253 & 0.13 & 222 & UCL \\
\object{HIP75647}	& 3.8275 & 0.25 & 22 & UCL \\
\object{HIP76297}	& 2.8081 & 0.10 & 97 & ULC \\
\object{HIP76503}	& 5.2766 & 0.33 & 86 & US \\
\object{HIP76600}	& 3.2907 & 0.28 & 114 & UCL \\
\object{HIP76945}        & 12.26 & 0.19 & 83 & UCL \\
\object{HIP77858}	& 1.9235 & 0.36 & 309 & US \\
\object{HIP77911}	& 1.264 & 0.61 & 330 & US \\
\object{HIP78104}	& 4.0031 & 0.27 & 231 & US \\
\object{HIP78168}	& 10.0535 & 0.58 & 340 & US \\
\object{HIP78265}	& 1.5701 & 0.15 & 25 & US \\
\object{HIP78820}	& 6.8281 & 0.28 & 38 & US \\
\object{HIP79404}	& 5.7805 & 0.19 & 115 & US \\
\object{HIP79374}	& 5.5521 & 0.11 & 267 & US \\
\object{HIP80112}	& 34.23 & 0.36 & 308 & US \\
\object{HIP80569}	& 138.8 & 0.44 & 325 & US \\

      \hline
      \hline
    \end{tabular}
    &
    \begin{tabular}{ll}
      \hline 
      \hline
      \multicolumn{2}{l}{Radial velocity variables}\\
      \\
      \hline
      \object{HIP67472} & 
      \object{HIP68245} \\ 
      \object{HIP68862} & 
      \object{HIP70300} \\ 
      \object{HIP73334} & 
      \object{HIP74100} \\ 
      \object{HIP75141} & 
      \object{HIP75304} \\ 
      \object{HIP76633} & 
      \object{HIP77635} \\ 
      \object{HIP77900} & 
      \object{HIP77939} \\ 
      \object{HIP78246} & 
      \object{HIP78384} \\ 
      \object{HIP78530} & 
      \object{HIP78549} \\ 
      \object{HIP78655} & 
      \object{HIP79031} \\ 
      \object{HIP79374} & 
      \object{HIP79739} \\ 
      \object{HIP80024} & 
      \object{HIP81266} \\ 
      \object{HIP82545} &  \\ 
      \\
      \\
      \\
      \\
      \hline
      \hline
    \end{tabular}  
    \\
  \end{tabular}
\end{table}

\end{document}